\documentclass[sigconf]{acmart}

\AtBeginDocument{%
  }

\copyrightyear{2026}
\acmYear{2026}
\setcopyright{cc}
\setcctype{by}
\acmConference[CHI '26]{Proceedings of the 2026 CHI Conference on Human Factors in Computing Systems}{April 13--17, 2026}{Barcelona, Spain}
\acmBooktitle{Proceedings of the 2026 CHI Conference on Human Factors in Computing Systems (CHI '26), April 13--17, 2026, Barcelona, Spain}
\acmDOI{10.1145/3772318.3791041}
\acmISBN{979-8-4007-2278-3/2026/04}

\acmSubmissionID{2464}

\usepackage{acmart-taps}

\usepackage{balance}
\usepackage{multirow}
\usepackage{enumitem}
\usepackage{bm}
\usepackage{amsmath}
\usepackage{makecell}
\usepackage{xparse}
\usepackage{soul}
\soulregister\cite7 
\soulregister\ref7
\soulregister\todo7 
\soulregister\toreview7 
\soulregister\rev7 
\soulregister\bad7 

\definecolor{actualColor}{HTML}{2e8113}   
\definecolor{adjustedColor}{HTML}{69a057} 
\definecolor{predColor}{HTML}{a83279}     

\newcommand{\todo}[1]{{\color{red} [#1]}}

\definecolor{darkred}{RGB}{220, 0, 0}
\newcommand{\ns}[1]{{\color{darkred}#1}}

\newcommand{\hlc}[2][yellow]{{\sethlcolor{#1}\hl{#2}}}
\definecolor{revision_color}{HTML}{FFFFFF}
\newcommand{\rev}[1]{#1}
\newcommand{\toreview}[1]{\hlc[yellow]{#1}}
\newcommand{\bad}[1]{\hlc[red]{#1}}

\newcommand{\trackmode}{1}

\usepackage{xcolor}
\usepackage{colortbl}

\newcommand{\actlabel}[1]{%
  \if\trackmode1
    \textcolor{actualColor}{#1}%
  \else
    #1%
  \fi
}
\newcommand{\adjlabel}[1]{%
  \if\trackmode1
    \textcolor{adjustedColor}{#1}%
  \else
    #1%
  \fi
}
\newcommand{\predlabel}[1]{%
  \if\trackmode1
    \textcolor{predColor}{#1}%
  \else
    #1%
  \fi
}

\newlist{enumerate-desiderata-counterfactual}{enumerate}{1}
\setlist[enumerate-desiderata-counterfactual,1]{
  label         = C\arabic*.,
  leftmargin    = 3.0em,                   
  labelsep      = 0.5em,                  
  itemsep       = 0.0\baselineskip     
}
\newlist{enumerate-desiderata-trace}{enumerate}{1}
\setlist[enumerate-desiderata-trace]{
  label         = T\arabic*.,
  leftmargin    = 3.0em,                   
  labelsep      = 0.5em,                  
  itemsep       = 0.0\baselineskip     
}

\begin{document}

\title{Comparables XAI: Faithful Example-based AI Explanations with Counterfactual Trace Adjustments}

\author{Yifan Zhang}
\affiliation{%
  \department{Department of Computer Science}
  \institution{National University of Singapore}
  \city{Singapore}
  \country{Singapore}
  }
\email{yifan.zhang\_@u.nus.edu}

\author{Tianle Ren}
\affiliation{%
  \department{Department of Computer Science}
  \institution{National University of Singapore}
  \city{Singapore}
  \country{Singapore}
}
\email{ren.tianle@u.nus.edu}

\author{Fei Wang}
\affiliation{%
  \department{Department of Computer Science}
  \institution{National University of Singapore}
  \city{Singapore}
  \country{Singapore}
}
\email{wang-fei@nus.edu.sg}

\author{Brian Y Lim}
\authornote{Corresponding author}
\affiliation{%
  \department{Department of Computer Science}
  \institution{National University of Singapore}
  \city{Singapore}
  \country{Singapore}
}
\email{brianlim@nus.edu.sg}

\renewcommand{\shortauthors}{Zhang, Ren, Wang, and Lim}

\begin{abstract}
Explaining with examples is an intuitive way to justify AI decisions. However, it is challenging to understand how a decision value should change relative to the examples with many features differing by large amounts. We draw from real estate valuation that uses Comparables—examples with known values for comparison. Estimates are made more accurate by hypothetically adjusting the attributes of each Comparable and correspondingly changing the value based on factors. We propose \textit{Comparables XAI} for relatable example-based explanations of AI with Trace adjustments that trace counterfactual changes from each Comparable to the Subject, one attribute at a time, monotonically along the AI feature space. In modelling and user studies, Trace-adjusted Comparables achieved the highest XAI faithfulness and precision, user accuracy, and narrowest uncertainty bounds compared to linear regression, linearly adjusted Comparables, or unadjusted Comparables. This work contributes a new analytical basis for using example-based explanations to improve user understanding of AI decisions.

\end{abstract}

\ccsdesc[500]{Human-centered computing~Empirical studies in HCI}
\ccsdesc[500]{Computing methodologies~Artificial intelligence}

\keywords{Explainable AI, Example-based Explanations, Comparable with Adjustments, Counterfactual explanations}

\maketitle

\section{Introduction}
As artificial intelligence (AI) has become increasingly prevalent~\cite{rane2023explainablehealthcare, rane2023explainablefinancial, scherer2019artificial}, the demand for explainability has grown correspondingly~\cite{EU_AI_Act_2024} to foster user understanding and adoption of AI~\cite{kocielnik2019will, tsai2021exploring,zhang2022debiased, deck2023critical}.  
While various XAI methods have shown promise in justifying AI decisions, most remain too technical and thus of limited practicality for non-technical users~\cite{van2021evaluating, Jesus2021How, Buçinca2020Proxy, kim2023help, ehsan2024xai, wang2025good}.
Providing explanations with examples is considered simple, intuitive, and thus preferred by non-technical users~\cite{darias2024empirical, cai2019effects}. 
However, if available examples differ too much, they cannot faithfully represent or explain how an AI would decide on the target case~\cite{chen2023understanding}.
This lack of faithfulness will hinder the user understanding and trust in the example-based explanations~\cite{papenmeier2022s}.

To address this gap, we propose to bridge example-based explanations to the target instance with counterfactual adjustments of attribute differences.
Inspired by real estate valuation, where appraisers estimate the price of a target property (Subject) by applying systematic adjustments to the known prices of similar properties (Comparables) according to their attribute differences and price factors~\cite{Yeh2018Building, Kang1991An}. 
For example, when estimating the price of a Subject house that has a Living Area larger by 1.5 ksqft but build quality Grade lower by 2 points than a Comparable, the appraiser adjusts the Comparable's price by
\$135k for the larger Living Area (at \$70k/ksqft) and 
$-$\$60k for the lower Grade (at \$30k/grade), 
to a combined adjustment of +\$65k.
The adjusted Comparable will be an estimate for the Subject's price.
This analytical \textit{Comparables with adjustments} method improves the accuracy of final estimates, making them transparent as well as convincing.

\begin{figure}
    \centering
    \includegraphics[width=7.2cm]{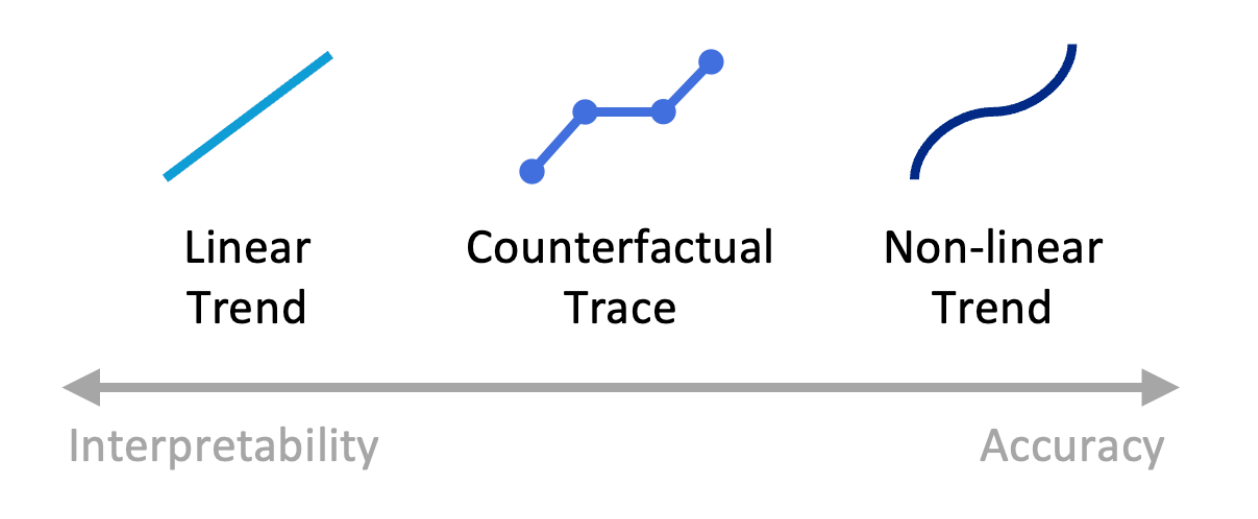}
    \caption{
        Trend-based explanations along the accuracy-interpretability trade-off.
        Linear Trends provide simple, highly interpretable relationships, but least accurate or faithful.
        Non-Linear Trends are more accurate, but more complex and less interpretable. 
        Counterfactual Traces strike a balance with a step-by-step, incremental trace of non-linear trends.
    }
    \label{fig:trade_off_general}
    \Description{
    The figure illustrates three types of trends arranged from left to right along an interpretability–accuracy spectrum. A horizontal axis runs beneath the images, with Interpretability labeled on the left and Accuracy labeled on the right.
    On the left, a Linear Trend is depicted as a simple straight line rising upward at a diagonal. It represents highly interpretable but less flexible relationships.
    In the center, a Counterfactual Trace is shown as a stepwise connected path with circular nodes, symbolizing a piecewise approximation that captures local non-linear behavior while remaining interpretable.
    On the right, a Non-Linear Trend is drawn as a smooth, curved line bending upward, representing more accurate but less transparent model behavior.
    Overall, the visual communicates that linear trends maximize interpretability, non-linear trends maximize accuracy, and counterfactual traces sit in the middle by offering locally faithful non-linear explanations in an interpretable, stepwise form.
    }
\end{figure}

Even so, these adjustments can still be inaccurate if the Comparables are very different due to a lack of available similar ones, because the adjustments are done linearly, but the trends are non-linear.
Instead, we propose \textbf{Comparables XAI} to provide more faithful explanations of adjusted Comparables.
Rather than making all multivariate adjustments at once from the Comparables, 
it uses the decision surface learned by the AI model and traces a path of counterfactuals, incrementally adjusting from the Comparable to the Subject.
Users can examine small changes between each counterfactual step, thus limiting cognitive load, and improving trust.
We modeled this trace as a piecewise linear function that traverses the feature space between the Comparable and Subject.
Hence, this Comparables Trace is more accurate than linear trends and more interpretable than non-linear trends, striking a balance along the accuracy-interpretability trade-off~\cite{murdoch2019definitions,arrieta2020explainable} (Fig.~\ref{fig:trade_off_general}).
To further improve the usability of the trace, we proposed and implemented desiderata to balance accuracy (faithfulness) and interpretability (sparsity, disjointness, monotonicity, and evenness).

We conducted a modeling study with sensitivity analysis to understand the accuracy and performance of Trace Adjustments.
In a formative user study, we examined how participants used, understood, or struggled with Comparables with Trace Adjustments and other 
baseline methods of Comparables Only, Comparables with Linear Regression, and Comparables with Linear Adjustments.
Subsequently, we conducted a summative user study to evaluate its usefulness.
In all experiments, we found that the Comparables with Trace Adjustments outperforms other baselines,
achieving the highest XAI faithfulness and precision, user accuracy, and the narrowest uncertainty bounds.
In summary, our \textbf{contributions} toward new HCI artifacts~\cite{wobbrock2016research} are:
\begin{itemize}
    \item \textbf{Comparables XAI} as an explanation paradigm, inspired from valuation, for users to analytically compare examples for AI decision understanding.
    \item \textbf{Comparables with Trace Adjustments} as a technical approach to provide incremental counterfactuals to improve faithfulness in example-based and counterfactual XAI.
\end{itemize}
We conclude with a discussion on the generalization, context, and limitations of our work.

\section{Background and Related Work}
We review explainable AI techniques, highlighting the intuitiveness and accessibility of example-based explanations, noting their limitations, which can be addressed with counterfactual explanations.

\subsection{Many AI Explanations are Too Technical}
Explainable AI (XAI) is designed to enhance the transparency and trustworthiness of AI systems~\cite{Vilone2021Classification}.
Focusing on tabular data, many techniques have been developed to attribute important features~\cite{ribeiro2016should, lundberg2017unified}, show non-linear partial dependence plots~\cite{lou2012intelligible, abdul2020cogam, friedman2001greedy}, and explain with logical rules~\cite{lakkaraju2016interpretable, ribeiro2018anchors}.
Although they satisfy users' need for explainability~\cite{lim2009assessing, liao2020questioning, wang2019designing}, these techniques are very technical, requiring graph and visualization literacy, numeracy, and familiarity with logical symbolic structures.
They were developed primarily to support AI engineers and data scientists, not lay users who prefer simpler explanations~\cite{cai2019effects}.
Example-based XAI is naturally simple, making the explanation relatable to examples and easy to understand.

\subsection{Example-based Explanations are More Accessible But Imprecise} 
Example-based explanations let users comprehend AI decisions by examples of other cases.
Typical methods include explaining with prototypes~\cite{Rada2023A,Liu2009A,filho2023explainable}, nearest neighbors~\cite{poche2023natural}, and alternative, hypothetical counterfactual examples~\cite{wachter2017counterfactual, zhang2022towards}.

Such explanations have been considered intuitive and thus preferred by users, especially with less domain knowledge or less technical skill~\cite{darias2024empirical}.
They are mostly adopted for intuitive or perception tasks, such as sketches~\cite{cai2019effects}.
User studies found that:
example-based explanations improve human-AI collaboration over attribute-based explanations~\cite{chen2023understanding}, comparative examples reveal AI limitations more effectively than normative examples~\cite{cai2019effects},
interactive editing of neighboring examples improved user understanding of AI strengths and limitations~\cite{suresh2022intuitively}, and
counterfactual examples improve understanding of visual~\cite{bhattacharya2023directive} and auditory explanations~\cite{zhang2022towards}.

However, for complex instances with many semantic attributes (e.g., predicting house price using a bunch of attributes like size, position, age, etc.), the reasoning becomes more cognitively demanding.
Areas such as real estate valuation~\cite{aamodt1994case}, health technology assessment~\cite{phillippo2019population}, energy usage estimation~\cite{kim2022derivation} have encountered this challenge and adopt `a Comparable with Adjustment' paradigm. In real estate valuation, appraising a property by selecting similar ones and adjusting their prices to account for differences is an established strategy that supports analytical comparison. A similar strategy exists in case-based reasoning, where past cases are retrieved as references, and their solutions are adapted to fit the subject~\cite{aamodt1994case,Chua2001Case-based,sormo2005explanation}.

Inspired by the work from real estate valuation, we propose to explain by examples with adjustments in XAI to enhance intuitiveness. 
Notice that making adjustments is simulating counterfactual Comparables that are closer to the Subject, we review counterfactual explanations in the following section.

\subsection{Counterfactual Explanations are Analytical But Mostly for Recourse}
Counterfactual explanations have been well-studied in XAI research. 
They answer the ``what if'' question by making comparison between Subject and generated instances~\cite{bhattacharya2023directive}.
Since counterfactuals are hypothetical and imaginary, preferred ones can be chosen, based on various desiderata:
monotonicity~\cite{hamer2024simple}, diversity~\cite{mothilal2020explaining}, stability~\cite{artelt2021evaluating}, minimal edits~\cite{wachter2017counterfactual, wexler2019if}, etc. 
Most works focus on providing single counterfactuals or a set of them.
StEP~\cite{hamer2024simple} is similar to our technical approach by providing a sequential series of counterfactuals. 
However, it focuses on providing plausible recourse with small, actionable steps, whereas \textit{Comparables XAI} generates faithful traces as explanations. 
Such a difference results in different desiderata priorities.

Given their importance for actionable insights, the use of counterfactuals has been well-studied, finding that
machine-generated counterfactuals differ from human-generated ones~\cite{delaney2023counterfactual}, counterfactual and prefactual explanations were equally helpful for predictions and diagnoses~\cite{dai2022counterfactual}, and users better understood counterfactuals with categorical feature changes than with numeric features~\cite{warren2024categorical}.
Bove et al. presented plural counterfactuals with comparative cards to improve user understanding and satisfaction~\cite{bove2023investigating}.
We similarly handle multiple counterfactuals, but as a sequential trace for incremental understanding.
While these studies focus on recourse, in this work, we focus on improving value estimation using faithful counterfactual traces.

Although some interactive tools exist for counterfactual XAI (e.g., Google's What-If Tool~\cite{wexler2019if}), they focus on data scientists with advanced visual analytic skills to interpret scatter plots across many dimensions. We focus on exploring the use of sales comparison grids in XAI that present examples in more accessible tables or spreadsheets. 
ViCE provides a simpler user-centric visualization to show the minimal-change counterfactual to a sample with interactive sensitivity checks~\cite{gomez2020vice}, but this focuses on one counterfactual instead of multiple like we do.
\begin{figure*}[t]
    \centering
    \includegraphics[width=0.86\linewidth]{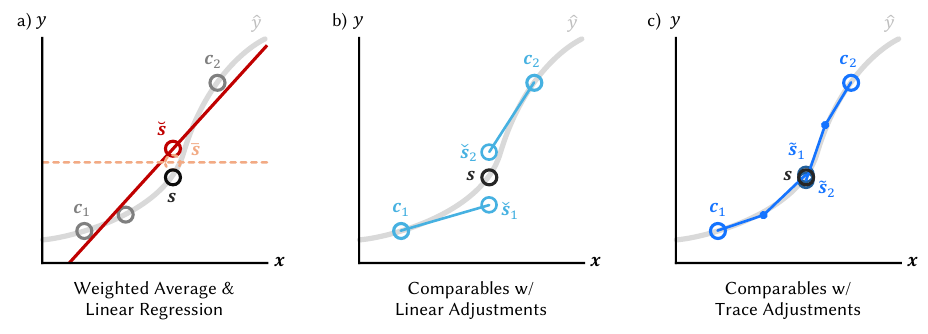}
    \caption{Conceptual examples of XAI types with univariate (1D) data shown for simplicity. 
    The background gray curve indicates the original AI system prediction $\hat{y}$, which is nonlinear with respect to input $x$. Given a subject case $s$ to be explained:
    a) Weighted averages the outcomes of Comparables ($c_1$, $c_2$, etc.), yielding an estimate $\bar{s}$ (dashed line), while the linear regression fits a single linear model across all Comparables (solid line), producing a prediction $\breve{s}$. 
    b) Comparables with Linear Adjustments fits a local linear regression between each Comparable and the subject case $s$, producing adjusted estimates such as $\check{s}_1$ and $\check{s}_2$. 
    c) Comparables with Trace Adjustments starts from each Comparable and stepwise adjustments are traced along the AI's underlying function ($\hat{y}$), yielding more consistent adjusted estimates $\tilde{s}_1$ and $\tilde{s}_2$ that more faithfully follow the nonlinear shape of the AI system.}
    \Description{
    The figure contains three conceptual diagrams, labeled a through c, showing different explanation types for a subject case. All panels have the same setup: the horizontal axis represents the input, the vertical axis represents the output, and a curved gray line shows the true AI system’s prediction, which is nonlinear. A black circle marks the subject case that needs explanation.
    a) Weighted Average and Linear Regression. Two comparable cases are shown as gray circles along the curved prediction line. Their outcomes are averaged to produce a dashed horizontal estimate. A red straight line is fitted across all comparables, representing a global linear regression model. The intersection of this straight line with the subject case provides a predicted estimate, which does not fully follow the curved shape of the AI system.
    b) Comparables with Linear Adjustments. Two comparable cases are again shown along the curved prediction line. Instead of fitting one global line, a separate local straight line is drawn between each comparable and the subject case. These local lines produce adjusted estimates for the subject, which better account for the comparables individually but still approximate the nonlinear curve only in straight segments.
    c) Comparables with Trace Adjustments. The same two comparables are shown. Starting from each comparable, small stepwise adjustments are traced along the underlying curved prediction line until they reach the subject case. This produces adjusted estimates that closely follow the true nonlinear curve of the AI system, resulting in more consistent and faithful estimates.
    }
    \label{fig:Conceptual_fig}
\end{figure*}

\section{Technical Approach}
In this section, we derive our explanatory approach starting with baseline methods to progressively increase the faithfulness and interpretability of example-based explanations.
\begin{enumerate}
    \item [1)]\textbf{Example-Based Explanations.} Assuming that similar cases should have similar values, in Section~\ref{sec:approach_comparables_only}, we can estimate the outcome of a Subject by \textit{averaging} the values of nearby Comparables. 
    \textbf{But} these examples may still be too different or irregularly dissimilar.
    
    \item [2)]\textbf{Comparables with Linear Regression.} Instead of flat averaging, in Section~\ref{sec:comparables_w_linear_reg}, we can fit a straight trend line (value by attributes) across the Comparables and estimate the Subject's value at its attributes position. 
    \textbf{But} the Comparables may still be too different, leading to an unreliable trend.
    
    \item [3)]\textbf{Comparables with Linear Adjustments.} Instead of referencing each Comparable directly, in Section~\ref{sec:comparable_w_linear_adjustments}, we can use linear interpolation to adjust the Comparable's value to get closer ``examples'', albeit hypothetical Counterfactual ones. 
    \textbf{But} such interpolation may be too simplistic for complex domains with non-linear trends.

    \item[4)]\textbf{Comparables with Trace Adjustments.} In Section~\ref{sec:comparable_w_trace_adjustments}, we trace the non-linear trend with step-by-step Counterfactual examples, until the last one has the same attributes as the Subject. Each Comparable has a final Counterfactual that we \textit{average} to estimate the Subject's value.
    \textbf{But} some methods to trace are more complex and less interpretable. 
    Hence, in Section~\ref{sec:desiderata}, we tune our tracing based on several \textbf{Desiderata}: faithfulness, sparsity, disjointness, monotonicity, and evenness.
\end{enumerate}
In the following subsections, we provide the mathematical details of our technical approach.

\subsection{Example-Based Explanations}
\label{sec:approach_comparables_only}

Consider an AI model $f$ 
that predicted $\hat{y}$ from an instance with attributes $\bm{x}$, i.e., $\hat{y} = f(\bm{x})$.
To determine if a specific prediction for a Subject instance $s$ is trustworthy, a user may seek examples.
An example-based explanation would first select a similar example (Comparable) $c$ and simply report its prediction $\hat{y}_c = f(\bm{x}_c)$, where $\bm{x}_c$ is the attributes of the Comparable.
The user would see that the attribute values are similar ($\bm{x}_s \approx \bm{x}_c$), and by the principle of similarity~\cite{tversky1977features, gentner1983structure} expect that the Subject would have a similar prediction value ($\hat{y}_s \approx \hat{y}_c$) or even actual value ($y_s \approx y_c$).

To improve confidence, multiple Comparables can be provided instead of one. However, how could one \textit{reconcile} a single value from a set of Comparables?
A basic way is to average among the examples, but one may want to emphasize Comparables that are closer to the Subject.
As shown with the horizontal dashed line in Fig.~\ref{fig:Conceptual_fig}, a more informed approach is a \textit{weighted average} based on similarity, i.e.,
\begin{equation}
    \label{eq:weighedavg}
    \bar{y}_{\text{s}} = \frac{\sum_{c} \rho_c \, y_c}{\sum_{c} \rho_c},
\end{equation}
where $\rho_c$ is the similarity of the $c$-th Comparable to the subject, and $y_c$ its actual value. We can accordingly compute this for prediction values too.
%
Next, we describe more advanced and accurate methods to estimate and reconcile the value of a Subject from Comparables.

\subsection{Comparables with Linear Regression}
\label{sec:comparables_w_linear_reg}
Another popular method to reconcile among multiple Comparables is to fit a \textit{linear regression} using the Comparables~\cite{rosen1974hedonic,palmquist1984estimating,case1987prices}.
In the context of local model agnostic XAI, this is similar to the approach in LIME~\cite{ribeiro2016should} to train a linear model from instances neighboring the Subject instance.
As shown with the solid red line in Fig.~\ref{fig:Conceptual_fig}, the linear regression to estimate the Subject's label value based on its attribute values $\bm{x}_s = (x^{(1)}, x^{(2)}, \dots)^\top$ is
\begin{equation}
    \breve{y}_s (\bm{x}_s) = \sum_{r} {w}^{(r)} \, x_{s}^{(r)} + b_s,
\end{equation}
 where ${x_s}^{(r)}$ is $r$-th attribute value of instance $s$, and ${w}^{(r)}$ is the corresponding weight, and $b_s$ the bias term.
A large weight means that a small change in attribute $x$ can significantly change the value, and positive (or negative) weights increase (or decrease) the value.

\subsection{Comparables with Adjustments} \label{sec:approach_comparables_with_Adjustments}
While Comparables provide useful references for estimation, they often differ much from the Subject in multiple attributes. 
Thus uncertainty persists with this estimation XAI Type.
For example, a Comparable house valued at \$500k has a smaller living area by 0.5 ksqft and is be closer by 3 miles to downtown than the Subject. 
But how much would the house be valued if it was larger by 0.5 ksqft and farther by 3 miles just like the Subject?
Valuators and real estate appraisers perform \textit{adjustments} to change each attribute of the house and adjust the corresponding price value based on linear factors{~\cite{appraisal_of_real_estate_15th_2020}.
This adjustment hypothesizes a \textit{counterfactual} Comparable that is similar in attributes to the Subject.
For example, if value changes by factors \$20k/ksqft of living area and $-$\$5k/mile from downtown, then the Comparable house will be adjusted by $\$20\text{k} \times 0.5 + (-\$5\text{k}) \times 3 = \$5\text{k}$, suggesting that the counterfactual Subject value of \$505k.

In practice, the linear factors are normally determined from market data or domain knowledge.
Although this information is complementary and essential, our approach uses the AI model only to estimate these factors.

\subsubsection{Comparables w/ Linear Adjustments}
\label{sec:comparable_w_linear_adjustments}
Specifically, for this baseline XAI Type, we use LIME~\cite{ribeiro2016should} to approximate a linear model near each Comparable instance based on sampling instances by sampling from the neighboring feature space of the AI model's predictions. The adjusted price of Comparable $c$ with linear adjustments to Subject $s$ is:
\begin{equation}
    \breve{y}_s (\bm{x}_s) = \sum_{r} w_c^{(r)} x_{s}^{(r)} + b_c,
\end{equation}
where $w_c^{(r)}$ and $b_c$ are the weights and bias, respectively, learned in the local linear model around Comparable $c$.
The adjustments are done for each Comparable separately, and each will estimate slightly different values for Subject $s$ after adjustment.
Fig. \ref{fig:Conceptual_fig}b shows that each \textit{Comparable w/ Linear Adjustment} estimator starts from its Comparable $c_i$ 
and extends to Subject $s$ to estimate its value.

\subsection{Comparables w/ Trace Adjustments}
\label{sec:comparable_w_trace_adjustments}
Linear adjustment assumes that the transition from Comparable to Subject is a straight path, but the values may change nonlinearly by attributes. 
This is especially so if Comparables are not very similar due to limited availability.
To improve faithfulness in the adjustment, we propose \textit{Trace Adjustments} to counterfactually adjust attributes along the nonlinear decision surface learned by the AI model.
Fig.~\ref{fig:Conceptual_fig}c shows how piecewise tracing allows counterfactual adjustments (blue) to be closer to the AI decision curve (grey).
To explain the adjustments incrementally, we model this curved trace as a piecewise linear function in the feature space. Each segment $\tau$ is defined with a linear equation, i.e.,
\begin{equation}
    \tilde{y}_{\tau}(\bm{x}) = \bm{w}_{\tau}^{\top} \bm{x} + b_{\tau},
\end{equation}
where $\bm{x}$ are the attributes of an instance within the $\tau$-th segment ($\bm{x} \in [\bm{\chi}_{\tau-1}, \bm{\chi}_\tau]$), and $\bm{w}_{\tau}$ and $b_{\tau}$ are the weights and bias of that linear segment. Fully expressed, the piecewise function for Trace Adjustments is:
\begin{equation}
    \tilde{y}(\bm{x}) = \sum_\tau \mathbb{I}(\bm{\chi}_{\tau-1} \le \bm{x} < \bm{\chi}_\tau) (\bm{w}_{\tau}^{\top} \bm{x} + b_{\tau}),
    \label{eq:traceadjustment}
\end{equation}
where $\mathbb{I}(P)$ is the indicator function which is 1 when $P$ is true and 0 otherwise.

We define the trace as starting from Comparable $c$, so $\bm{\chi}_{0} = \bm{x}_{c}$ and ending at the Subject $s$.
Since adjacent segments are continuous, $\tilde{y}_{\tau}(\bm{\chi}_\tau) = \tilde{y}_{\tau + 1}(\bm{\chi}_\tau)$, we can solve bias terms of each segment, i.e.,
\begin{equation}
    \begin{split}
    b_{\tau} &= b_{\tau - 1} - (\bm{w}_{\tau} - \bm{w}_{\tau - 1})^{\top} \bm{\chi}_\tau \\
    \equiv 
    b_{\tau > 0} &= b_1 + \sum_{t=1}^{\tau - 1} (\bm{w}_{t} - \bm{w}_{t + 1})^{\top} \bm{\chi}_t
    \end{split}
\end{equation}
This Trace piecewise function is trained using gradient descent with $(\bm{w}_{\tau}, \bm{\chi}_\tau), \forall \tau > 0$ and $b_0$ as parameters and supervised with the AI model predictions $\hat{y}$.

\begin{figure}
    \centering
    \includegraphics[width=6.2cm]{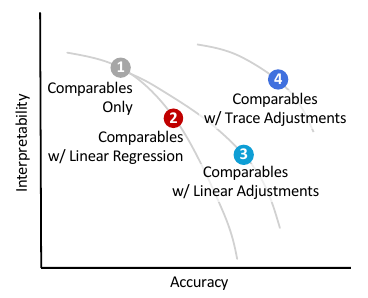}
    \caption{
        Trade-off between accuracy and interpretability for different example-based explanations:
        1) Comparables Only averaging is simplest and most interpretable, 
        2) Comparables w/ Linear Regression has more sophisticated aggregating which improves accuracy but at the cost of interpretability,
        3) Comparables w/ Linear Adjustments further trade-off accuracy for interpretability, and
        4) Comparables w/ Trace Adjustments leverages non-linearity for the highest accuracy, but in steps to retain interpretability.
        The grey solid line shows the Pareto front of similarly performing methods.
        Positions are illustrative only rather than factual.}
    \label{fig:trade_off_xai_type}
    \Description{
    The figure shows a conceptual graph with Accuracy on the horizontal axis and Interpretability on the vertical axis. A curved dotted line slopes downward from high-interpretability/low-accuracy (top-left) toward high-accuracy/low-interpretability (bottom-right), representing the typical trade-off between the two.
    Four numbered circles mark the positions of different explanation methods:
    1. Comparables Only (Number 1, pale peach) is placed near the top-left, indicating high interpretability but lower accuracy.
    2. Comparables with Linear Regression (Number 2, dark red) sits slightly to the right and lower, indicating higher accuracy but reduced interpretability.
    3. Comparables with Linear Adjustments (Number 3, blue) is farther right and further down, reflecting a greater gain in accuracy but even less interpretability.
    4. Comparables with Trace Adjustments (Number 4, dark blue) appears to the far right but relatively high on the vertical axis. This suggests that Trace Adjustments improve accuracy while preserving interpretability, positioning it above the traditional trade-off curve.
    Overall, the visual communicates how each method balances the trade-off differently, with Trace Adjustments aiming to push beyond the usual accuracy–interpretability boundary.
    }
\end{figure}

\subsubsection{Desiderata}
\label{sec:desiderata}

There are many ways that intermediate counterfactuals can be chosen; some are more plausible, trustworthy, and interpretable than others.
To select more desirable counterfactuals, we identify 5 desiderata determined from the literature and frame them regarding counterfactual cases $\chi_\tau$ (C) and the counterfactual trace (T).

\aptLtoX[graphic=no,type=html]{\begin{enumerate}
    \item[C1] \textbf{Faithfulness} 
    to ensure that the explanation predicts just as the AI model~\cite{ribeiro2016should, spitzer2025don}.
    \\ \rev{In our case,} each counterfactual has a label matching the AI model prediction for that case.
    \\ \rev{Formally,} this loss is modeled for the primary explanation task, $\mathcal{L}_\text{F} = |\hat{y}(\bm{x}) - \tilde{y}(\bm{x})|$.
    
    \item[C2] \textbf{Sparsity} 
    to \rev{reduce cognitive load} by minimizing the number of \rev{terms in the explanation~\cite{zeng2017interpretable, lou2016sparse, treviso2020explanation}}.
    \\ In our case, we aim to reduce the number of attribute changes between adjacent counterfactuals.
    \\ Formally, we penalize this with sparsity (L1 norm) loss $\mathcal{L}_\text{S} = \sum_{\tau} ||\bm{\chi}_\tau - \bm{\chi}_{\tau-1}||_1$.
    Unlike Disjointness (T1), this does not specify which attribute remains unchanged, and may have the same attribute change again along the counterfactual trace.
\end{enumerate}}{\begin{enumerate-desiderata-counterfactual}
    \item \textbf{Faithfulness} 
    to ensure that the explanation predicts just as the AI model~\cite{ribeiro2016should, spitzer2025don}.
    \\ \rev{In our case,} each counterfactual has a label matching the AI model prediction for that case.
    \\ \rev{Formally,} this loss is modeled for the primary explanation task, $\mathcal{L}_\text{F} = |\hat{y}(\bm{x}) - \tilde{y}(\bm{x})|$.
    
    \item \textbf{Sparsity} 
    to \rev{reduce cognitive load} by minimizing the number of \rev{terms in the explanation~\cite{zeng2017interpretable, lou2016sparse, treviso2020explanation}}.
    \\ In our case, we aim to reduce the number of attribute changes between adjacent counterfactuals.
    \\ Formally, we penalize this with sparsity (L1 norm) loss $\mathcal{L}_\text{S} = \sum_{\tau} ||\bm{\chi}_\tau - \bm{\chi}_{\tau-1}||_1$.
    Unlike Disjointness (T1), this does not specify which attribute remains unchanged, and may have the same attribute change again along the counterfactual trace.
\end{enumerate-desiderata-counterfactual}}

\aptLtoX[graphic=no,type=html]{\begin{enumerate}
    \item [T1]\textbf{Disjointness} 
    to reduce redundancy in the explanation~\cite{skulmowski2022understanding}. 
    \\ In our case, we aim for each attribute to only change once along the trace.
    \\ Formally, we denote a change in the $r$-th attribute of the $\tau$-th Comparable as $\mathbb{I}(|\Delta\chi_\tau^{(r)}| > \delta)$, where $\Delta\chi_\tau^{(r)} = \chi_\tau^{(r)} - \chi_{\tau-1}^{(r)}$ and $\delta$ is a threshold limit.
    The number of times attribute $r$ changes across the trace is $n_\Delta^{(r)} = \sum_\tau \mathbb{I}(|\Delta\chi_\tau^{(r)}| > \delta)$.
    We penalize the highest number of times of any attribute change with $\mathcal{L}_\text{D} = \max_r(n_\Delta^{(r)})$.
    
    \item[T2] \textbf{Monotonicity} 
    to reduce the surprise of unexpected direction change of values~\cite{hamer2024simple, delosh1997extrapolation, carvalho2019machine}.
    \\ In our case, we aim to choose counterfactuals such that their attribute values and label values are consistently increasing (or decreasing) only along the direction from Comparable to Subject.
    \\ Formally, given the change of attribute values ($\Delta\bm\chi_\tau = \bm\chi_\tau - \bm\chi_{\tau-1}$) and label values ($\Delta\tilde{y}_\tau = \tilde{y}_\tau - \tilde{y}_{\tau-1}$):
    if $\bm\chi_\tau$ direction changes, then $\Delta\bm\chi_\tau \Delta\bm\chi_{\tau-1} < 0$; 
    similarly, if $\tilde{y}_\tau$ direction changes, then $\Delta\tilde{y}_\tau \Delta\tilde{y}_{\tau-1} < 0$.
    We penalize this with the loss term $\mathcal{L}_\text{M} = \sum_{\tau} (\text{ReLU}(-\Delta\bm\chi_\tau \Delta\bm\chi_{\tau-1}) + \text{ReLU}(-\Delta\tilde{y}_\tau \Delta\tilde{y}_{\tau-1}))$.
    
    \item[T3]\textbf{Evenness} 
    to ensure stability in explanations~\cite{artelt2021evaluating}.
    \\ In our case, we aim to select counterfactuals along a trace that are evenly spaced by their value.
    \\ Formally, given the value of counterfactuals $\hat{y}_c$, we penalize the variance of consecutive label differences, i.e., $\mathcal{L}_{\mathrm{E}} = \sum_{\tau}(\Delta \tilde{y}_\tau - \mu_{\Delta \tilde{y}})^2$, where $\Delta \tilde{y}_\tau = \tilde{y}_\tau(\bm \chi_\tau) - \tilde{y}_\tau(\bm \chi_{\tau - 1})$ and $\mu_{\Delta \tilde{y}}$ is the average of $\Delta g_\tau$ across segments $\tau$.    
\end{enumerate}}{\begin{enumerate-desiderata-trace}
    \item \textbf{Disjointness} 
    to reduce redundancy in the explanation~\cite{skulmowski2022understanding}. 
    \\ In our case, we aim for each attribute to only change once along the trace.
    \\ Formally, we denote a change in the $r$-th attribute of the $\tau$-th Comparable as $\mathbb{I}(|\Delta\chi_\tau^{(r)}| > \delta)$, where $\Delta\chi_\tau^{(r)} = \chi_\tau^{(r)} - \chi_{\tau-1}^{(r)}$ and $\delta$ is a threshold limit.
    The number of times attribute $r$ changes across the trace is $n_\Delta^{(r)} = \sum_\tau \mathbb{I}(|\Delta\chi_\tau^{(r)}| > \delta)$.
    We penalize the highest number of times of any attribute change with $\mathcal{L}_\text{D} = \max_r(n_\Delta^{(r)})$.
    
    \item \textbf{Monotonicity} 
    to reduce the surprise of unexpected direction change of values~\cite{hamer2024simple, delosh1997extrapolation, carvalho2019machine}.
    \\ In our case, we aim to choose counterfactuals such that their attribute values and label values are consistently increasing (or decreasing) only along the direction from Comparable to Subject.
    \\ Formally, given the change of attribute values ($\Delta\bm\chi_\tau = \bm\chi_\tau - \bm\chi_{\tau-1}$) and label values ($\Delta\tilde{y}_\tau = \tilde{y}_\tau - \tilde{y}_{\tau-1}$):
    if $\bm\chi_\tau$ direction changes, then $\Delta\bm\chi_\tau \Delta\bm\chi_{\tau-1} < 0$; 
    similarly, if $\tilde{y}_\tau$ direction changes, then $\Delta\tilde{y}_\tau \Delta\tilde{y}_{\tau-1} < 0$.
    We penalize this with the loss term $\mathcal{L}_\text{M} = \sum_{\tau} (\text{ReLU}(-\Delta\bm\chi_\tau \Delta\bm\chi_{\tau-1}) + \text{ReLU}(-\Delta\tilde{y}_\tau \Delta\tilde{y}_{\tau-1}))$.
    
    \item \textbf{Evenness} 
    to ensure stability in explanations~\cite{artelt2021evaluating}.
    \\ In our case, we aim to select counterfactuals along a trace that are evenly spaced by their value.
    \\ Formally, given the value of counterfactuals $\hat{y}_c$, we penalize the variance of consecutive label differences, i.e., $\mathcal{L}_{\mathrm{E}} = \sum_{\tau}(\Delta \tilde{y}_\tau - \mu_{\Delta \tilde{y}})^2$, where $\Delta \tilde{y}_\tau = \tilde{y}_\tau(\bm \chi_\tau) - \tilde{y}_\tau(\bm \chi_{\tau - 1})$ and $\mu_{\Delta \tilde{y}}$ is the average of $\Delta g_\tau$ across segments $\tau$.    
\end{enumerate-desiderata-trace}}

Together, we constrain the Trace Adjustments with the addition of loss terms to regularize training:
\begin{equation}
    L
    = \lambda_F\mathcal{L}_{\mathrm{F}}
    + \lambda_S \mathcal{L}_{\mathrm{S}}
    + \lambda_D\mathcal{L}_{\mathrm{D}}
    + \lambda_M\mathcal{L}_{\mathrm{M}}
    + \lambda_E\mathcal{L}_{\mathrm{E}},
    \label{eq:loss}
\end{equation}
where each $\lambda_*$ term is a hyperparameter to control the importance of each desideratum.

While optimizing toward Sparsity, Disjointness, Monotonicity, and Evenness aim to improve the \textit{interpretability} of Comparables w/ Trace Adjustments, this comes at a cost to explanation Faithfulness (\textit{accuracy}).
Prioritizing among them depends on the application and user goals to balance the accuracy-interpretability trade-off~\cite{arrieta2020explainable}.

\subsubsection{Implementation Details}
We implemented Trace Adjustment as a single PyTorch\footnote{\url{https://pytorch.org/}} module that learns a piecewise-linear model along the trace between each Comparable-subject pair. The implementation adopted PyTorch 2.5.1 and NumPy 1.26.4. 
Training employed the Adam optimizer~\cite{kingma2015adam} with an initial learning rate of 0.8, which adapts dynamically during training. 
We initialized model parameters with small random values to promote stable convergence.
The training loss incorporated all desiderata; see Section~\ref{sec:approach_sensitivity_analysis} for the hyperparameter details.

\section{Evaluation}
We conducted several experiments to examine:
i) their accuracy toward the actual ground truth and their faithfulness toward the AI prediction in a modeling study,
ii) their usage strategies and usefulness in estimating the actual value through a qualitative formative user study, and
iii) their impact on decision-making performance in a quantitative summative user study.

\subsection{XAI Types}
We evaluated Comparables w/ Trace Adjustments explanations against a set of baseline XAI Types.

\subsubsection{Comparables Only}
This baseline explanation uses raw Comparables and estimates the Subject's value by reconciling a weighted average of the values of the Comparables as described in Section~\ref{sec:approach_comparables_only} Eq.~\ref{eq:weighedavg}. 
All XAI Types use this reconciliation method unless explicitly stated otherwise.

\subsubsection{Comparables w/ Linear Regression} 
Instead of weighted averaging, this baseline fits a linear regression on the Comparables as described in Section~\ref{sec:comparables_w_linear_reg}.

\subsubsection{Comparables w/ Linear Adjustments}
This approach first adjusts the Comparable values and reconciles based on the adjustments.
For this baseline, LIME-based linear adjustments are performed as described in Section~\ref{sec:comparable_w_linear_adjustments}.

\subsubsection{Comparables w/ Trace Adjustments}
Our proposed approach that adjusts each Comparable based on counterfactually tracing the AI decision surface from the Comparable to the Subject as described in Section~\ref{sec:comparable_w_trace_adjustments} Eq.~\ref{eq:traceadjustment}. It is also fully constrained by desiderata described in Section~\ref{sec:desiderata}.

\subsection{Modeling Study}
We conducted a modeling study to evaluate the \textit{accuracy} and \textit{precision} of the explanations across different settings and datasets. 
We evaluated how each explanation performs in three respects: (i) how accurately it approximates the ground-truth labels, (ii) how faithfully it reflects the underlying AI predictions, and (iii) how narrow its uncertainty bounds are. 
We also performed sensitivity analyses that examine how each desiderata may influence the faithfulness and interpretability.

\subsubsection{Applications and Datasets} 
We evaluated on regression prediction tasks, which are relevant to valuation.
For generality, we selected datasets across five application domains that employ Comparables or case-based reasoning.
We describe each application task, the dataset chosen, and the performance of each trained AI model (all trained on 80\% of the dataset).}
\begin{itemize}
    \item \textbf{House Price Valuation} 
    to estimate the price of a residential property. 
    Real-estate agents estimate the price of a subject property by tabulating attributes against comparable properties (\textit{Comparables})~\cite{farkas2020multi, kim2020applying}.
    \\ Dataset: House Sales in King County, USA dataset~\cite{harlfoxem2019house}, 
    with selected 6 attributes (to limit cognitive load): \# Bathrooms, Living Area, Grade, Age, Condition, and converted (latitude, longitude) coordinates into a single attribute representing Distance to Downtown. 
    \\ AI Performance: mean absolute error (MAE) = \$107.8K and $R^2$ = 0.78.
    
    \item \textbf{Salary Estimation} 
    to identify an appropriate market-aligned compensation for a job role by referencing similar positions across organizations as \textit{Comparables}, thereby ensuring fairness and supporting talent acquisition and retention~\cite{cullen2022s}.
    \\ Dataset: Stack Overflow Developer Survey dataset~\cite{berkayalan2024stackoverflow},
    with selected 6 attributes: country, education level, work experience (years), work mode, organization size, and job title.
    \\ AI Performance: MAE = \$22K and $R^2$ = 0.68.
    
    \item \textbf{Energy Consumption Estimation} 
    \rev{to forecast energy demand in a building to support planning, optimization, and resource management, commonly referencing similar buildings or the building’s own historical performance as \textit{Comparables}~\cite{seyedzadeh2018machine, ke2013analysis}.}
    \\ \rev{Dataset:} ASHRAE Great Energy Predictor III dataset~\cite{ashrae2019gep3} with selected 5 attributes: primary use, building size, air temperature, hour of the day, and whether it is a weekday.
    \\ \rev{AI Performance:} MAE = 0.22kWh and $R^2$ = 0.95.
    
    \item \textbf{\rev{Drug Sensitivity Analysis}} 
    \rev{to estimate drug sensitivity on cancer cell lines based on biomarkers, where similar cell lines serve as \textit{Comparables} for inferring likely treatment outcomes~\cite{wang2017improved, liu2018anti}. This can support personalized patient treatment~\cite{azuaje2017computational}.}
    \\ \rev{Dataset: Genomics of Drug Sensitivity in Cancer dataset~\cite{Alipour_2024_GDSC} with selected 7 attributes: drug name, DNA repair status (stable or not), screening medium, cell growth properties, molecular target of the drug, and target pathway.} 
    \\ \rev{AI Performance: MAE = 75.3 $\mu M$ and $R^2$ = 0.82.}
    
    \item \textbf{Crop Yield Estimation}
    to predict expected agricultural production, which helps farmers make informed economic and management decisions, and supports famine-prevention efforts and global food security~\cite{chang2023data}. In this context, farmlands with similar conditions serve as valuable \textit{Comparables} for inferring yield~\cite{liu2021analogy}.
    \\ Dataset: Crop Yield in Indian States dataset~\cite{gupta2024crop} with selected 7 attributes: crop type, season, land area, annual rainfall, fertilizer usage, pesticide usage, and state.
    \\ AI Performance: MAE = 157K tons and $R^2$ = 0.98.
\end{itemize}

\begin{figure*}[t]
    \centering
    \includegraphics[width=0.84\linewidth]{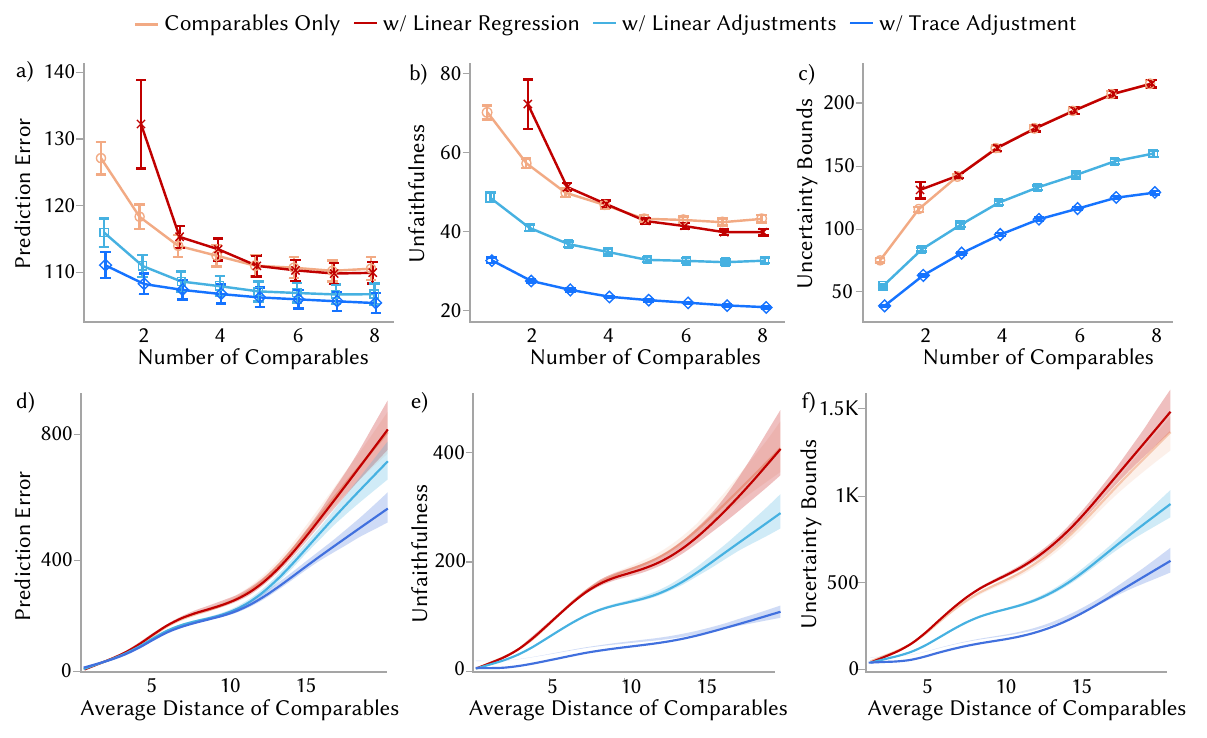}
    \caption{Modeling study results for different XAI types in the House Price domain, varying the Number of Comparables (a--c) and the Average Distance of Comparables (d--f), distance was computed using the Manhattan metric with standardized features.
     Increasing the Number of Comparables reduces a) prediction error and b) unfaithfulness across all methods, while c) uncertainty bounds become broader. 
     Increasing the Average Distance of Comparables worsens all metrics (d--f). Trends are smoothed with a cubic spline (smoothing parameter $\lambda=1000$). 
    Across all conditions, Comparables w/ Trace Adjustments consistently yield lower prediction error (a, d), lower unfaithfulness (b, e), and narrower uncertainty bounds (c, f).}
    
    \Description{
    The figure contains six line charts, labeled a through f, showing results of a modeling study in the House Price domain. Each panel compares four explanation methods: Comparables Only,  Comparables with Linear Regression, Comparables with Linear Adjustments, and Comparables with Trace Adjustments. Error bars or shaded regions represent uncertainty.
    a) Prediction Error by Number of Comparables. On the horizontal axis, the number of comparables ranges from 2 to 8. Prediction error decreases for all methods as the number of comparables increases. Linear Regression starts with the highest error, around 135, then drops sharply but remains above the other methods. Comparables Only begins around 125 and gradually decreases. Both Linear Adjustments and Trace Adjustments start lower, around 115 and 110 respectively, and maintain the lowest errors across all numbers of comparables.
    b) Unfaithfulness by Number of Comparables. Unfaithfulness decreases as the number of comparables increases. Linear Regression begins highest, near 90, and steadily declines but remains above the others. Comparables Only starts near 70 and decreases. Linear Adjustments and Trace Adjustments are lower throughout, starting at about 50 and 40, and steadily declining to the lowest levels.
    c) Uncertainty Bounds by Number of Comparables. Uncertainty bounds increase as the number of comparables increases. Linear Regression has the broadest bounds, rising from about 130 to over 200. Comparables Only increases from around 120 to 170. Linear Adjustments and Trace Adjustments show narrower bounds, beginning near 90 and 50, then rising more moderately, with Trace Adjustments always the narrowest.
    d) Prediction Error by Average Distance of Comparables. The horizontal axis shows the average distance from 0 to 20. Prediction error increases steadily as the distance grows. Linear Regression shows the highest error curve, followed by Comparables Only. Linear Adjustments and Trace Adjustments are consistently lower, with Trace Adjustments yielding the smallest errors at all distances.
    e) Unfaithfulness by Average Distance of Comparables. Unfaithfulness also increases with distance. Linear Regression is highest, Comparables Only is slightly lower, and Linear Adjustments and Trace Adjustments remain much lower, with Trace Adjustments the most faithful.
    f) Uncertainty Bounds by Average Distance of Comparables. Uncertainty bounds increase with distance. Linear Regression shows the broadest growth, reaching nearly 1,500 at the farthest distances. Comparables Only is somewhat lower. Linear Adjustments and Trace Adjustments show much narrower bounds, with Trace Adjustments the most consistent and tightest.
    }
    \label{fig:modeling_study}
    
\end{figure*}

\subsubsection{Evaluation Measures} 
We measured 
\begin{enumerate}
    \item \textbf{XAI Prediction Error $\downarrow$} with the mean Absolute Error (AE) between the XAI estimate and the ground truth value (lower AE is better performance).
    \item \textbf{XAI-AI Unfaithfulness $\downarrow$} with the mean Absolute Error (AE) between the XAI estimate and the AI prediction.
    \item \textbf{XAI Uncertainty Bounds $\downarrow$} as the range from the \textit{minimum} Comparable's value to the \textit{maximum} among the set of Comparables.
    Adjusted values are used for Comparables with Linear and Trace Adjustments.    
\end{enumerate}

\subsubsection{Results}
\label{sec:modeling_res}
Fig.~\ref{fig:modeling_study} shows the results for the House Price domain.  (a–c) vary the Number of Comparables from 1 to 8, and d–f vary the Average Distance of Comparables. Distances were computed by standardizing each attribute in the feature set and then calculating the Manhattan distance between attribute vectors, capturing the similarity (consistency) of Comparables to the Subject.  
Comparables w/ Trace Adjustments achieved the best performance, with the lowest XAI Prediction Error, highest faithfulness (lowest XAI–AI Unfaithfulness), and greatest confidence (tightest XAI Uncertainty Bounds).  
This was followed by Comparables w/ Linear Adjustments, then Comparables w/ Linear Regression, which performed about as poorly as Comparables Only.  
Across all methods, performance improved with  
i) more Comparables ($\uparrow$ Number of Comparables), and  
ii) more similar Comparables ($\downarrow$ Average Distance of Comparables).  
Results for the Salary and Energy Consumption domains are provided in Appendix ~\ref{sec:modeling_domain_generalization}, showing similar trends across datasets. 

\subsubsection{Sensitivity Analysis of Desiderata Hyperparameters}
\label{sec:approach_sensitivity_analysis}
Improving interpretability through desiderata (C2, T1--T3) may sacrifice Faithfulness too much, so we performed a sensitivity analysis to examine this accuracy-interpretability trade-off. 
For brevity, we focused on the Housing Price application.
We examined several evaluation measures of: \# Adjustments for Sparsity and Disjointness,
\# Reversals for Monotonicity, and variance of adjustments for Evenness.
Formal definitions of each evaluation metric are detailed in Appendix~\ref{app:sensitivity_analysis}.
Specifically, we varied hyperparameters to regularize Sparsity $\lambda_S$, Disjointness $\lambda_D$, Monotonicity $\lambda_M$, Evenness $\lambda_E$. 
This also helps to inform our selection of hyperparameter settings.

Fig.~\ref{fig:sensitivity_analysis} shows that the desiderata are effectively controlled:
\begin{itemize}
    \item $\uparrow$ Sparsity $\lambda_S$ $\implies \downarrow$ \# Adjustments and $\uparrow$ Unfaithfulness (Fig.~\ref{fig:sensitivity_analysis}a). 
    Since \# Adjustments reduce slowly beyond 10, we select $\lambda_S$ = 10.
    \\
    \item $\uparrow$ Disjointness $\lambda_D$ $\implies \downarrow$ \# Adjustments and $\uparrow$ Unfaithfulness (Fig.~\ref{fig:sensitivity_analysis}b).
    Since increasing $\lambda_D$ beyond 10 further causes a sharp sacrifice in faithfulness, we select $\lambda_D = 10$.
    \item $\uparrow$ Monotonicity $\lambda_M$  $\implies$ Unchanged \# Reversals and $\uparrow$ Unfaithfulness (Fig.~\ref{fig:sensitivity_analysis}c). 
    Since \# Reversals do not practically change, we focus on limiting increasing unfaithfulness, and we select $\lambda_M = 1$.
    Perhaps the dataset attributes are already quite monotonic.
    \item \rev{$\uparrow$ Evenness $\lambda_E$  $\implies \downarrow$ Unevenness and $\uparrow$ Unfaithfulness (Fig.~\ref{fig:sensitivity_analysis}d). 
    Since Evenness improves significantly at 1, and higher values gain little but harm faithfulness, we select $\lambda_E = 1$.}
\end{itemize}

\subsection{Formative User Study}
We examined how users use, interpret, or struggle with explainable Comparables. We recruited 25 participants from a local university who were on average 23 years old (range: 20–29 years), and 15 identified as female (10 male). 
They were undergraduate or graduate students from diverse disciplines, including Computer Science, Electrical Engineering, Industrial System Engineering, Data Science, Economics, Law, Medicine, Chemistry, Physics, Business, and Real Estate.
We paid special attention to two participants (E2, E13) in Real Estate considered to have relevant domain expertise, and one participant with valuation experience (E11).
\subsubsection{Experiment Method}
We conducted a within-subjects experimental design where each participant was exposed to 2--3 XAI Types, and not all to limit overwhelming them and as time permits.
We used the same user interface for all three domains (House Price, Salary, Energy Consumption), which we describe next.
\begin{figure*}[t]
    \centering
    \includegraphics[scale=0.57]{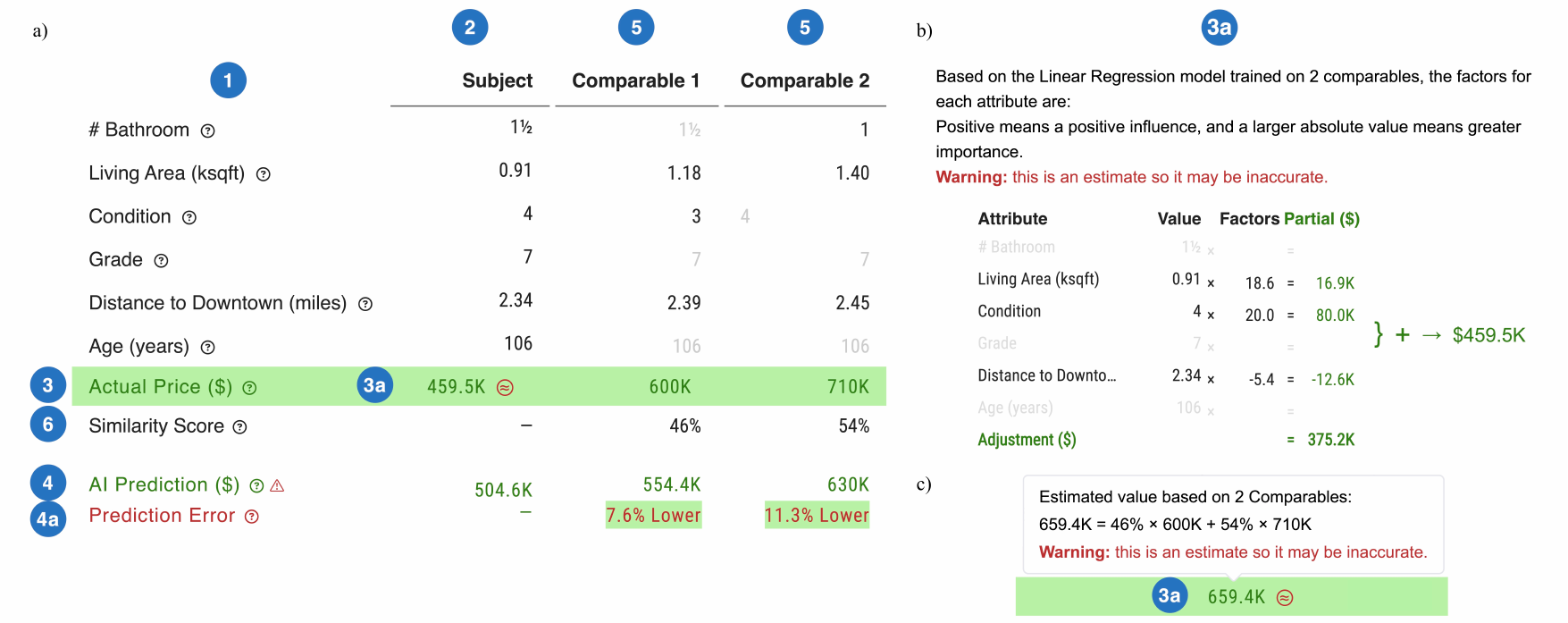}
    \caption{\textit{Comparables Only and Comparables w/ Linear Regression}: 
    a) The shared interface used by both conditions. 
    (1) rows listing all attributes; 
    (2) a column displaying the Subject property; 
    (5) columns showing the Subject's Comparables, annotated with similarity scores (6); 
    (3) a row showing the Actual Price, where the red $\approx$ (3a) indicates an estimated value for the Subject; (4) shows the AI Prediction and the corresponding Prediction Error (4a) compared to Actual Price. b) The tooltip for Comparables w/ Linear Regression when hovering over (3a), presenting regression coefficients and the regression-based estimate. 
    c) The tooltip for Comparables Only when hovering over (3a), showing a weighted-average estimate based on similarity scores.}
    \Description{
    The figure presents three interface panels illustrating Comparables Only and Comparables with Linear Regression explanations in the House Price prediction task.
    Panel a (left): Comparables Only interface.
    The table shows one subject case and two comparable houses side by side. Rows list features: number of bathrooms, living area in square feet, condition, grade, distance to downtown, and age. The subject has 1.5 bathrooms, 0.91 living area, condition 4, grade 7, 2.34 miles distance, and age 106 years. The actual price for the subject is 459.5K. Comparable 1 is priced at 600K with 46]\% similarity, and Comparable 2 is priced at 710K with 54\% similarity. Below, the AI system predicts a value of 504.6K, which differs from the actual price. Prediction error is highlighted.
    Panel b (top right): Linear Regression explanation pop-up. This panel shows how the AI regression model combines attribute values into the prediction. Each feature is listed with its contribution in dollars. For example, living area contributes about 16.9K, condition contributes 80K, grade adds nothing, distance to downtown subtracts 12.6K, and age contributes 375.2K. These factors sum to a regression-based estimate of 459.5K. A warning indicates the estimate may be inaccurate.
    Panel c (bottom right): Weighted average of comparables. This panel shows the subject’s value estimated directly from the two comparables, using their similarity weights. The calculation combines 46\% of Comparable 1’s price and 54\% of Comparable 2’s price, giving an estimate of 659.4K. A warning again indicates the estimate may be inaccurate.}
    \label{fig:linear_reg}
\end{figure*}

\begin{figure*}[t]
    \centering
    \includegraphics[scale=0.9]{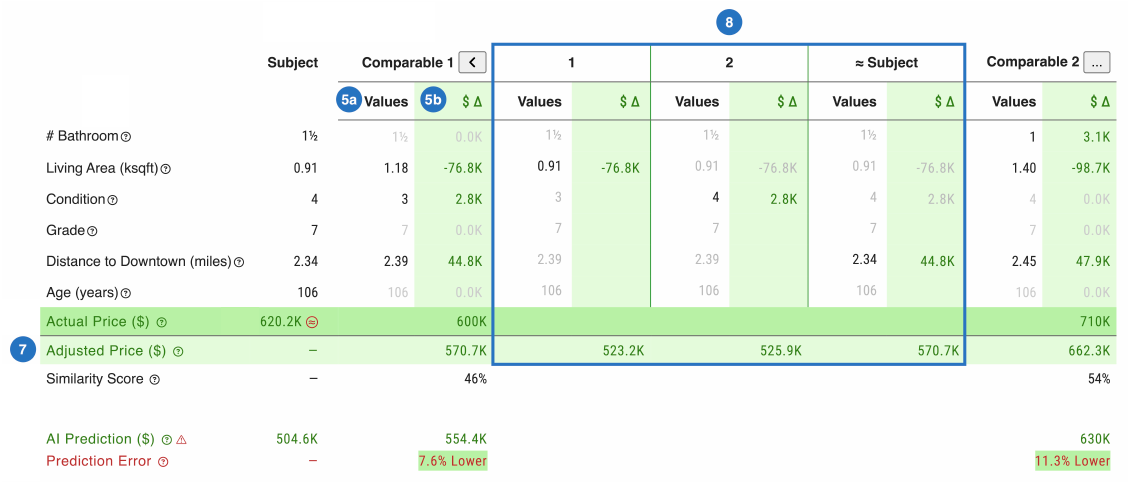}
    \caption{\textit{Comparables w/ Linear Adjustments and w/ Trace Adjustments}: The interface keeps the same information as illustrated in Fig.~\ref{fig:linear_reg}, but with adjustments (5b), and final Adjusted Price (7). For Comparables w/ Trace Adjustments, clicking the “$\dots$” button reveals (8) a trace of intermediate hypothetical cases, with one column for each step (1, 2, …) and subcolumns detailing adjusted attributes and their corresponding adjusted prices. 
    For example, in this case, we adjust Comparable 1 one attribute at a time. 
    First, we change only the Living Area from 1.18 (ksqft) to 0.91 (ksqft) to match the Subject, and the \$$\Delta$ column shows that this reduces the price by \$76.8K. 
    Next, we adjust Condition from 3 to 4, which adds +\$2.8K, and so on. 
    Each step updates only one attribute, and the adjusted price accumulates these deltas. 
    By the final ``$\approx$ Subject'' column, all attributes match the subject, and the last adjusted price reflects the final adjusted price. The \textit{Comparables w/ Linear Adjustments} interface is similar except that the table is not expandable (without (8)).}
    \Description{The figure shows an interface for Comparables with Linear Adjustment and Comparables with Trace Adjustment. It extends the earlier Comparables table with additional columns showing how adjustments are applied step by step.
    Main table (left side): The subject case is displayed alongside Comparable 1 and Comparable 2. Rows list the same house attributes as in earlier figures: number of bathrooms, living area, condition, grade, distance to downtown, and age. The subject’s actual price is 620.2K. Comparable 1 is priced at 600K with a similarity score of 46\%, and Comparable 2 is priced at 710K with a similarity of 54\%. The AI system predicts 504.6K for the subject, which is lower than the actual value. Prediction errors and adjusted prices are shown below.
    Adjustment details (center): Clicking to expand Comparable 1 reveals intermediate hypothetical cases in separate numbered columns. For example, one column shows attribute values adjusted closer to the subject’s, with corresponding price adjustments. Living area is adjusted downward by about 76.8K, condition contributes +2.8K, and distance to downtown adds about +44.8K. These adjustments result in new intermediate adjusted prices: 523.2K in step 1, 525.9K in step 2, and 570.7K when all adjustments are applied, aligning with the subject’s attributes.
    Comparable 2 (right side): Comparable 2 is also displayed with its own attribute differences. Living area adjustment reduces value by nearly 98.7K, distance to downtown adds about 47.9K, and bathroom adds +3.1K. This produces an adjusted price of 662.3K, compared to the original 710K.}
    \label{fig:trace}
\end{figure*}
\subsubsection{Apparatus (Comparables User Interface)} 
Our user interface was inspired by the Sales Comparison Grid~\cite{Grayson2024CompGridModel}, which is a spreadsheet table that enables appraisers to compare multiple properties side by side across key attributes.
Basic components show attributes 
(1), the Subject's attribute values (2) and estimated value (3a), and 
(5) for each Comparable, their attribute values and actual value (3).
The estimated value is based on weighted sum reconciliation, and users can see the formula and calculation (see Fig. \ref{fig:linear_reg}b) by hovering their mouse on the estimated value.
We adapted it to add in rows for each Comparable on AI prediction (4) and error (4a), and similarity (6) to the Subject (see Fig. \ref{fig:linear_reg}a). 

For Comparables with Linear regression, when hovering over the estimated value (3a), instead of the weighted average, the popup shows a table explaining the linear regression with attribute values and factors, summed to the reconciled weighted sum (see Fig. \ref{fig:linear_reg}c).

To show the incremental counterfactuals of Trace Adjustments, we added columns (5) that users can see on clicking an expansion button (see Fig. \ref{fig:trace}).
Each counterfactual has two columns for adjusted attributes (5a) and the corresponding value adjustments (5b).

\subsubsection{Experiment Procedure}
After consenting, participants first went through tutorials introducing them to the system’s user interface, followed by screening questions to ensure correct understanding. 
The main task required them to predict the actual value (Price, Salary, or Energy Consumption) for a Subject (House, Employee, or Building), given the explanations generated by the system. 
To help participants become familiar with the system and form a baseline sense of the influencing factors, we provided 2--3 practice cases during which the actual values of the subjects were revealed. 
In the subsequent main session, each participant evaluated 3--5 trials without being shown the actual values in advance.
We asked participants to think aloud to elicit their thought processes so that we could understand how they perceived the explanations. 
Participants could ask clarification questions at any time.
They ended by answering demographics questions and were compensated.

With the consent of the participants, we recorded both the audio and the screen interactions. Each session lasted 60 minutes, and each participant received $\$15\ \text{USD}$ equivalent in local currency as compensation.

\subsubsection{Findings}
\label{sec:formative_findings}
We analyzed participants' utterances and interactions and report key findings organized by XAI Type.

\textbf{Comparables Only} fosters estimations based on \textbf{few similar examples} and \textbf{error bias}.
By examining similarity scores and manually inspecting attribute differences, participants determined which Comparables were more similar or distant to the Subject.
They may make a decision on the closest 1--2 Comparables, rather than all provided examples. 
P19 explained that \textit{``the more similar the houses are, the closer their prices should be... and [considering] 2 is enough because [I think] the third closest one has so many different [attributes]''}.

Some participants realized that multiple Comparables shared the same direction of AI error (e.g., AI prediction tends to be higher than the actual value) 
P5 \textit{``made my adjustments higher[than the AI prediction] because all of these [AI predictions for other Comparables] are lower [than their actual prices]''}.
However, this became challenging when the error was balanced across Comparables, \textit{``because I’ve seen these results being a mix of higher and lower than the actual values, so it is difficult to really infer from these''} (P6).
Instead, they \textit{``looked at and attribute values and made own judgments''} (P18).

Notably, without explanatory support in Comparables Only, participants with domain expertise in Valuation chose \textit{``to rely on my own judgment because I don’t feel the attribute values are unreliable; they are all true values rather than reconciled or AI-generated predictions''} (E11).

\textbf{Comparables with Linear Regression} are \textbf{overly complex} for non-technical users and can be \textbf{spurious}.
While some participants appreciated the transparency of the factors in reconciling the value estimate with Comparables, these tended to have more technical backgrounds (e.g., E11 Real Estate, P6 Computer Science, P1 Business).
\textit{``I think [Linear Regression] is quite reasonable. Yes, I can tell... the factor values align well with what I know''} (E11).
Other less technical participants struggled to grasp abstract the linear factors (Fig. \ref{fig:linear_reg}b), and preferred the tangibility of concrete examples and reasoning by the weighted average (Fig. \ref{fig:linear_reg}c).
P8 \textit{``found the factors not intuitive... so I would still prefer to rely on the previous method (weighted average).''}
Some factors were also spurious and deviated from participant expectations. For instance,
\textit{``the model puts more weight on the education level [in this case], which should matter less, because...there's no way someone looks at whether or not you have a degree after 25 years of working''} (P6). 

\textbf{Comparables with Linear Adjustments} were more \textbf{aligned} with domain expertise, and more trustworthy than unadjusted Comparables, but \textbf{unintelligible} in derivation.
Participants with background in Real Estate felt that the adjustments were familiar and intuitive. 
E13 described the experiment task as \textit{``very similar''} to her daily tasks, since she would \textit{``start from finding similar houses... focus only on the most important attributes ... then adjust the prices of the Comparables accordingly, and [finally] take an average [of the adjusted prices]''}. 
Non-expert participants also trusted the adjustments more than the unadjusted values. 
P8's strategy was to simply \textit{``check whether the adjustment direction makes sense''} and \textit{``follow the reconciled price``} as given by the system.
However, some wanted to know how the adjustments were determined. 
P14 \textit{``wonder[ed] how exactly these adjustments are obtained, and what calculations or logic is behind them.''}

\textbf{Comparables w/ Trace Adjustments} are perceived as more \textbf{tightly consistent} and usably \textbf{interpretable}, but can be \textbf{overwhelming}.
Due to each Trace Adjustment being faithful to the AI prediction, their adjustments were closer.
P5 thought \textit{``the differences [between each Comparables] are quite small, so it gives me greater confidence.''}
Participants appreciated expanding on the adjustments to see the explanatory counterfactual trace. 
P19 \textit{``like[d] the fact that I can see the details''}.
P1 felt that \textit{``the adjustments from Trace Adjustments makes a lot more sense''} and this \textbf{plausibility} made him \textit{``trust it more''}. 
Participants \textbf{appreciated some desiderata} that we constrained on Trace Adjustments. 
P9 found the \textit{``step by step [trace] were even more helpful [compared to one step adjustment] because....there are fewer things to look at each step''}; this illustrates the benefit of \textbf{Sparsity}.
E2 noted that \textit{``changing only one attribute value at a time fits well with the appraisal practice in Real Estate, which made it much easier for me to accept''}; 
this indicates the usefulness of \textbf{Sparsity} and \textbf{Disjointness}. 
P21 noticed that 
\textit{``[when using Comparable w/ Linear Adjustment,] there was an example where the house was only 3 [ksqft], but the adjustment step was already 1.5 [ksqft]... I felt the value was not very trustworthy... but in [Comparable w/] Trace [Adjustments] condition, such extreme adjustments did not appear.''}
Thus emphasizing the expectation for \textbf{Evenness}.

Despite its usefulness, Trace Adjustments UI can be \textbf{overwhelming}. Some participants found there to be \textit{``too much information to look at''} (P17) and \textit{``overwhelming''} (P17, P14, P5).
Nevertheless, they would typically keep the counterfactual columns collapsed and were happy to have them available. 
P1 \textit{``became convinced of the adjusted prices after a few [practice] cases``} and in later cases \textit{``feel[s] lazy looking at the expanded table anymore ... especially when there are 3 or 4 Comparables''}.
\subsection{Summative User Study} 
We conducted a summative user study to evaluate the usefulness of different XAI types, focusing on how various explanations affected participants’ decision accuracy and confidence. The following sections outline the experimental design, hypotheses, procedure, analysis methods, and results.

\subsubsection{Experiment Design} 
We designed a 4×4 factorial mixed-design experiment with independent variables (IVs): 
\begin{enumerate}
    \item \textbf{XAI type} (Comparables Only, Comparables w/ Linear Regression, Comparables w/ Linear Adjustments, Comparables w/ Trace Adjustments),
    in \textit{between-subjects} arrangement to avoid the learning effect where participants tend to stick to the first explanation style they are trained on.
    
    \item \textbf{Number of Comparables} (1 to 4), \textit{within-subjects}, with 100 instances sampled from the full dataset.
\end{enumerate} 

For generality, we included random variable (RV):
\begin{itemize}
    \item \textbf{Domain Dataset} (House Price, Salary), \textit{between-subjects} to limit cognitive load of learning multiple domains.
    We excluded the Energy Consumption domain, since we found in a pilot study that participants struggled to form a good mental model of the domain due to its obscurity.
\end{itemize}

We measured decision accuracy and confidence primarily by asking users to indicate a minimum value $\mathring{y}_\mathrm{min}$ and a maximum value $\mathring{y}_\mathrm{max}$ within which they are 90\% confident that the Actual value of the Subject will be in (see Fig.~\ref{fig:double_range_slider}).
From this, we derive the following objective dependent variables (DVs):

\begin{figure}[t]
    \centering
    \includegraphics[scale=0.6]{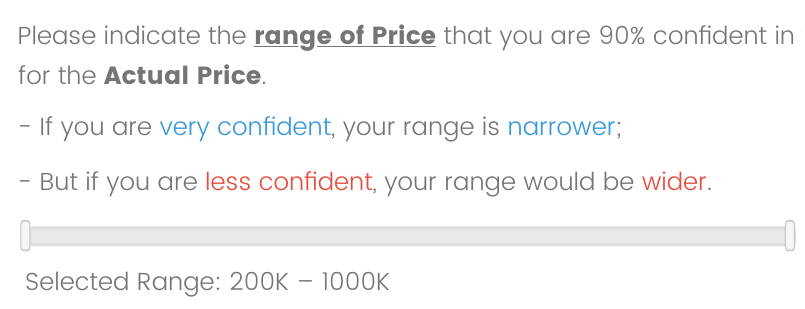}
    \caption{Double range slider bar to input responses.}
    \Description{
    The figure shows a user interface element for collecting confidence intervals about house prices. At the top, the instruction reads: “Please indicate the range of Price that you are 90\% confident in for the Actual Price.” Two guiding notes follow: If the participant is very confident, the range should be narrower. If the participant is less confident, the range should be wider.
    Below the instructions, there is a horizontal slider bar with two movable handles—one on the left and one on the right. This allows the participant to select both a lower and an upper bound, creating a double-ended range.
    Under the slider, the currently selected range is displayed in text. In the example shown, the selected range is from 200K to 1000K.
    }
    \label{fig:double_range_slider}
\end{figure}

\begin{itemize}
   \item \textbf{Decision Credible Interval $\downarrow$}
   represents a range of probable values based on a Bayesian probability distribution\footnote{In contrast, a confidence interval represents a range that is likely to contain the true population parameter for a certain percentage of repeated samples.}. 
   Here, we define the interval for 90\% and assume a Gaussian probability distribution. 
   This is calculated from the user response as $\mathring{y}_\mathrm{max} -  \mathring{y}_\mathrm{min}$. 
   A narrower interval indicates greater certainty and higher \textit{precision}.
   Due to a long tail effect, we apply a logarithmic transform to increase its normality, i.e., 
   $\log(\mathring{y}_\mathrm{max} -  \mathring{y}_\mathrm{min})$.

   \item \textbf{Decision Mean Error $\downarrow$} 
   captures how accurately participants estimated the actual value. 
   Given the participant's credible interval $[\mathring{y}_\mathrm{min},  \mathring{y}_\mathrm{max}]$, we define 
   the participants point estimate mean as $\mathring{y}_\mathrm{mean} = (\mathring{y}_\mathrm{min} + \mathring{y}_\mathrm{max})/2$.
   We calculate Decision Mean Error as the absolute error between the mean $\mathring{y}_\mathrm{mean}$ and ground truth $y$ and log transformed to increase normality, i.e., 
   $\log(\lvert \mathring{y}_\mathrm{mean} - y \rvert)$.

   \item \textbf{Correctness Probability Density $\uparrow$} 
   estimates how likely the actual (Correct) value is given the participant's credible interval.
   Assuming that the participant's belief is a Gaussian distribution with 90\% credible interval from $\mathring{y}_\mathrm{max}$ to $\mathring{y}_\mathrm{min}$, we can derive the Correctness Probability Density as:
   \begin{equation}
   \label{eq:probdensity}
   \begin{aligned}
       p(y \mid \mathring{y}_\mathrm{min}, \mathring{y}_\mathrm{max}) 
       &= \frac{1}{\sigma \sqrt{2\pi}} 
       \exp\!\left(-\frac{(y - \mathring{y}_\mathrm{mean})^2}{2\sigma^2}\right), \\
       \sigma &= \frac{ \mathring{y}_\mathrm{max} - \mathring{y}_\mathrm{min}}{z_{0.95} - z_{0.05}}, 
    \end{aligned}
   \end{equation}
   where $\sigma$ is the standard deviation of the participant's belief, 
   calculated from $z_{0.95}$ and $z_{0.05}$ which are the $z$-scores of the standard normal distribution corresponding to the 95th and 5th percentiles, respectively. 
%
   This measure captures both the \textit{accuracy} (of the participant's mean estimate to the actual value) and the \textit{precision} (narrowness of the credible interval). 
   Higher probability means that the participant is more correct.
  
\end{itemize}

We measured perceived ratings of different components---Comparables, Adjustments, Linear Factors, and Trace Details---on a 7-point Likert scale (-3 Strongly Disagree to +3 Strongly Agree) for the following subjective dependent variables (DVs):
    \begin{itemize}
        \item \textbf{Perceived Helpfulness} 
        \item \textbf{Perceived Ease-of-Use} 
    \end{itemize}
    
\textbf{Summary:} see Table~\ref{tab:hypo_and_finding} for our hypotheses and the corresponding findings from our results, described later. 

\begin{table*}[t]
\centering
\caption{Hypotheses and findings of the summative user study regarding different dependent variables for IV1: various XAI types and IV2: the Number of Comparables: Comparables Only (O), Comparables w/ Linear Regression (LR), Comparables w/ Linear Adjustments (L$\Delta$), Comparables w/ Trace Adjustments (T$\Delta$).}
\label{tab:hypo_and_finding}
\begin{tabular}{@{}lcllll@{}}
\toprule
Dependent Variable & Metric & IV & Hypothesis & Finding & Evidence \\ 
\midrule

\multirow{2}{*}{Decision Mean Error}
& \multirow{2}{*}{$\mathrm{log}\!(|\mathring{y}_\mathrm{mean} - y|)$}
& XAI Type   & O $\approx$ LR $<$ L$\Delta$ $<$ T$\Delta$ & O $\approx$ LR $\ns{\approx}$ L$\Delta$ $<$ T$\Delta$  & Fig.~\ref{fig:summative_objective_res}a\\ 
&  & \# Comps     & 1 $<$ 4 $<$ 2 $\approx$ 3 & 1 $<$ 4 $\ns{\approx}$ 2 $<$ 3  & Fig.~\ref{fig:summative_objective_res}d \\
\hline

\multirow{2}{*}{Decision Credible Interval} 
& \multirow{2}{*}{$\mathrm{log}(|\mathring{y}_\mathrm{max} - \mathring{y}_\mathrm{min}|)$} 
& XAI Type   
& O $\approx$ LR $<$ L$\Delta$ $<$ T$\Delta$ 
& O $\approx$ LR $\ns{\approx}$ L$\Delta$ $<$ T$\Delta$  & Fig.~\ref{fig:summative_objective_res}b\\ 
&  & \# Comps     
&  4 $<$ 1 $<$ 2 $\approx$ 3  
& 1 $<$ 2 $\ns{\approx}$ 3 $\approx$ 4  & Fig.~\ref{fig:summative_objective_res}e \\
\hline

\multirow{2}{*}{Correctness Probability Density} 
& \multirow{2}{*}{\makecell[c]{$p(y \mid \mathring{y}_\mathrm{min}, \mathring{y}_\mathrm{max}) $ \\ (Eq.~\eqref{eq:probdensity})}} 
 
& XAI Type   & O $\approx$ LR $<$ L$\Delta$ $<$ T$\Delta$ &  O $\approx$ LR $\ns{\approx}$ L$\Delta$ $<$ T$\Delta$ & Fig.~\ref{fig:summative_objective_res}c \\ 
&  & \# Comps    & 1 $<$ 4 $<$ 2 $\approx$ 3  & 1 $<$ 2 $\approx$ 3 $\approx$ 4 & Fig.~\ref{fig:summative_objective_res}f \\
\hline

Helpfulness of Comparables
&\multirow{2}{*}{7-pt Likert scale} 
& XAI Type   & O $<$ LR $<$ L$\Delta$ $<$ T$\Delta$ & O $\ns{\approx}$ LR $\ns{\approx}$ L$\Delta$ $<$ T$\Delta$ &\multirow{2}{*}{Fig.~\ref{fig:summative_subjective_res}}  \\

Helpfulness of Adjustments/Factors
& 
& XAI Type   & LR $<$ L$\Delta$ $<$ T$\Delta$ & LR $\ns{\approx}$ L$\Delta$ $\ns{\approx}$ T$\Delta$ &  \\

\arrayrulecolor{lightgray}\midrule
\arrayrulecolor{black}

Ease-of-use of Comparables
& \multirow{2}{*}{7-pt Likert scale} 
& XAI Type   & O $<$ LR $<$ L$\Delta$ $<$ T$\Delta$ & O $\ns{\approx}$ LR $\ns{\approx}$ L$\Delta$ $\ns{\approx}$ T$\Delta$  &\multirow{2}{*}{Fig.~\ref{fig:summative_subjective_res}}  \\

Ease-of-use of Adjustments/Factors
& 
& XAI Type   & LR $<$ L$\Delta$ $<$ T$\Delta$ & LR $\ns{\approx}$ L$\Delta$ $\ns{\approx}$ T$\Delta$ \\

\bottomrule
\end{tabular}
\end{table*}

\subsubsection{Experiment Apparatus}
The user interface 
was identical to that used in the formative study (Section 4.3.2). 

\subsubsection{Experiment Procedure}

Each participant was engaged in the following procedure:

\aptLtoX[graphic=no,type=html]{\begin{enumerate}
    \item[1)] Introduction to the study (see Appendix Fig.~\ref{fig:survey_application}).
    \item [2)]Consent to participant. The study was reviewed and approved by the university’s Institutional Review Board (IRB).
    \item [3)]Tutorial on application task and user interface (UI). 
    Different components are introduced depending on the specific conditions, with screening questions to ensure that participants can understand the task and use the explanation correctly (see Appendix Figs.~\ref{fig:tut_decision_tool}--\ref{fig:tut_trace_details}).
    \item[4)] Practice session of 4 cases (ranging from 1 Comparable to 4 Comparables), for each case:
\end{enumerate}
    \begin{enumerate}
         \item[i)] Participants were asked to estimate the actual value of a Subject (e.g., a house) using the decision tool together with the assigned explanations (see Appendix Fig.~\ref{fig:trace_practice}). 
        \item [ii)]The actual value was then revealed as feedback, enabling participants to identify potential mistakes or overly broad estimate intervals (see Appendix Fig.~\ref{fig:ans}). 
    \end{enumerate}
\begin{enumerate}
    \item[5)] Main testing session. Participants used the same explanations and decision tool, but no answers were revealed after each case (see Appendix Fig.~\ref{fig:trace_main}). 
    \item[6)]Answer ratings questions on Perceived helpfulness and ease-of-use for each component.
    \item[7)] Answer demographics questions (see Appendix Fig.~\ref{fig:post_survey}).
    \item[8)]Acknowledge bonus calculations and exit.
\end{enumerate}}{\begin{enumerate}[label=\arabic*)]
    \item Introduction to the study (see Appendix Fig.~\ref{fig:survey_application}).
    \item Consent to participant. The study was reviewed and approved by the university’s Institutional Review Board (IRB).
    \item Tutorial on application task and user interface (UI). 
    Different components are introduced depending on the specific conditions, with screening questions to ensure that participants can understand the task and use the explanation correctly (see Appendix Figs.~\ref{fig:tut_decision_tool}--~\ref{fig:tut_trace_details}).
    \item Practice session of 4 cases (ranging from 1 Comparable to 4 Comparables), for each case:
    \begin{enumerate}[label=\roman*)]
         \item Participants were asked to estimate the actual value of a Subject (e.g., a house) using the decision tool together with the assigned explanations (see Appendix Fig.~\ref{fig:trace_practice}). 
        \item The actual value was then revealed as feedback, enabling participants to identify potential mistakes or overly broad estimate intervals (see Appendix Fig.~\ref{fig:ans}). 
    \end{enumerate}
    \item Main testing session. Participants used the same explanations and decision tool, but no answers were revealed after each case (see Appendix Fig.~\ref{fig:trace_main}). 
    \item Answer ratings questions on Perceived helpfulness and ease-of-use for each component.
    \item Answer demographics questions (see Appendix Fig.~\ref{fig:post_survey}).
    \item Acknowledge bonus calculations and exit.
\end{enumerate}}

We provided an incentive bonus of £0.10 for each accurate response, defined as cases where the participant’s selected range contained the correct value and the relative error of the range width was within 10\%.

\aptLtoX[graphic=no,type=html]{\begin{table}[t]
\renewcommand{\arraystretch}{0.94}
\caption{Statistical analysis of responses using linear mixed-effects models. Each row reports one effect, including fixed effects, random effects, and their interactions. $F$ and $p$ values are from ANOVA tests.}
\label{tab:summative_statistical_ana}
\begin{tabular}{llrr}
\hline
Response & \makecell[l]{Linear Effect Model \\ (Random Effects: \\Participant, Case)} & $F$ & $p > F$  \\
\hline
\multirow{6}{*}{\makecell[l]{Decision \\Mean Error}}    
& XAI Type +  & 24.2 & \textless{}.0001  \\
& \# Comps +  & 38.5  & \textless{}.0001  \\
& Log(Comps Interval) + & 139.2 & \textless{}.0001  \\
& Domain + & 13.1  & .0003  \\
& \color{lightgray} Domain $\times$ XAI Type $+$ & \color{lightgray}1.7 & \color{lightgray}{n.s.}  \\
& \color{lightgray} Domain $\times$ \# Comps $+$ & \color{lightgray}1.4  & \color{lightgray}{n.s.}   \\
\hline


\multirow{6}{*}{\makecell[l]{Decision \\Credible Interval}}    
& XAI Type $+$  & 4.8 & .0030  \\
& \# Comps $+$  & 4.7  & .0031  \\
& Log(Comps Interval) $+$ & 260.6 & \textless{}.0001    \\
& \color{lightgray}Domain $+$ & \color{lightgray}0.3  & \color{lightgray}n.s.  \\
& \color{lightgray}Domain $\times$ XAI Type $+$ & \color{lightgray}1.7  & \color{lightgray}{n.s.}   \\
& \color{lightgray}Domain $\times$ \# Comps $+$& \color{lightgray}1.4 & \color{lightgray}{n.s.} \\
\hline


\multirow{6}{*}{\makecell[l]{Correctness \\Probability Density}}    
& XAI Type $+$  & 8.7 & \textless{}.0001   \\
& \# Comps $+$  & 7.6  & \textless{}.0001  \\
& Log(Comps Interval) $+$ & 30.4 & \textless{}.0001    \\
& \color{lightgray}Domain + & \color{lightgray}2.8  & \color{lightgray}n.s.  \\
& \color{lightgray}Domain $\times$ XAI Type $+$ & \color{lightgray}0.6  & \color{lightgray}{n.s.}   \\
& \color{lightgray}Domain $\times$ \# Comps & \color{lightgray}0.8  & \color{lightgray}{n.s.}    \\
 \hline 
 \multirow{2}{*}{\makecell[l]{Perceived Helpfulness \\ (Comparables)}}    
& \color{lightgray}XAI Type +  & \color{lightgray}8.7 & \color{lightgray}.0412  \\
& \color{lightgray}Domain +  & \color{lightgray}7.6  & \color{lightgray}.0086 \\
\hline
  \multirow{2}{*}{\makecell[l]{Perceived Helpfulness \\ (Adjustments / Factors)}}    
& \color{lightgray}XAI Type +  & \color{lightgray}1.9 & \color{lightgray} {n.s.}   \\
& \color{lightgray} Domain +  & \color{lightgray}1.5  & \color{lightgray} {n.s.}  \\
\hline
 
\multirow{2}{*}{\makecell[l]{Perceived Ease-of-Use \\ (Comparables)}}    

& \color{lightgray}XAI Type +  & \color{lightgray}1.1 & \color{lightgray} {n.s.}  \\

& Domain +  & 10.5  & .0014  \\
\hline

\multirow{2}{*}{\makecell[l]{Perceived Ease-of-Use \\ (Adjustments / Factors)}}    

& \color{lightgray}XAI Type +  & \color{lightgray}2.1 & \color{lightgray} n.s. \\

& \color{lightgray}Domain  & \color{lightgray}4.3  & \color{lightgray}.0394 \\

\hline
\end{tabular}
\end{table}}{{\setlength{\tabcolsep}{3pt}
\begin{table}[t]
\renewcommand{\arraystretch}{0.95}
\caption{Statistical analysis of responses using linear mixed-effects models. Each row reports one effect, including fixed effects, random effects, and their interactions. $F$ and $p$ values are from ANOVA tests.}
\label{tab:summative_statistical_ana}
\begin{tabular}{llrr}
\toprule
Response & \makecell[l]{Linear Effect Model \\ (Random Effects: \\Participant, Case)} & $F$ & $p > F$  \\
\hline
\multirow{6}{*}{\makecell[l]{Decision \\Mean Error}}    
& XAI Type +  & 24.2 & \textless{}.0001  \\
& \# Comps +  & 38.5  & \textless{}.0001  \\
& Log(Comps Interval) + & 139.2 & \textless{}.0001  \\
& Domain + & 13.1  & .0003  \\
& \color{lightgray} Domain $\times$ XAI Type $+$ & \color{lightgray}1.7 & \color{lightgray}n.s.  \\
& \color{lightgray} Domain $\times$ \# Comps $+$ & \color{lightgray}1.4  & \color{lightgray}n.s.   \\
\arrayrulecolor{lightgray}\midrule
\arrayrulecolor{black}

\multirow{6}{*}{\makecell[l]{Decision \\Credible Interval}}    
& XAI Type $+$  & 4.8 & .0030  \\
& \# Comps $+$  & 4.7  & .0031  \\
& Log(Comps Interval) $+$ & 260.6 & \textless{}.0001    \\
& \color{lightgray}Domain $+$ & \color{lightgray}0.3  & \color{lightgray}n.s.  \\
& \color{lightgray}Domain $\times$ XAI Type $+$ & \color{lightgray}1.7  & \color{lightgray}n.s.   \\
& \color{lightgray}Domain $\times$ \# Comps $+$& \color{lightgray}1.4 & \color{lightgray}n.s. \\
\arrayrulecolor{gray}\midrule
\arrayrulecolor{black}

\multirow{6}{*}{\makecell[l]{Correctness \\Probability Density}}    
& XAI Type $+$  & 8.7 & \textless{}.0001   \\
& \# Comps $+$  & 7.6  & \textless{}.0001  \\
& Log(Comps Interval) $+$ & 30.4 & \textless{}.0001    \\
& \color{lightgray}Domain + & \color{lightgray}2.8  & \color{lightgray}n.s.  \\
& \color{lightgray}Domain $\times$ XAI Type $+$ & \color{lightgray}0.6  & \color{lightgray}n.s.   \\
& \color{lightgray}Domain $\times$ \# Comps & \color{lightgray}0.8  & \color{lightgray}n.s.    \\
 \midrule 
 \multirow{2}{*}{\makecell[l]{Perceived Helpfulness \\ (Comparables)}}    
& \color{lightgray}XAI Type +  & \color{lightgray}8.7 & \color{lightgray}.0412  \\
& \color{lightgray}Domain +  & \color{lightgray}7.6  & \color{lightgray}.0086 \\
 \arrayrulecolor{lightgray}\midrule
  \multirow{2}{*}{\makecell[l]{Perceived Helpfulness \\ (Adjustments / Factors)}}    
& \color{lightgray}XAI Type +  & \color{lightgray}1.9 & \color{lightgray} n.s.   \\
& \color{lightgray} Domain +  & \color{lightgray}1.5  & \color{lightgray} n.s.  \\
\arrayrulecolor{lightgray}\midrule
 
\multirow{2}{*}{\makecell[l]{Perceived Ease-of-Use \\ (Comparables)}}    

& \color{lightgray}XAI Type +  & \color{lightgray}1.1 & \color{lightgray} n.s.  \\

& Domain +  & 10.5  & .0014  \\
 \arrayrulecolor{lightgray}\midrule

\multirow{2}{*}{\makecell[l]{Perceived Ease-of-Use \\ (Adjustments / Factors)}}    

& \color{lightgray}XAI Type +  & \color{lightgray}2.1 & \color{lightgray} n.s. \\

& \color{lightgray}Domain  & \color{lightgray}4.3  & \color{lightgray}.0394 \\

\arrayrulecolor{black}
\bottomrule
\end{tabular}
\end{table}
}}
\newpage
\subsubsection{Participants} 
We recruited 330 participants from Prolific.co.
They had a median age 38 years old (19 to 78), and were 49\% female. Participants completed the survey in a median time of 52 min, and were compensated with a base of £6.00 and a Median bonus of \pounds 0.6.

\subsubsection{Statistical Analysis}
We performed linear mixed-effects model fits for each dependent variable, with XAI type, Number of Comparables, Domain, and Comparables Interval\footnote{Our formative study revealed that participants looked at the range of adjusted values from Comparables to influence their range estimates. We applied log transform to control the long tail of this measure, i.e., Log(Comparables Interval).} as fixed effects, and Case and Participant as random effects. 
See Table~\ref{tab:summative_statistical_ana} for detailed results\footnote{Example linear mixed effects model equation (Table~\ref{tab:summative_statistical_ana} first row): 
Decision Mean Error $\sim$ XAI Type + \# Comps + Log(Comparables Interval) + Domain + Domain $\times$ XAI Type + Domain $\times$ \# Comps + 1$\mid$Participant + 1$\mid$Case.}
 To ensure generalizability across datasets, both domains were standardized before analysis, allowing estimated influences to be directly comparable across domains.

\subsubsection{Quantitative Results}
We describe
i) how accurately participants estimated the actual values, 
ii) their precision in these decisions, and 
iii) their perceptions of the helpfulness and ease of use.
Table~\ref{tab:summative_statistical_ana} summarizes the significant effects found. 
We consider results with $p \leq .005$ as statistically significant to account for up to 10 multiple comparisons up and avoid Type I error.
For specific comparisons, we performed post-hoc contrast tests.

\begin{figure}[t]
    \centering
    \includegraphics[width=1\linewidth]{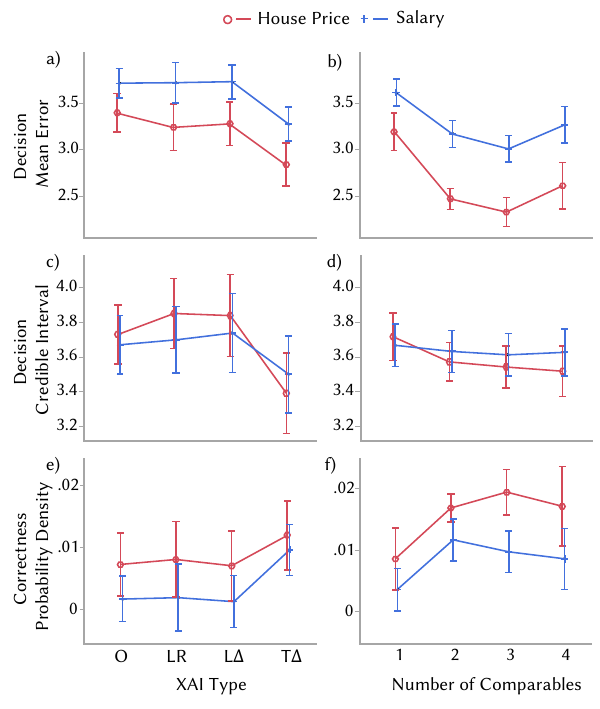}
    \caption{Results from the summative user study trials evaluating participants’ ability to estimate actual values across two domains (House Price and Salary). We compare how different XAI types (O: Comparables only, LR: Comparables with Linear Regression, L$\Delta$: Comparables with Linear Adjustments, and T$\Delta$: Comparables with Trace Adjustments.) and Number of Comparables influence the Decision Accuracy (a--b), Decision Credible Interval (c--d), and Correctness Probability Density (e--f).
    Error bars indicate 95\% confidence interval.
    }
    \Description{
    This figure contains six line charts labeled a through f. Each chart compares two domains—House Price and Salary—on participants’ decision-making outcomes. The vertical axis shows the dependent variable, while the horizontal axis shows XAI type or Number of Comparables. Error bars represent 95\% confidence intervals.
    a) Decision Mean Error by explanation type. Decision error values range between 2.8 and 3.8. For House Price, error decreases from about 3.3 under Comparables Only to around 3.0 under Trace Adjustments. For Salary, error begins higher at about 3.7, stays near 3.6 under Linear Regression and Adjustments, then drops to 3.3 under Trace Adjustments. Overall, House Price tasks show lower error than Salary, and both domains improve slightly with more advanced explanations.
    b) Decision Mean Error by number of comparables. The horizontal axis ranges from 1 to 4 comparables. For House Price, error starts around 3.2 with one comparable, drops sharply to 2.3 at 3 comparables, then rises slightly to 2.6 at 4 comparables. For Salary, error starts higher at about 3.6, decreases steadily to 3.0 at 3 comparables, and increases slightly to 3.2 at 4 comparables. In both domains, the lowest error occurs when participants see 3 comparables.
    c) Decision Credible Interval by explanation type. Values range between 3.5 and 4.1. For House Price, intervals begin near 3.7, rise to 4.0 under Linear Adjustments, then drop to 3.5 under Trace Adjustments. For Salary, values start lower at about 3.6, rise slightly to 3.9 under Linear Adjustments, and fall to 3.5 with Trace Adjustments. Both domains show narrower intervals with Trace Adjustments.
    d) Decision Credible Interval by number of comparables. Intervals remain stable between 3.5 and 3.8 across all conditions. House Price begins slightly higher at 3.8 with one comparable, then decreases to 3.5 by four comparables. Salary follows a similar pattern, staying around 3.6 to 3.7. Overall, credible intervals shrink slightly as the number of comparables increases.
    e) Correctness Probability Density by explanation type. Values range from 0.0 to 0.02. For House Price, density rises gradually from about 0.008 with Comparables Only to nearly 0.015 with Trace Adjustments. For Salary, density is close to zero under Comparables and Linear Regression, rises to 0.01 under Linear Adjustments, and 0.012 with Trace Adjustments. House Price consistently shows a higher density than Salary.
    f) Correctness Probability Density by number of comparables. For House Price, density starts around 0.01 with one comparable, increases to 0.02 at three comparables, then levels off near 0.018 at four comparables. For Salary, density begins near 0.002, rises to 0.012 at three comparables, then declines slightly at four comparables. Both domains show the highest probability density at three comparables, with House Price higher overall.
    }
    \label{fig:summative_objective_res}
\end{figure}

\textbf{Decision Mean Error.} 
Fig.~\ref{fig:summative_objective_res}a shows that participants using Comparables w/ Trace Adjustments had significantly lower error than in all baseline conditions (contrast test: $p<.0001$), 
%

\begin{figure}[t]
    \centering
    \includegraphics[width=0.98\linewidth]{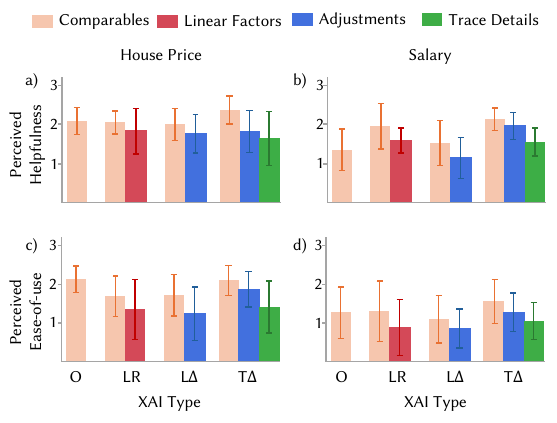}
    \caption{Perceived helpfulness (a, c) and perceived ease-of-use (b, d) of different XAI types across two domains: house price (a, b) and salary (c, d). Bars show mean ratings with error bars indicating 95\% confidence interval. Comparables-only (light orange), Linear Factors (red), Adjustments (blue), and Trace Details (green) were evaluated across four explanation settings: O: Comparables only, LR: Comparables with Linear Regression, L$\Delta$: Comparables with Linear Adjustments, and T$\Delta$: Comparables with Trace Adjustments.}
    \Description{
    This figure contains four bar charts labeled a through d. Each chart compares four explanation types—Comparables Only, Linear Regression, Linear Adjustments, and Trace Adjustments—on participants’ subjective ratings. The vertical axis represents ratings on a scale from 0 to 3, with higher values meaning greater helpfulness or ease-of-use. Error bars represent variation across participants.
    a): Perceived Helpfulness for House Price. All explanation types are rated between 1.7 and 2.1. Comparables Only and Linear Regression reach about 2.0, while Linear Adjustments and Trace Adjustments are slightly lower, around 1.8. Overall differences are small, with no clear standout method.
    b): Perceived Helpfulness for Salary. Ratings range from about 1.2 to 2.1. Comparables Only is lowest, around 1.3. Linear Regression is around 1.7, Linear Adjustments is about 1.1, and Trace Adjustments is highest, about 2.0. Here, Trace Adjustments are rated noticeably more helpful than other methods.
    c): Perceived Ease-of-Use for House Price. Comparables Only scores around 2.1, the highest among the four. Linear Regression is lowest at about 1.5. Linear Adjustments and Trace Adjustments both fall around 1.8 to 2.0. Participants generally find Comparables Only easiest to use, with regression the hardest.
    d): Perceived Ease-of-Use for Salary.Ratings range from 1.0 to 1.6. Comparables Only and Trace Adjustments are both around 1.4. Linear Regression is lowest at about 1.0, while Linear Adjustments is about 1.2. Overall ratings are lower here than in the House Price domain, with none of the methods perceived as highly easy to use.
    }
    \label{fig:summative_subjective_res}
\end{figure}
Fig.~\ref{fig:summative_objective_res}b shows that error decreased from one to two Comparables ($p<.0001$), and remained low from two to three Comparables ($p=.0132$), but then increased from three to four Comparables ($p=.0014$).
Notably, participants were more accurate on the House Price domain than on the Salary domain ($p=.0003$), suggesting that the house valuation task may have been relatively easier.

\textbf{Decision Credible Interval.} 
Fig.~\ref{fig:summative_objective_res}b shows that participants using Comparables w/ Trace Adjustments indicated significantly narrower intervals than baseline XAI Types (contrast test: p = .0008).
Fig.~\ref{fig:summative_objective_res}d shows that participants were more precise with two Comparables compared to one ($p<.0029$), but there was no further improvement with more Comparables. 

\textbf{Correctness Probability Density.} 
Fig.~\ref{fig:summative_objective_res}e shows that participants using Comparables w/ Trace Adjustments were significantly better than baseline XAI Types (contrast test: $p < .0001$)
For the Number of Comparables, increasing from one to two Comparables significantly improved correctness ($p<.0001$), but no more with more Comparables. 

\textbf{Perceived Ratings.} 
Fig.~\ref{fig:summative_subjective_res} shows the Perceived Helpfulness and Perceived Ease-of-Use ratings for each explanatory feature.
Generally, there were no significant differences across features, but ratings were mostly positive.
Though not statistically significant, Trace Details (showing the columns of counterfactuals in the trace) are less helpful and easy to use. 
This was also remarked by participants in the formative study.
\subsubsection{Summary of results}
\textbf{Comparables with Trace Adjustment} are most helpful to improve decision accuracy and precision.
Effect is most pronounced at 2--3 Comparables, and may begin to taper off from 3--4 Comparables.
All XAI Types are perceived as helpful and somewhat easy to use.

\subsection{Qualitative Expert User Study}
Having shown that Comparables XAI improves user understanding for lay users, we next investigate if expert users who regularly handle Comparables would need such explanations.
Based on the housing price prediction task, we conducted a user study with real-estate professionals (experts) to investigate:
1) how they employ Comparables and adjustments in their day-to-day operations,
2) what explanations they would need if they had an AI decision support tool, and
3) whether Comparables XAI with Trace Adjustments would be useful for their workflow.

\subsubsection{Experiment Procedure} 
Experts were first invited to describe their current property valuation practices, including the typical procedures and challenges encountered.
We then explored whether and how they would integrate AI into their appraisal workflow and their needs for AI explanations.
Next, we presented different explanation variants (No Explanation, Comparables w/ Linear Adjustments, Comparables w/ Trace Adjustments w/o Sparsity, Trace Adjustments w/o Disjointness, Trace Adjustments w/o Monotonicity, Trace Adjustments w/o Evenness, and Comparables w/ Trace Adjustments), and the experts evaluated their clarity and practical usefulness, suggesting improvements or alternatives.
The interviews were conversational, with flexible follow-up questions to probe deeper. 
Sessions were recorded and fully transcribed for thematic analysis.

\subsubsection{Participants}
We recruited 5 real estate professionals (E1–E5) with an average of 6 years of valuation experience (range: 3–7 years).
E1 specializes in residential valuation and has additional experience assessing restaurant properties. 
E2 and E3 focus on residential valuation, with occasional involvement in commercial properties. 
E4 conducts residential valuations for banks to support mortgage approvals and ensure alignment between loan amounts and assessed property values, and additionally performs investment-focused valuations for external family offices.
E5 is a business analyst in the real estate sector, collaborating closely with valuation professionals.

\subsubsection{Findings} \label{sec:expert_user_study_finding}
We report key findings by our research objectives: 
1) experts’ challenges in using Comparables in valuation, 
2) their views on using AI in valuation and the need for XAI, and 
3) opinion regarding the counterfactual and trace desiderata of Section~\ref{sec:desiderata}.

\textbf{Comparables} are \textbf{widely used} in valuation, and \textbf{adjustments} are \textbf{necessary} but often \textbf{complex}.
\textit{``Real estate has inherent heterogeneity… each unit differs in numerous attributes,''} making it \textit{``impossible to determine price from a universal rule''} (E1), and \textit{``even close Comparables still differ in meaningful ways''}(E3).
As a result, \textbf{adjustments are necessary}, but there are \textit{``no strict rules, different companies use different calculation methods''} (E2), and these \textbf{adjustment rules can be complex and opaque}, as E4 mentioned, \textit{``we only use them, not derive them''}.
Moreover, adjustments \textit{``can be non-linear, context-dependent, and sometimes require subjective judgment''} (E4). 
In practice, applying adjustments to Comparables requires nuanced judgment and substantial cognitive effort.

Experts are \textbf{receptive to AI}, but \textbf{explanations} are \textbf{desired}. 
All interviewed experts indicated willingness to try AI-supported tools, recognizing the growing integration and usage of AI in the valuation process. As several experts remarked, \textit{``AI makes predictions based on data, and we also conduct valuation using data. But AI can examine far more data than I can, so the information it provides could serve as a useful reference''} (E3) and \textit{``I always struggle to find enough relevant examples ... wish AI could help me find more''} (E1).
However, they also expressed the demand of explanations for AI predictions. E4 \textit{``would not trust an AI prediction without any explanations.''} 
Furthermore, the experts expressed a preference for AI systems that justify valuations using Comparables in a manner consistent with their current practice, as remarked by E2, \textit{``explaining the AI’s prediction through comparables and adjustments is ideal, as it aligns with our professional practice and enables us to quickly verify and understand the reasoning''}.
Next, we discuss detailed attitudes toward the explanation desiderata we implemented for Comparables with Trace Adjustment.

\textbf{Trace adjustments} are \textbf{helpful}, but the hypothetical cases add \textbf{uncertainty} to valuation.
Experts appreciated that the in-between hypothetical cases can serve as additional reference points. 
E3  \textit{``often struggle to find sufficient Comparables. The trace provides hypothetical properties with AI-generated labels, which can serve as additional references.''}
Experts also valued the trace adjustments' ability to capture \textbf{non-linear relationships}:  
\textit{``it captures non-linear changes, which feels more accurate''} (E2).
However, E1 emphasized that:
\textit{``these in-between hypothetical cases lack real market prices and aren’t reliable. Adjusting the next step based on them may increase uncertainty in real practice. ''}

Trace adjustments are expected to be \textbf{both Faithful and simple}, and the experts especially appreciated \textbf{Sparsity}, as E3 noted that \textit{``less information makes it easier to compare''}. They also valued \textbf{Disjointness} in the trace: adjusting one attribute per step, with no attribute revisited, was seen as \textit{``clear and aligns with our real workflow''} (E4).   
However, relaxing \textbf{Disjointness} for faithfulness is also preferred; e.g., E3 noted that
\textit{``if one attribute is adjusted several times to show its more detailed non-linear influence, that would be fine.''}
Experts acknowledged that \textit{``[\textbf{Monotonicity}] feels natural as adjustments progressively move closer to the Subject property''} (E5). However, \textit{``actual adjustments often include both positive and negative components, making strict monotonic behavior unrealistic.''} 
Thus, non-monotonic adjustments are also common and acceptable to experts.
Regarding \textbf{Evenness}, experts expressed reliability concerns when adjustments were too large:  
\textit{``when an [attribute] adjustment is too large, it becomes questionable ... [this] suggests that the comparables are not similar enough''} (E2).

In summary, experts saw trace adjustments as a useful way to explain AI reasoning, especially for visualizing in-between Comparables (Counterfactuals) and non-linear relationships. However, they emphasized that these traces should not be interpreted as actual market data and should be clearly presented as hypothetical. 
Experts prefer small, simple, and one-attribute-at-a-time adjustments that match their professional practice. 
Overall, participants welcomed the trace adjustment as an explainable decision support tool that complements their professional reasoning.

\section{Discussion}
We introduce the paradigm of \textit{Comparables XAI}, which seeks to generate more faithful example-based explanations while supporting more accurate and confident AI-supported decision-making. 
Here, we discuss our key findings, implications, generalizability, and limitations of Comparables XAI.

\subsection{Key Findings and Implications}

We summarize our key findings and corresponding implications as follows.

\subsubsection{Desire for Trace of Counterfactuals}
Our results show that users preferred closer Comparables and viewed distant ones as less reliable. They were willing to consider closer hypothetical counterfactuals than farther real examples, but with some reduced trust.
Future work can compare how close Counterfactuals against distant real Comparables influence users’ trust, cognitive load, and decision-making.

Furthermore, instead of only one, participants also appreciated multiple counterfactuals arranged as a desiderata-optimized trace. This was helpful for revealing nonlinear trends while keeping the explanation simple.
We found that some desiderata were more effective and preferable than others. 
Participants valued Sparsity and Disjointness for reducing information load, whereas Monotonicity was seen as less important.
This conflicts with the axiomatic considerations in XAI (e.g., \cite{hamer2024simple, chakraborty2021picking, maiti2025counterfactual}), which emphasize Monotonicity.
This could be due to differences in decision goals (estimation vs. recourse).
Further work is needed to evaluate these desiderata through controlled experiments across different tasks and goals.

\subsubsection{Analytical Paradigm for Example-Based XAI}
Example-based explanations are popular due to their intuitive representation of cases accessible to lay users.
Indeed, prior studies showed that example-based explanations outperform other formats (e.g., attribution-based) due to their intuitiveness~\cite{fel2023craft, humer2022comparing, naiseh2023different}.
Most user research on example-based explanations focuses on their intuitive appeal.
Chen et al.~\cite{chen2023understanding} found that incorrectly predicted examples prompt users to compare similarities and override both the AI’s prediction and their own initial intuition.
Cai et al.~\cite{cai2019effects} examined different kinds of visual examples, showing that prototype example improves understanding and increase trust, while counterfactuals reveal model limitations.
Additionally, Kuhl et al.~\cite{kuhl2023better} showed that upward, or better than fact, counterfactual examples are especially intuitive for guiding decisions.
Overall, these findings focus on the usage of examples in perceptual judgments.

In contrast, we examine the analytical usage of examples by presenting counterfactual traces between example and the target.
Our results show that users valued this analytical framing, gaining a more accurate understanding.
Future work can compare intuitive and analytical example-based explanations to investigate how cognitive demands differ and when each is better for different tasks.

\subsection{Generalizing Comparables XAI}

We have focused on valuation tasks, such as estimating house prices and salaries. This paradigm could naturally extend to other valuation domains, including company valuation~\cite{koller2010valuation}, energy consumption estimation~\cite{kim2022derivation}, and insurance premium prediction~\cite{nissim2013relative}. 
Across valuation domains, the use of Comparables is a fundamental principle, as prices are typically determined by reference to similar cases in order to ensure fairness and alignment with market standards~\cite{damodaran2007valuation}.

Beyond valuation, we also test on drug sensitivity prediction and agricultural crop yield estimation, both of which similarly rely on analogical reasoning~\cite{azuaje2017computational,liu2021analogy}. 
More broadly, Comparables also play an important role in other areas of decision-making. For example, in health technology assessment, indirect comparisons via a common comparator (anchored comparisons) are widely used~\cite{phillippo2019population}, often requiring careful statistical adjustments. Similarly, in public policy and policy evaluation, the assessment of new policy implementations typically relies on comparisons across different regions or time periods~\cite{zeldow2021confounding}. 
In sum, indirect comparisons and careful adjustments are widely used across many domains to ensure fairness and validity. \textit{Comparables XAI}, grounded in reasoning by Comparables with Adjustments, can therefore support not only valuation tasks but also a broad range of non-valuation decision-making contexts.

Our approach had also focused on regression tasks where the goal is precise numerical estimation, and the adjustment values are expressed as continuous values that align with the units of the predicted outcome. 
Nevertheless, it can be adapted for classification tasks by operating on class probabilities as the continuous value, rather than discrete class labels.
For example, a disease risk score can indicate if a patient will likely have a disease~\cite{lloyd2019use}.

\subsection{Scope of Comparables XAI}
\textit{Comparables XAI} is targeted toward domains where decisions require analytic, multi-attribute comparison using examples, rather than intuitive or perceptual judgment based on examples.
Example-based explanations are effective when similarity can be perceived effortlessly, as in image or audio recognition, people intuitively match patterns without explicit reasoning~\cite{renkl2014toward,cai2019effects,zhang2022towards}.
In contrast, many domains \cite{phillippo2019population,zeldow2021confounding} require decisions based on careful comparison across multiple attributes and criteria, where intuitive matching alone is insufficient. 
In such cases, examples must be examined analytically and compared systematically, precisely the analytic mode that \textit{Comparables XAI} supports.
However, just as visual analytic tools are designed for professional or expert users, we do not argue for lay users to use \textit{Comparables XAI} for everyday tasks.
Indeed, our formative results found that less technically savvy users may not be familiar with the calculations communicated in the spreadsheet.

Our approach is model-agnostic and applicable to any discriminative AI model.
While we focused on generating counterfactuals from the decision surface of such models, generative AI models could also be used.
Generative AI models like GAN~\cite{goodfellow2014generative}, VAE~\cite{kingma2013auto} model the relationships between attributes, so they can generate more plausible examples (counterfactuals) than discriminative models.
Moreover, advanced generative models like LLMs~\cite{achiam2023gpt, gat2023faithful} support users to interactively prompt for Comparable examples using natural language. This can leverage the large pre-trained or fine-tuned knowledge base to generate alternative Comparables and enable user-driven specification of Comparables.
Future work can investigate how this creative freedom can support critical analysis.

\subsection{Limitations and Future Work}
Our investigation was based on several assumptions and comes with certain limitations.

\subsubsection{Assumption of Preselected Comparables}

Our Comparables XAI focuses on interpolating Comparables via counterfactual traces with the assumption that these real Comparables have already been given. However, in practice, selecting Comparables is a non-trivial task, especially when many candidate cases exist.
The interviewed experts expressed a wish that AI could help find suitable ones, given the difficulty of finding sufficiently similar Comparables (Section \ref{sec:expert_user_study_finding}).
Improving user interaction, future work can explore user interactions with technical methods, such as example organization and visualization techniques that retrieve examples in response to user queries~\cite{hsu2011case}.
On the other hand, future technical work can also investigate Comparable selection algorithms considering criteria such as increasing the number of Comparables and reducing the average distance of Comparables (as we show in Fig.~\ref{fig:modeling_study} and in \cite{vandell1991optimal,koller2010valuation,nissim2013relative}),
minimizing the variance of the combined prediction~\cite{vandell1991optimal,mccluskey2017theory}, 
preferring recent rather than outdated Comparables~\cite{RICS2019ComparableEvidence,IAAO2013Standard}, and 
selecting representative Comparables that provide good coverage of the overall instance distribution~\cite{IAAO2013Standard,Damodaran2012Relative}.

\subsubsection{Limited Experimentation on Confounders}
In our studies, we primarily examined the effects of the number and distance of Comparables. 
However, other confounders have been shown to affect the usefulness of explanations.
For example, \textit{prediction errors} on Comparables significantly affect the decision~\cite {chen2023understanding} and trust~\cite{papenmeier2022s, jeung2023correct}.
Thought we found evidence in our qualitative study (Section~\ref{sec:formative_findings}) that such errors diminish user confidence.
In addition, the \textit{directionality} of counterfactuals (e.g., all worse, all better, or mixed) also impacts user decisions~\cite{kuhl2023better}.
Given the increased complexity of showing multiple Comparables and counterfactuals, these issues could be amplified, and future work should examine them.

\subsubsection{Theoretical Guarantees on Explanation Faithfulness}
Our research approach follows the user-centered research principles in HCI with determining user needs or practice (via literature), technology solution, and evaluation with users.
We substantiate the benefits of our solution with empirical results from modeling, formative, and summative user studies.
However, unlike theoretical AI works, we did not formally prove the theoretical guarantees on the faithfulness of our Comparables trace adjustment XAI, given the complexity of the AI decision surface.
Such rigorous proofs are beyond the scope of this work.
Future work can investigate this with formal methods, such as bounding test error using training error as in~\cite{li2021}, theoretical analyses of piecewise linear approximations~\cite{stampfle2000optimal,singh2018mesh}, and error analyses of linear functions used to approximate nonlinear functions~\cite{althoff2008reachability,hsu2014random}.

\section{Conclusion}
We introduced \textit{Comparables XAI} as a new analytical paradigm that leverages example-based explanations to improve user understanding of AI decisions.
We proposed \textit{Comparables with Trace Adjustments} as a technical approach that generates incremental counterfactuals tracing from Comparables to the Subject, guided by desiderata aligned with human intuition. Piecewise linear formulations were adopted to fit the traces.
In the modeling study, we validated that \textit{Comparables with Trace Adjustments} achieved the highest faithfulness compared with unadjusted Comparables, 
Comparables with Linear Regression, and Comparables with Linear Adjustments.
In user studies, \textit{Comparables with Trace Adjustments} was perceived as exhibiting greater consistency and practical interpretability. Quantitatively, it not only enhanced users’ decision accuracy but also reinforced their confidence.
This work demonstrates the efficacy of trace-based adjustments to improve both faithfulness and intuitiveness, thereby fostering users’ confidence in AI-assisted decision-making through a deeper understanding of the AI system.

\begin{acks}
This research is supported by the National Research Foundation, Singapore and Infocomm Media Development Authority under its Trust Tech Funding Initiative (Award No: DTC-RGC-09) and the NUS Institute for Health Innovation and Technology (iHealthtech). Any opinions, findings and
conclusions or recommendations expressed in this material are those of the author(s) and do not reflect the views of National Research Foundation, Singapore and Infocomm Media Development Authority.
\end{acks}

\balance
\bibliographystyle{ACM-Reference-Format}
\bibliography{manuscript}

\onecolumn
\appendix
\section{Appendix}
\label{sec:modeling_domain_generalization}

\subsection{Assessing the Generalizability}
\begin{figure}[ht]
    \centering
    \includegraphics[width=0.78\linewidth]{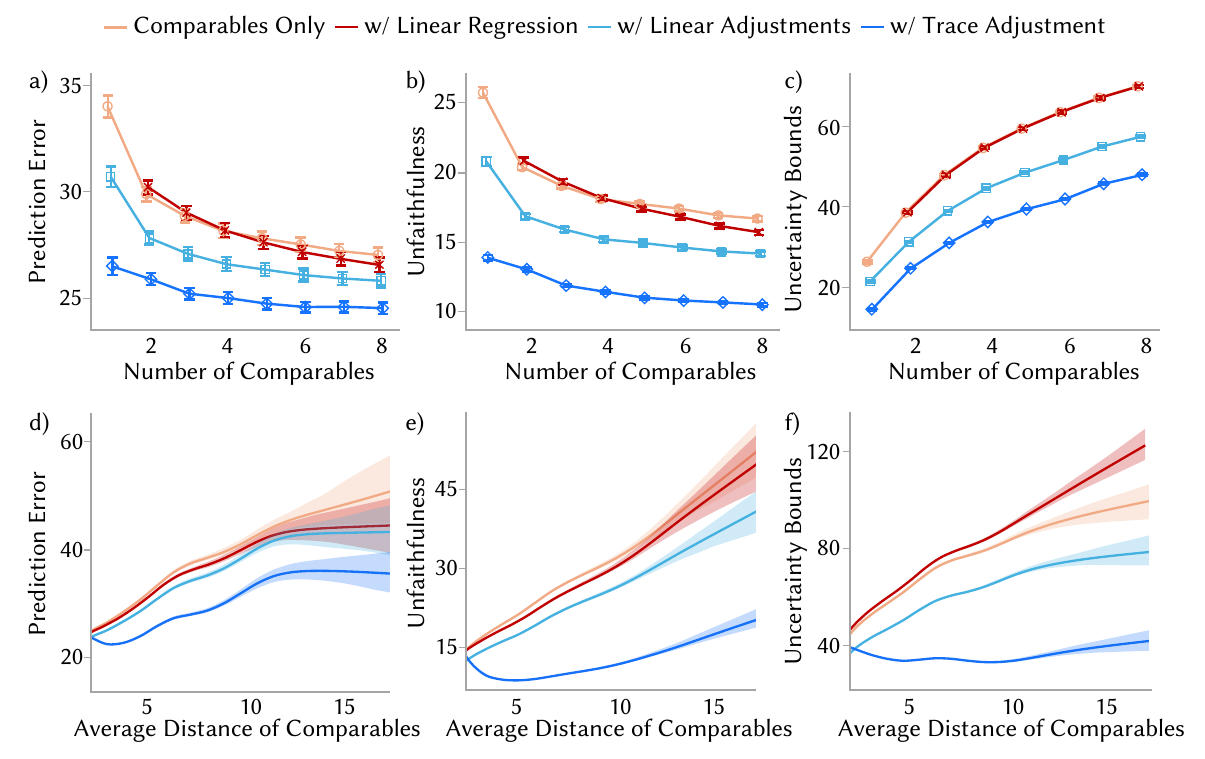}
    \caption{Modeling study results for different XAI types in the Salary domain, varying the Number of Comparables (a--c) and the Average Distance of Comparables (d--f).
     Increasing the Number of Comparables reduces a) prediction error and b) unfaithfulness across all methods, while c) uncertainty bounds become broader. 
     Increasing the Average Distance of Comparables worsens all metrics (d--f). Trends are smoothed with a cubic spline (smoothing parameter $\lambda=1000$). 
    Across all conditions, Comparables w/ Trace Adjustments consistently yield lower prediction error (a, d), lower unfaithfulness (b, e), and narrower uncertainty bounds (c, f).
    }
    \Description{
    This figure contains six charts labeled a through f. Each chart compares four explanation types: Comparables Only, Linear Regression, Linear Adjustments, and Trace Adjustments on quantitative modeling metrics in the Salary domain. Error bars or shaded regions represent variation across runs.
    a): Prediction Error vs. Number of Comparables. Error decreases for all methods as the number of comparables increases. Comparables Only starts highest around 34 and declines to about 27. Linear Regression begins near 31 and drops to about 26. Linear Adjustments fall from 28 to about 25. Trace Adjustments are consistently lowest, declining from 26 to around 23.
    b): Unfaithfulness vs. Number of Comparables. All methods decrease as the number of comparables grows. Comparables Only begins near 25 and ends near 18. Linear Regression follows closely, around 24 down to 18. Linear Adjustments start near 20 and end near 15. Trace Adjustments remain lowest, from about 15 down to 10.
    c): Uncertainty Bounds vs. Number of Comparables. All methods increase with more comparables. Comparables Only ranges from about 30 up to 55. Linear Regression spans about 35 to 65. Linear Adjustments rise from about 25 to 50. Trace Adjustments remain narrowest, from about 20 to 40.
    d): Prediction Error vs. Average Distance of Comparables. Error rises with distance for all methods. Linear Regression and Comparables Only increase steeply, reaching 50 or higher at large distances. Linear Adjustments climb more moderately, peaking around 45. Trace Adjustments remain lowest, around 35 to 40 even at high distances.
    e): Unfaithfulness vs. Average Distance of Comparables. All methods rise with distance. Linear Regression and Comparables Only reach 40 or more at high distances. Linear Adjustments peak near 35. Trace Adjustments rise least, from about 15 to 25.
    f): Uncertainty Bounds vs. Average Distance of Comparables. All methods widen substantially with distance. Linear Regression grows most steeply, exceeding 120. Comparables Only and Linear Adjustments reach about 100. Trace Adjustments expand more slowly, staying near 70 at the largest distances.
    }
    \label{fig:app_modeling_salary}
    
\end{figure}

\begin{figure}[H]

    \centering
    \includegraphics[width=0.72\linewidth]{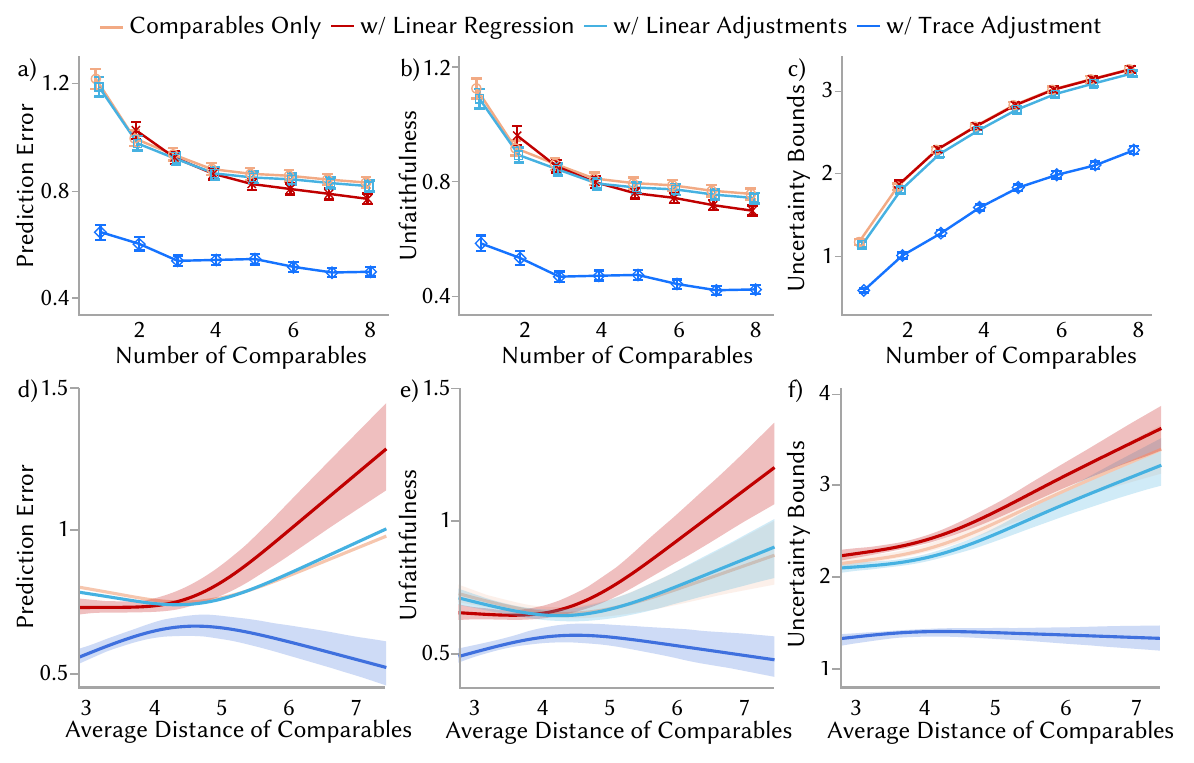}
    \caption{ Modeling study results for different XAI types in the Energy domain.
     Increasing the Number of Comparables reduces a) prediction error and b) unfaithfulness across all methods, while c) uncertainty bounds become broader. 
     Increasing the Average Distance of Comparables worsens all metrics, whereas Trace Adjustments remain stable and unaffected (d--f). Trends are smoothed with a cubic spline (smoothing parameter $\lambda=1000$).
    }
    \Description{
    This figure presents six modeling results for different XAI methods in the Energy domain. The top row (a--c) varies the Number of Comparables, and the bottom row (d--f) varies the Average Distance of Comparables, computed using Manhattan distance on standardized features. All plots compare four explanation types: Comparables Only, Linear Regression, Linear Adjustments, and Trace Adjustments, with shaded regions showing variability.

    (a) Prediction Error vs. Number of Comparables. All methods show decreasing error as more comparables are used. Comparables Only begins near 1.2 and declines to ~0.8. Linear Regression and Linear Adjustments follow similar trends. Trace Adjustments achieve the lowest error throughout, dropping from ~0.7 to ~0.45.
    (b) Unfaithfulness vs. Number of Comparables.
    Unfaithfulness declines for all methods as the number of comparables increases. Comparables Only and Linear Regression decrease from ~1.1 to ~0.8. Linear Adjustments fall from ~1.0 to ~0.8. Trace Adjustments remain lowest and most stable, decreasing from ~0.6 to ~0.4.
    (c) Uncertainty Bounds vs. Number of Comparables.
    Uncertainty widens as more comparables are added. Comparables Only, Linear Regression, and Linear Adjustments increase from ~1.5 to above 3.0. Trace Adjustments grow more slowly, from ~0.7 to ~2.2, maintaining the narrowest bounds.
    (d--f) Effects of Average Distance of Comparables.
    Increasing the average distance worsens prediction error (d), unfaithfulness (e), and uncertainty (f) for all methods except Trace Adjustments. Trace Adjustments remain stable and largely unaffected across distance levels. Trends are smoothed with a cubic spline (λ = 1000). Overall, Trace Adjustments consistently produce lower prediction error, lower unfaithfulness, and tighter uncertainty bounds across all conditions.}
    \label{fig:app_modeling_energy}
\end{figure}

\begin{figure}[H]
    \centering
    \includegraphics[width=0.72\linewidth]{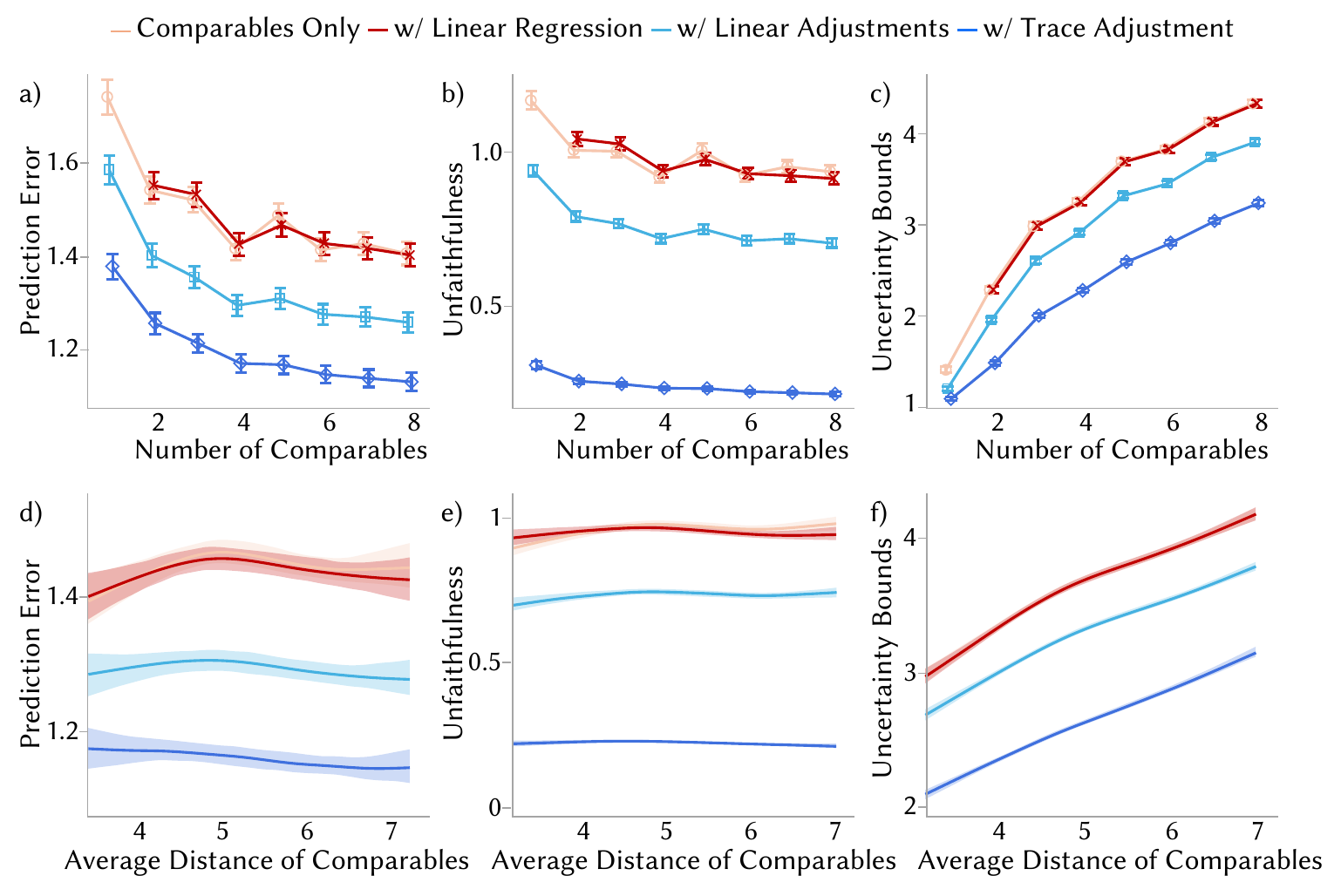}
    \caption{Modeling study results for different XAI types in the Drug Sensitivity Analysis.
     Increasing the Number of Comparables reduces a) prediction error and b) unfaithfulness across all methods, while c) uncertainty bounds become broader. 
     Increasing the average distance of comparables does not influence the prediction accuracy and faithfulness significantly (d--e), likely because all selected similar examples are sufficiently alike and therefore still provide useful reference. 
     However, it does widen the uncertainty bounds (f). Trends are smoothed with a cubic spline (smoothing parameter $\lambda=1000$). 
    }
    \Description{This figure contains six charts labeled a through f. Each chart compares four explanation types—Comparables Only, Linear Regression, Linear Adjustments, and Trace Adjustments—on quantitative modeling metrics in the Drug Sensitivity domain. Error bars (a–c) and shaded bands (d–f) indicate variation across runs. The top row varies the number of comparables, and the bottom row varies the average distance of comparables (Manhattan distance on standardized features).
    a) Prediction Error vs. Number of Comparables. Prediction error decreases as more comparables are used. Comparables Only starts highest (around 1.7) and declines to about 1.4. Linear Regression and Linear Adjustments begin around 1.55–1.6 and drop to roughly 1.4. Trace Adjustments remain lowest throughout, falling from about 1.38 to nearly 1.12. 
    b) Unfaithfulness vs. Number of Comparables. Unfaithfulness also decreases with additional comparables. Comparables Only and Linear Regression reduce from approximately 1.1–1.15 to about 0.9. Linear Adjustments drop from around 0.9 to roughly 0.7. Trace Adjustments are consistently lowest, decreasing from about 0.3 to near 0.2.
    c) Uncertainty Bounds vs. Number of Comparables. Uncertainty widens as the number of comparables increases. Comparables Only and Linear Regression rise from about 1.4 to above 4.0. Linear Adjustments increase from roughly 1.2 to around 3.9. Trace Adjustments grow more slowly, from about 1.1 to near 3.3, maintaining the narrowest bounds. 
    d) Prediction Error vs. Average Distance of Comparables. Prediction error remains relatively stable across distance. Comparables Only is highest (around 1.4--1.48), Linear Regression sits slightly lower, Linear Adjustments are lower still, and Trace Adjustments stay lowest (about 1.15–1.18) with minimal variation.
    e) Unfaithfulness vs. Average Distance of Comparables. Unfaithfulness is largely insensitive to distance. Comparables Only and Linear Regression remain near 0.95--1.0, Linear Adjustments around 0.7–0.75, and Trace Adjustments around 0.2–0.25, with small fluctuations.
    f) Uncertainty Bounds vs. Average Distance of Comparables. Uncertainty increases as comparables become more distant. Comparables Only and Linear Regression rise from about 3.0 to above 4.0, Linear Adjustments from roughly 2.7 to about 3.8, and Trace Adjustments from around 2.1 to near 3.2, again yielding the narrowest bounds.
    }
    \label{fig:app_modeling_drug}
    
\end{figure}

\begin{figure}[ht]

    \centering
    \includegraphics[width=0.8\linewidth]{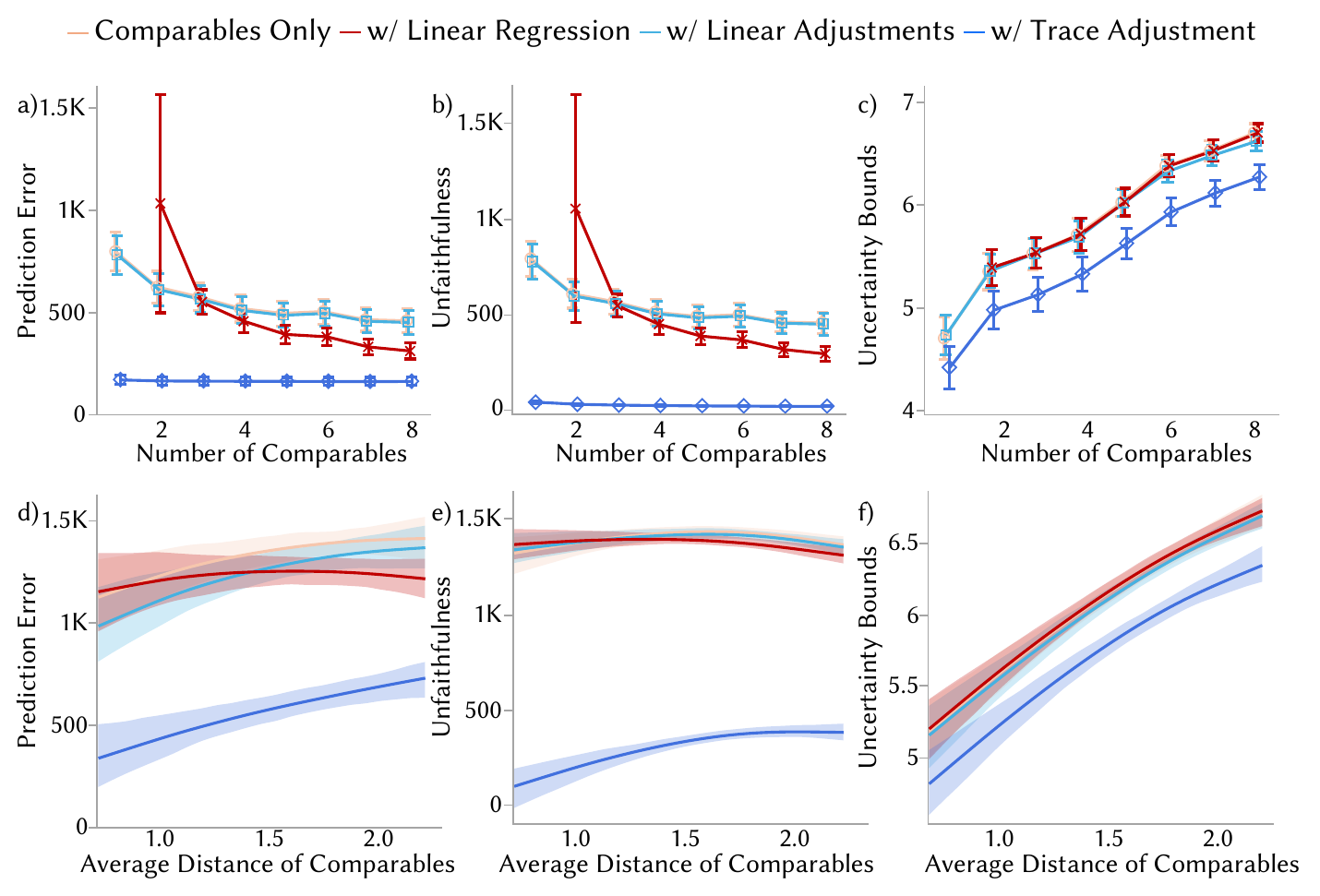}
    \caption{Modeling study results for different XAI types in the Agriculture domain, varying the Number of Comparables (a--c) and the Average Distance of Comparables (d--f), distance was computed using the Manhattan metric with standardized features.
     Increasing the Number of Comparables reduces a) prediction error and b) unfaithfulness across all methods, while c) uncertainty bounds become broader. 
     Increasing the Average Distance of Comparables worsens all metrics (d--f). Trends are smoothed with a cubic spline (smoothing parameter $\lambda=1000$). 
    Across all conditions, Comparables w/ Trace Adjustments consistently yield lower prediction error (a, d), lower unfaithfulness (b, e), and narrower uncertainty bounds (c, f).}
    \Description{
    This figure contains six charts labeled a through f. Each chart compares four explanation types—Comparables Only, Linear Regression, Linear Adjustments, and Trace Adjustments—on quantitative modeling metrics in the Agriculture domain. Error bars (a–c) and shaded bands (d–f) indicate variation across runs. The top row varies the number of comparables, and the bottom row varies the average distance of comparables (Manhattan distance on standardized features).
    a) Prediction Error vs. Number of Comparables. Prediction error decreases as more comparables are used. Comparables Only and Linear Adjustments fall from around 750–800 to about 450–500. Linear Regression drops sharply from about 1000 to below 400. Trace Adjustments remain much lower throughout, staying near 150–170 with only minor changes.
    b) Unfaithfulness vs. Number of Comparables. Unfaithfulness declines with additional comparables. Comparables Only and Linear Adjustments decrease from roughly 750–800 to around 450–480. Linear Regression drops from about 1000 to near 300. Trace Adjustments are consistently lowest, decreasing from about 40 to near zero.
    c) Uncertainty Bounds vs. Number of Comparables. Uncertainty increases as the number of comparables grows. Comparables Only, Linear Regression, and Linear Adjustments rise from about 4.7–5.0 to nearly 6.7. Trace Adjustments increase more slowly, from around 4.3 to about 6.2, retaining the narrowest bounds.
    d) Prediction Error vs. Average Distance of Comparables. Prediction error increases as comparables become more distant. Comparables Only and Linear Regression rise from about 1150 to near 1300–1400. Linear Adjustments increase from approximately 1000 to about 1400. Trace Adjustments grow from roughly 300 to around 700, remaining substantially lower than other methods.
    e) Unfaithfulness vs. Average Distance of Comparables. Unfaithfulness worsens as distance increases. Comparables Only, Linear Regression, and Linear Adjustments all remain high, rising slightly from about 1300 to around 1400 before tapering. Trace Adjustments increase from roughly 100 to around 400 but are still far lower than the others.
    f) Uncertainty Bounds vs. Average Distance of Comparables. Uncertainty widens steadily with distance. All methods increase from around 5.0 to above 6.5, with Trace Adjustments exhibiting the smallest absolute values and the slowest growth.
    }
    \label{fig:app_modeling_agriculture}
    
\end{figure}

To assess generalizability, we also evaluate Comparables with Trace Adjustments in four additional domains: salary, energy, drug sensitivity, and crop yield estimation. Fig.~\ref{fig:app_modeling_salary} presents results for the salary domain, Fig.~\ref{fig:app_modeling_energy} presents results for the energy domain, Fig.~\ref{fig:app_modeling_drug} presents results for the drug sensitivity analysis and  Fig.~\ref{fig:app_modeling_agriculture} presents results for the crop yield estimation which show the same trend as in the house price domain illustrated in Section~\ref{sec:modeling_res}.

\newpage

\subsection{Sensitivity Analysis} \label{app:sensitivity_analysis}
\begin{figure}[ht]
    \centering
    \includegraphics[width=0.84\linewidth]{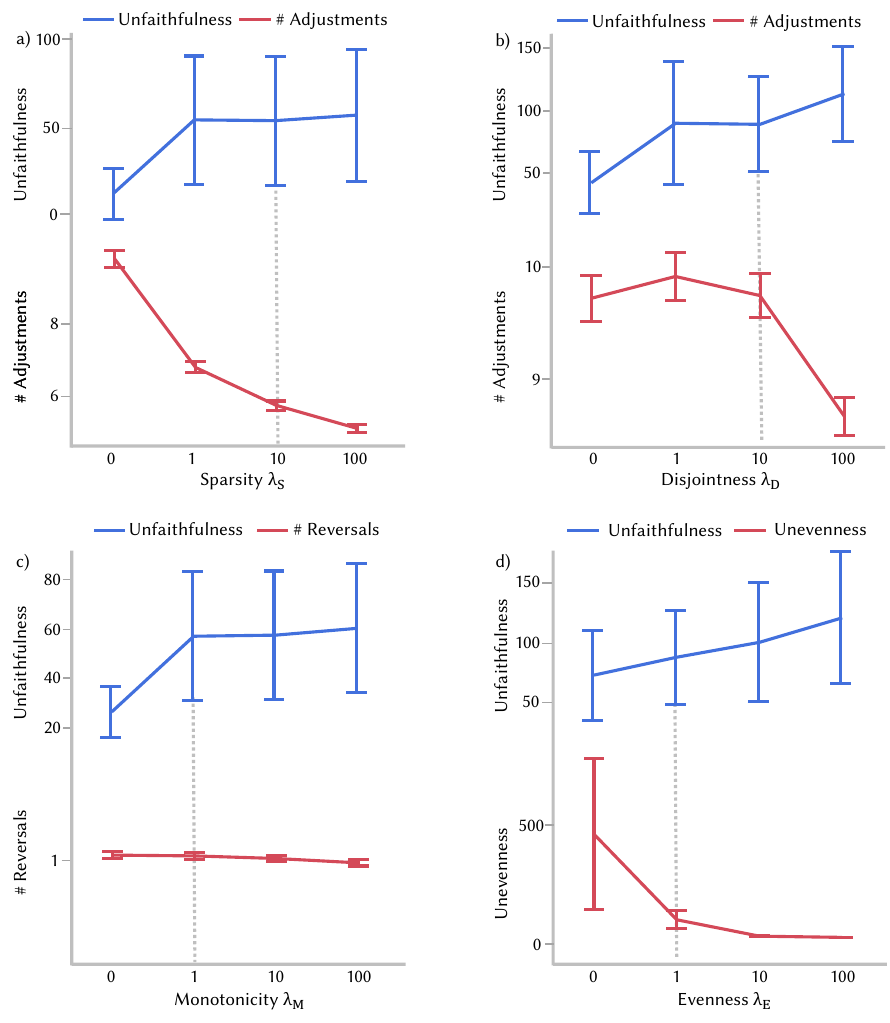}
    \caption{Sensitivity analysis results for the House Price domain with Faithfulness fixed at $\lambda_F = 1$. 
    (a--b) Increasing Sparsity $\lambda_S$ and Disjointness $\lambda_D$ significantly reduces cognitive load. 
    (c) Increasing Monotonicity $\lambda_M$ decreases direction-reversal occurrences. 
    (d) Increasing Evenness $\lambda_E$ produces more uniformly distributed explanations, increasing these $\lambda$ parameters sacrifices faithfulness to different extents (blue line).
    The gray dashed lines denote the most optimal hyperparameters that balance unfaithfulness with the desiderata.}
    \label{fig:sensitivity_analysis}
    \Description{
    This figure contains four charts labeled a through d, showing a sensitivity analysis of the House Price domain with Faithfulness fixed at $\lambda_F$ = 1. Each chart varies one desideratum parameter—Sparsity ($\lambda_S$), Disjointness ($\lambda_D$), Monotonicity ($\lambda_M$), or Evenness ($\lambda_E$)—and reports its effect on Unfaithfulness (blue) and a secondary measure (red). Error bars denote variation across runs.
    a) Sparsity $\lambda_S$ vs. Unfaithfulness and Number of Adjustments. As sparsity increases from 0 to 100, the number of adjustments steadily decreases (from around 9 down to about 5), indicating simpler explanations. Unfaithfulness shows a mild upward trend with large variability, suggesting that highly sparse adjustments may sacrifice alignment with the underlying model.
    b) Disjointness $\lambda_D$ vs. Unfaithfulness and Number of Adjustments. Increasing disjointness initially raises the number of adjustments slightly before dropping at $\lambda_D$ = 100. Unfaithfulness generally increases with stronger disjointness constraints, reflecting the tradeoff between minimizing cognitive load and maintaining fidelity.
    c) Monotonicity $\lambda_M$ vs. Unfaithfulness and Number of Reversals. Higher monotonicity reduces direction-reversal occurrences (from slightly above 1.0 down to nearly 1.0), indicating smoother, more directionally consistent traces. Unfaithfulness increases gradually with $\lambda_M$, showing that enforcing strict monotonicity limits the faithfulness of local adjustments.
    d) Evenness $\lambda_E$ vs. Unfaithfulness and Unevenness.
    Strengthening evenness dramatically reduces unevenness (red) from above 500 to nearly zero, producing more evenly distributed adjustments. However, Unfaithfulness rises substantially with $\lambda_E$, demonstrating that uniform adjustment magnitudes can significantly constrain faithfulness.}
\end{figure}

To understand how different Desiderata affect the XAI faithfulness and interpretability, we perform a sensitivity analysis on hyperparameters: Sparsity $\lambda_S$, Disjointness $\lambda_D$, Monotonicity $\lambda_M$, and Evenness $\lambda_E$.
We measured:
\begin{enumerate}
    \item \textbf{Unfaithfulness $\downarrow$} measured as the absolute error (AE) between the XAI estimate and the AI prediction; lower values indicate higher faithfulness.
    \item \textbf{\# Adjustments $\downarrow$} representing the total number of adjustments within the trace; lower counts reflect lower cognitive load.
    \item \textbf{\# Reversals $\downarrow$} defined as the number of directional reversals along the trace; fewer reversals imply stronger monotonicity. 
    \item \textbf{Unevenness $\downarrow$}
    quantified as the variance of adjustment values along the trace, reported as $ \mathrm{Var}\left( \Delta  \tilde{y}_{1:T} \right)$; lower values indicate more even adjustments.  
\end{enumerate}
As shown in Fig.~\ref{fig:sensitivity_analysis}, increasing the $\lambda$ values generally improves interpretability (via reduced adjustments, reversals, and unevenness) but also increases unfaithfulness, reflecting the inherent trade-off. 
Notably, interpretability gains can sometimes be substantial while the loss in faithfulness remains small, indicating that moderate hyperparameter settings offer meaningful improvements in interpretability with minimal sacrifice of faithfulness.

\newpage
\subsection{User Interface}
\begin{figure}[H]
    \centering
    \includegraphics[width=1\linewidth]{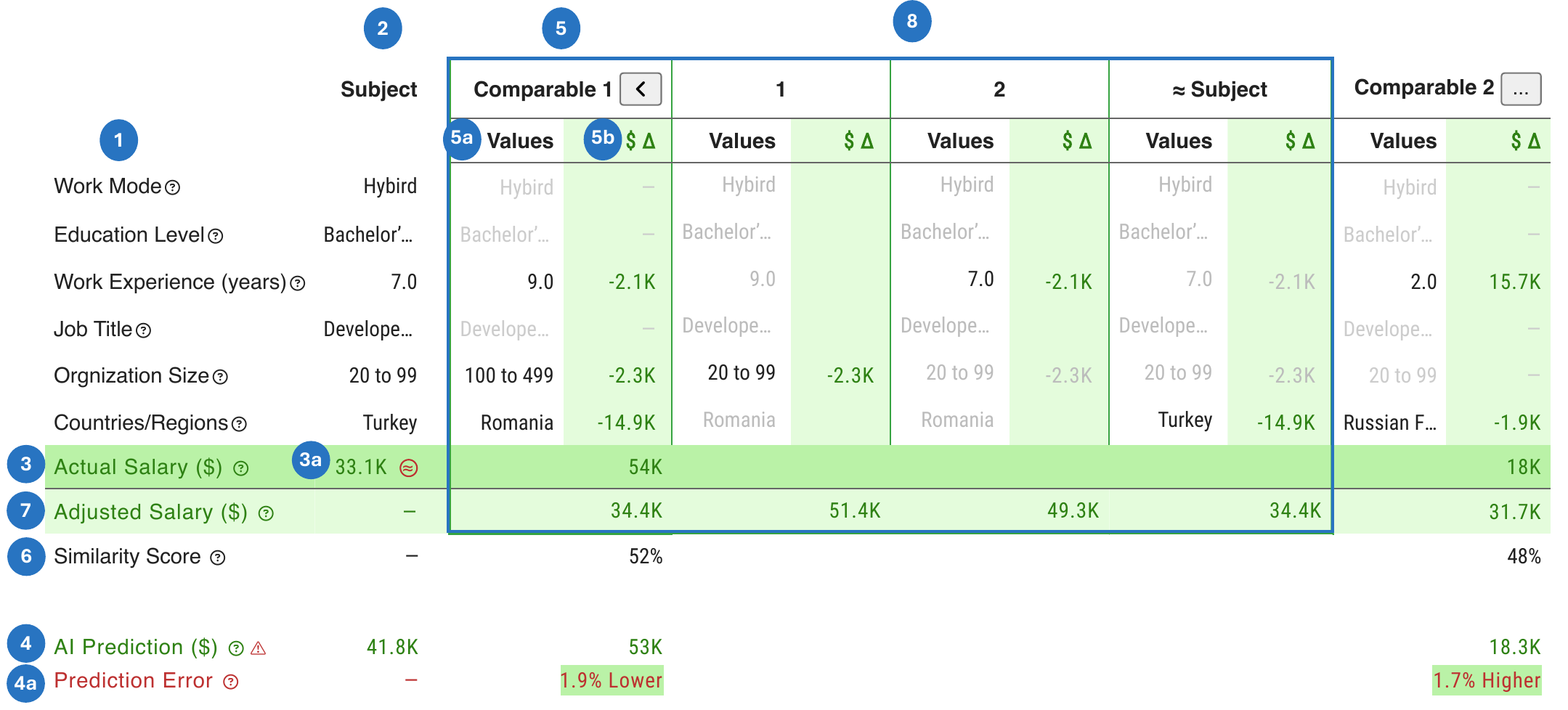}
    \caption{\textit{Comparables w/ Trace Adjustments}: with faithfully calculated adjustments and an extendable UI. Clicking on the “$\dots$” button reveals a trace of intermediate hypothetical cases, with one column for each step (1, 2, …) and subcolumns detailing adjusted attributes and their corresponding adjusted energy usage. For example, the given example indicates adjusting the Work Experience from 9 years to 7 years will decrease the salary by 2.1K. The \textit{Comparables w/ Linear Adjustments} interface is similar except that the table is not expandable.} 
    \Description{A screenshot of an interactive comparison table for a salary explanation interface labeled “Comparables with Trace Adjustment.” On the left, a column labeled Subject lists personal attributes such as work mode (Hybrid), education level (Bachelor’s), work experience (7 years), job title (Developer), organization size (20–99), and country (Turkey). Beneath these, three summary rows show Actual Salary (33.1K), AI Prediction (41.8K), and Prediction Error.
    To the right of the subject, a boxed region highlights a sequence of columns starting with Comparable 1 followed by intermediate steps labeled “1” and “2,” which represent a step-by-step trace of adjustments. Each step has two subcolumns: Values (showing the attribute values at that step) and $\Delta x$ (showing how much each one changes the salary). Green cells indicate positive or negative salary deltas. Example deltas include −2.1K for adjusting work experience from 9 to 7 years, −2.3K for organization size, and −14.9K for country. At the end of the trace, a column labeled “≈ Subject” shows the case after all adjustments, whose attributes now match the subject’s values.
    On the far right, another column labeled Comparable 2 displays a different example with its own values and deltas. Along the bottom of the table, rows report Adjusted Salary for each stage (e.g., 54K, 51.4K, 49.3K, and 34.4K) and a Similarity Score (e.g., 52\% for Comparable 1 and 48\% for Comparable 2). Small numbered callouts annotate parts of the interface, such as the “Actual Salary,” “AI Prediction,” the step trace, and the values and delta columns.}
    \label{fig:app_ui_salary}
\end{figure}

\begin{figure}[ht]
    \centering
    \includegraphics[width=1\linewidth]{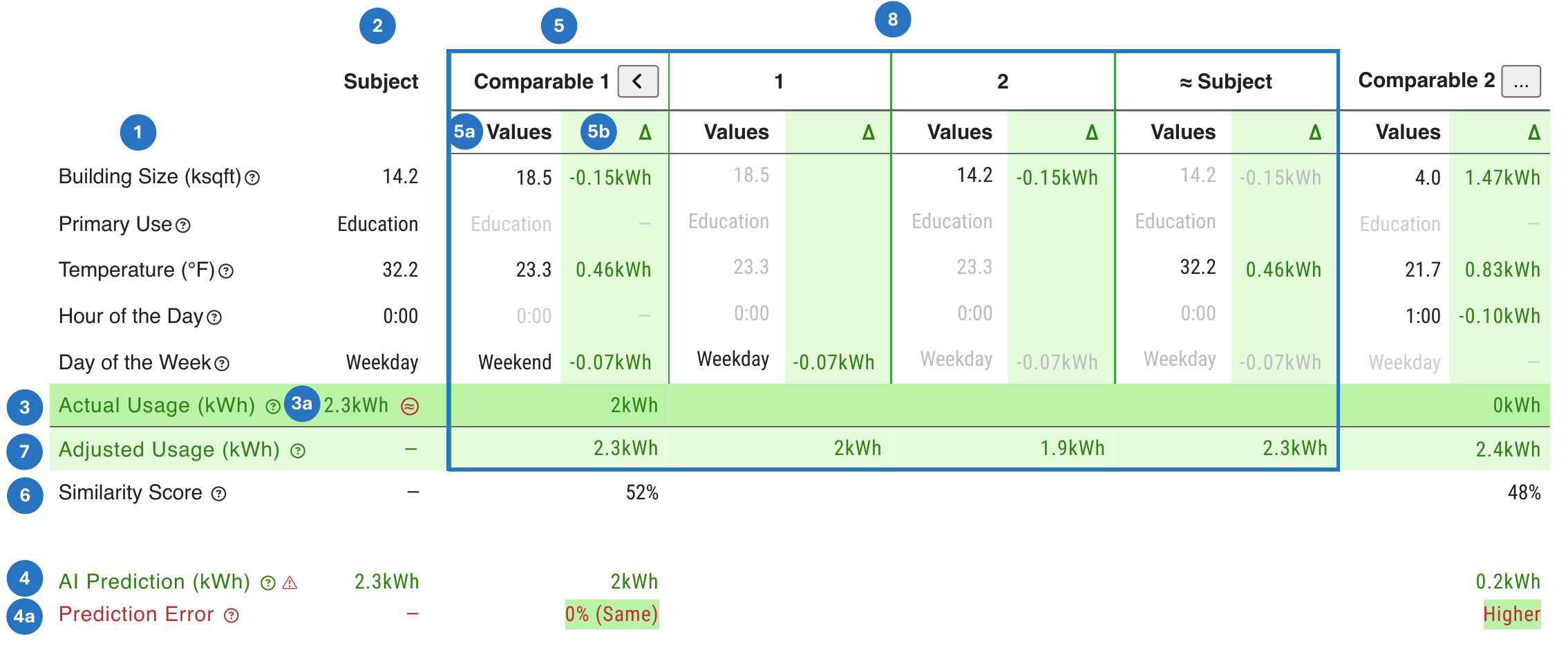} \caption{\textit{Comparables w/ Trace Adjustments}: with faithfully calculated adjustments and an extendable UI. Clicking the “$\dots$” button reveals a trace of intermediate hypothetical cases, with one column for each step (1, 2, …) and subcolumns detailing adjusted attributes and their corresponding adjusted salary. For example, the given example indicates adjusting the Building Size from 18.5ksqft to 14.25ksqft will decrease the energy usage by 0.15kWh. The \textit{Comparables w/ Linear Adjustments} interface is similar except that the table is not expandable.} 
    \Description{This figure is a screenshot of an interactive comparison table for an energy-usage explanation interface labeled Comparables with Trace Adjustment.
    On the left, a column labeled Subject lists building attributes including building size (14.2 ksqft), primary use (Education), temperature (32.2°F), hour of the day (0:00), and day of the week (Weekday). Beneath these, three summary rows show Actual Usage (2.3 kWh), AI Prediction (2.3 kWh), and Prediction Error.
    To the right of the subject, a boxed region highlights a sequence of columns starting with Comparable 1, followed by intermediate steps labeled “1” and “2,” which represent a step-by-step trace of adjustments. Each step has two subcolumns: Values (showing attribute values at that step) and $\Delta x$ (showing how much each change affects energy usage). Green cells indicate energy deltas. Example deltas include −0.15 kWh for reducing building size from 18.5 to 14.2 ksqft, +0.46 kWh for changing temperature, −0.07 kWh for switching from weekend to weekday, and +0.83 kWh for temperature differences in another comparable.
    At the end of the trace, a column labeled ≈ Subject shows the case after all adjustments, whose attributes now match the subject’s values. On the far right, another column labeled Comparable 2 displays a different example with its own values and deltas. Along the bottom of the table, rows report Adjusted Usage for each stage (e.g., 2.3 kWh, 2.0 kWh, 1.9 kWh, and 2.3 kWh) and a Similarity Score (e.g., 52\% for Comparable 1 and 48\% for Comparable 2). Small numbered callouts annotate parts of the interface, including the subject attributes, actual usage, AI prediction, adjustment trace, and delta columns.}
    \label{fig:app_ui_energy}
\end{figure}

\begin{figure}[ht]
    \centering
    \includegraphics[width=0.9\linewidth]{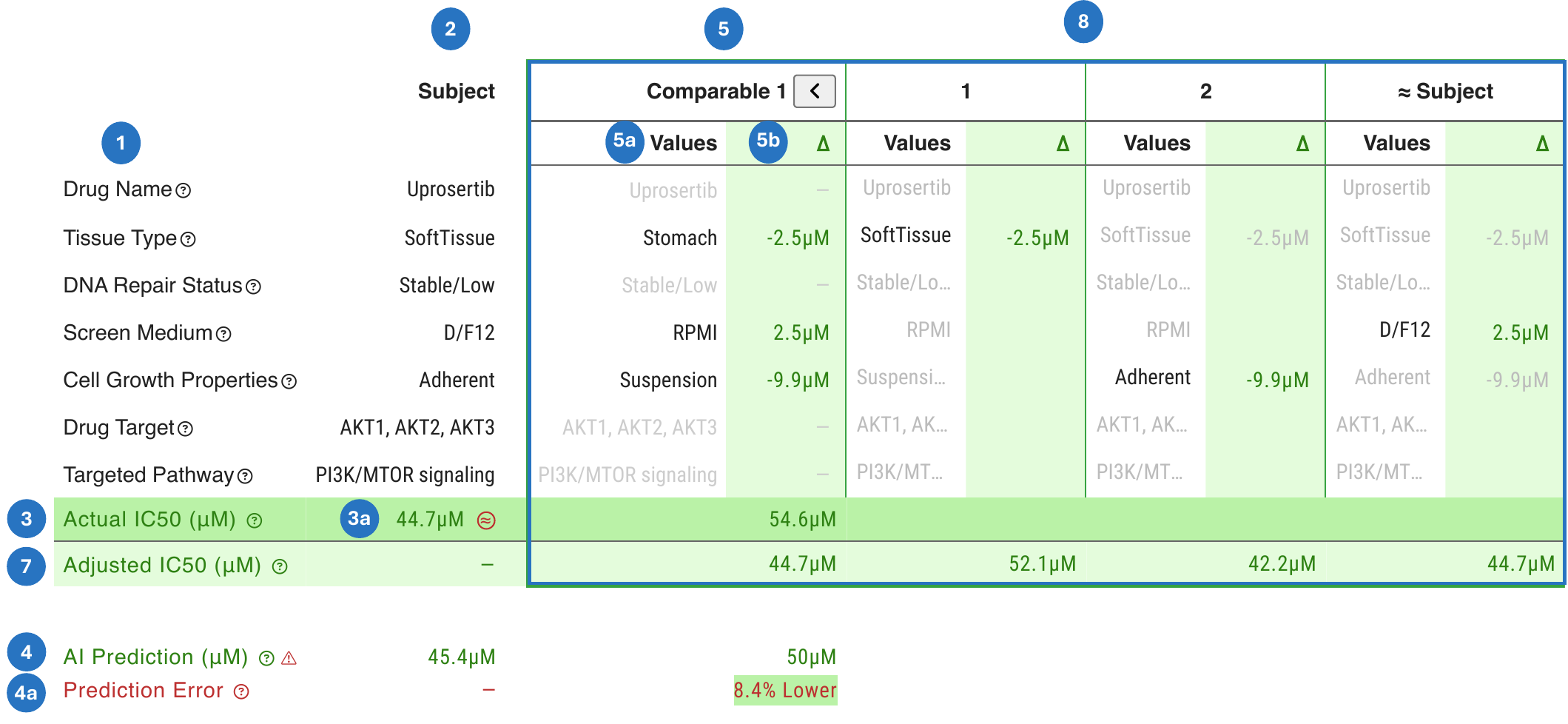}
    \caption{\textit{Comparables w/ Trace Adjustments}, with faithfully calculated adjustments and an extendable UI. Clicking the “$\dots$” button reveals a trace of intermediate hypothetical cases, with one column for each step (1, 2, …) and subcolumns detailing adjusted attributes and their corresponding adjusted drug sensitivity. For example, the given example indicates adjusting the Tissue Type from Stomach to SoftTissue will decrease the IC50 by 2.5$\mu$M. The \textit{Comparables w/ Linear Adjustments} interface is similar except that the table is not expandable.} 
    \Description{This figure is a screenshot of an interactive comparison table for a drug sensitivity explanation interface labeled Comparables with Trace Adjustment. On the left, a column labeled Subject lists biological and experimental attributes, including drug name (Uprosertib), tissue type (SoftTissue), DNA repair status (Stable/Low), screen medium (D/F12), cell growth properties (Adherent), drug target (AKT1, AKT2, AKT3), and targeted pathway (PI3K/MTOR signaling). Beneath these, summary rows show Actual IC50 (44.7 $\mu$M), AI Prediction (45.4 $\mu$M), and Prediction Error. To the right of the subject, a boxed region displays a sequence starting with Comparable 1, followed by intermediate columns labeled “1” and “2,” which represent a step-by-step trace of counterfactual adjustments. Each step contains two subcolumns: Values (showing the attribute state at that step) and $\Delta$ (showing how each change affects IC50 in micromolar units). Green cells indicate dosage changes. Example adjustments include −2.5 $\mu$M for changing tissue type from Stomach to SoftTissue, +2.5 µM for modifying the screen medium to RPMI, and −9.9 $\mu$M for switching growth properties from Suspension to Adherent. At the end of the trace, a column labeled ≈ Subject shows the final adjusted case, where all attributes now match the subject. Along the bottom, a row labeled Adjusted IC50 reports values for each stage (e.g., 44.7 $\mu$M, 52.1 $\mu$M, and 42.2 $\mu$M). Below the table, the AI prediction for Comparable 1 is shown as 50 $\mu$M with an annotation “8.4\% lower.” Small numbered markers indicate interface components such as subject attributes, the adjustment trace, the delta column, and prediction output.}
    \label{fig:app_ui_drug}
\end{figure}

\begin{figure}[H]

    \centering
    \includegraphics[width=0.9\linewidth]{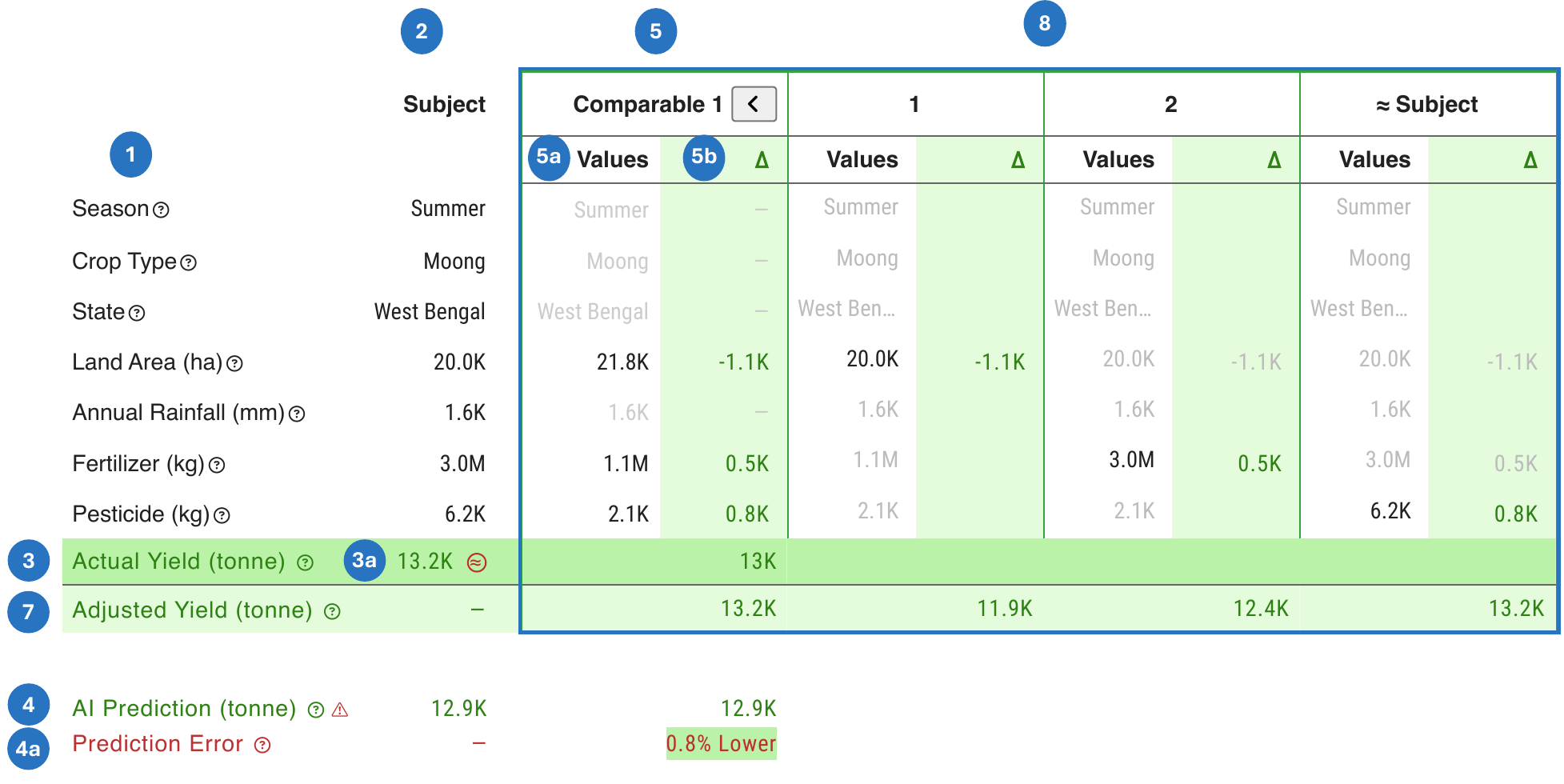}
    \caption{\textit{Comparables w/ Trace Adjustments}, with faithfully calculated adjustments and an extendable UI. Clicking the “$\dots$” button reveals a trace of intermediate hypothetical cases, with one column for each step (1, 2, …) and subcolumns detailing adjusted attributes and their corresponding adjusted yield. For example, the given example indicates adjusting the Land Area from 21.8K to 20K will decrease the Actual Yield by 1.1K. The \textit{Comparables w/ Linear Adjustments} interface is similar except that the table is not expandable.}
    \Description{This figure is a screenshot of an interactive comparison table for an agricultural yield explanation interface labeled Comparables with Trace Adjustment.
    On the left, a column labeled Subject lists farming attributes including season (Summer), crop type (Moong), state (West Bengal), land area (20.0K hectares), annual rainfall (1.6K millimeters), fertilizer amount (3.0M kilograms), and pesticide amount (6.2K kilograms). Summary rows below display Actual Yield (13.2K tonnes), AI Prediction (12.9K tonnes), and Prediction Error.
    To the right, a highlighted region shows Comparable 1, followed by intermediate trace columns labeled “1” and “2,” which visualize a stepwise adjustment process. Each column contains Values and $\Delta$ subcolumns. $\Delta$ shows how yield changes with each attribute modification. Example deltas include −1.1K tonnes for reducing land area from 21.8K to 20.0K hectares, +0.5K tonnes for increasing fertilizer, and +0.8K tonnes for pesticide differences.
    The rightmost column labeled ≈ Subject shows the final state after all attributes align with the subject’s profile. A row labeled Adjusted Yield reports values for each stage (13.2K, 11.9K, 12.4K, and 13.2K tonnes). Below, the AI’s predicted yield is shown as 12.9K tonnes with a label “0.8\% lower.”
    Blue numbered callouts annotate major areas of the interface such as subject attributes, actual yield, delta values, prediction, and the trace view.}
    \label{fig:app_ui_agriculture}
\end{figure}

Figs.~\ref{fig:app_ui_salary}, \ref{fig:app_ui_energy}, \ref{fig:app_ui_drug}, and \ref{fig:app_ui_agriculture} illustrate the interface demonstrations for the salary, energy, drug sensitivity, and agriculture domains.

\newpage
\subsection{Survey for the summative user study}
\begin{figure*}[ht]
    \centering
    \includegraphics[scale=0.5]{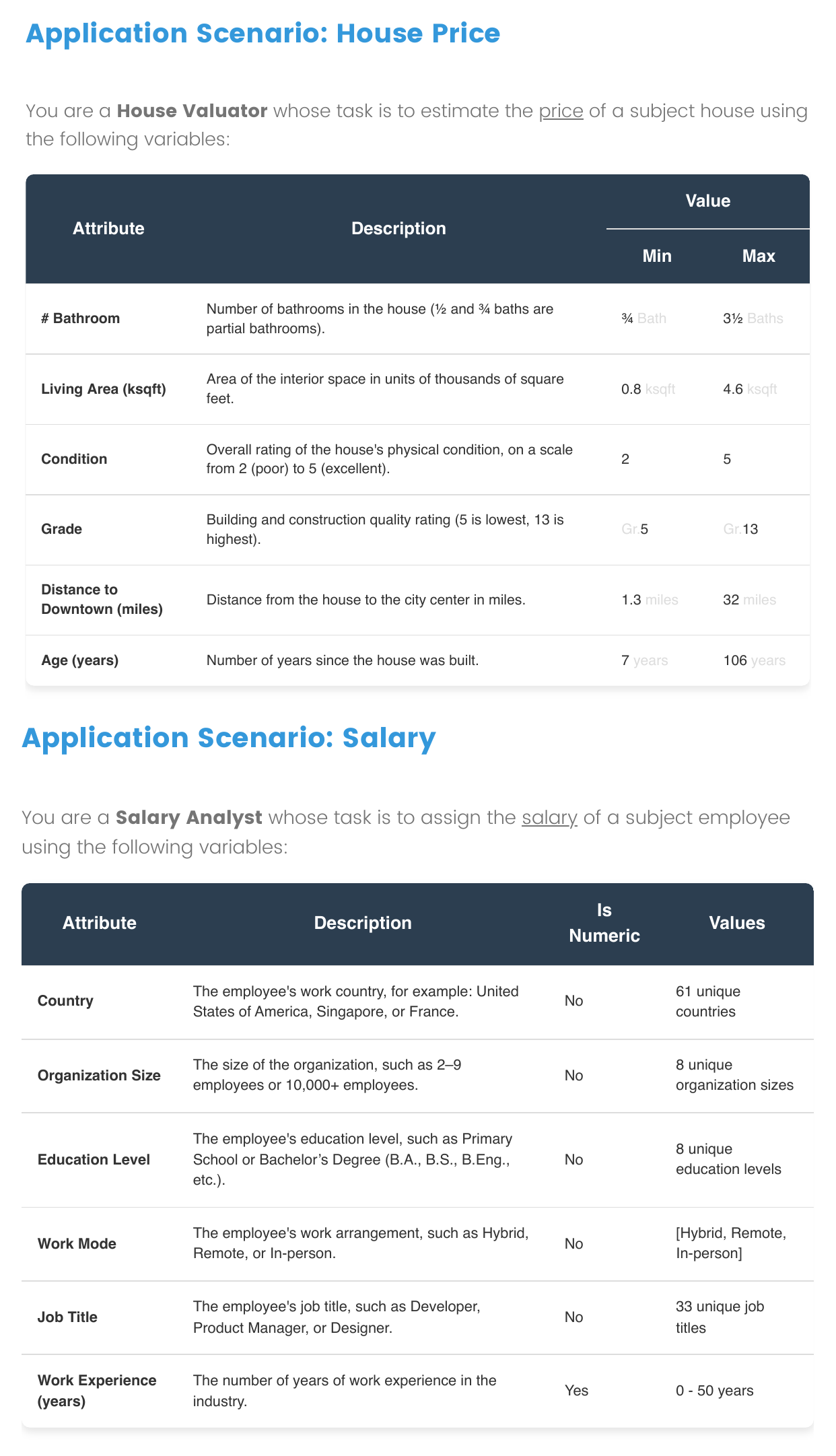}
    \caption{Introduction of attributes on house price and salary domain.}
    \Description{
    This figure presents two application scenarios, House Price and Salary, each with a table of attributes used for prediction tasks.
    House Price: The user acts as a House Valuator tasked with estimating the price of a subject house. The table lists six attributes. Number of bathrooms ranges from three-quarter bath to three and a half baths. Living area spans 0.8 to 4.6 thousand square feet. Condition is rated from 2 to 5 on a five-point scale. Grade refers to construction quality, ranging from grade 5 to grade 13. Distance to downtown ranges from 1.3 to 32 miles. Age ranges from 7 to 106 years.
    Salary: The user acts as a Salary Analyst tasked with assigning the salary of a subject employee. The table lists six attributes. Country includes 61 unique values. Organization size includes 8 categories ranging from small to very large. Education level has 8 categories, from primary school to graduate degrees. Work mode has 3 categories: hybrid, remote, and in-person. Job title has 33 unique values, such as developer or product manager. Work experience is numeric, ranging from 0 to 50 years.
    }
    \label{fig:survey_application}
\end{figure*}

\begin{figure*}[t]
    \centering
    \includegraphics[scale=0.6]{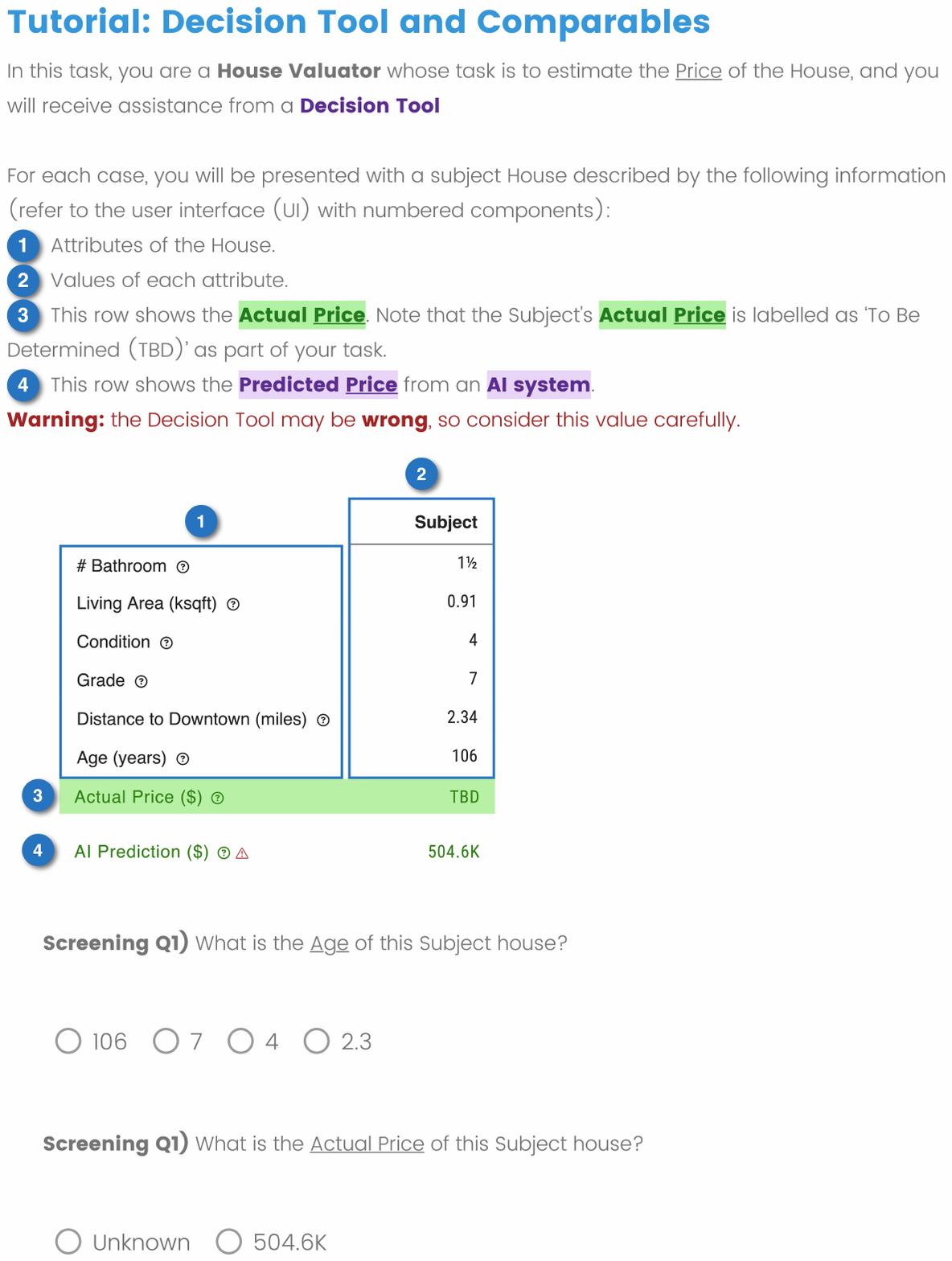}
    \caption{Tutorial on the Decision Tool and Screening Questions.}
    \Description{
    This figure shows the tutorial screen for the Decision Tool and Comparables in the House Price task. The user acts as a House Valuator, estimating the price of a subject house with assistance from the Decision Tool. The interface highlights four numbered components. 
    1: Attributes of the house are listed in a table, including number of bathrooms, living area, condition, grade, distance to downtown, and age. 
    2: Values of these attributes for the subject house are displayed, such as 1.5 bathrooms, 0.91 thousand square feet, condition 4, grade 7, distance to downtown 2.34 miles, and age 106 years. 
    3: The row for Actual Price is shown but labeled as To Be Determined, indicating that the true price is hidden for the task. 
    4: The row for AI Prediction shows a predicted price of 504.6K dollars, marked with a warning that the decision tool may be wrong. 
    Below the interface, two screening questions test comprehension. The first asks for the age of the subject house, with multiple-choice options 106, 7, 4, and 2.3. The second asks for the actual price, with options Unknown and 504.6K. 
    }
    \label{fig:tut_decision_tool}
\end{figure*}

\begin{figure*}[t]
    \centering
    \includegraphics[scale=0.68]{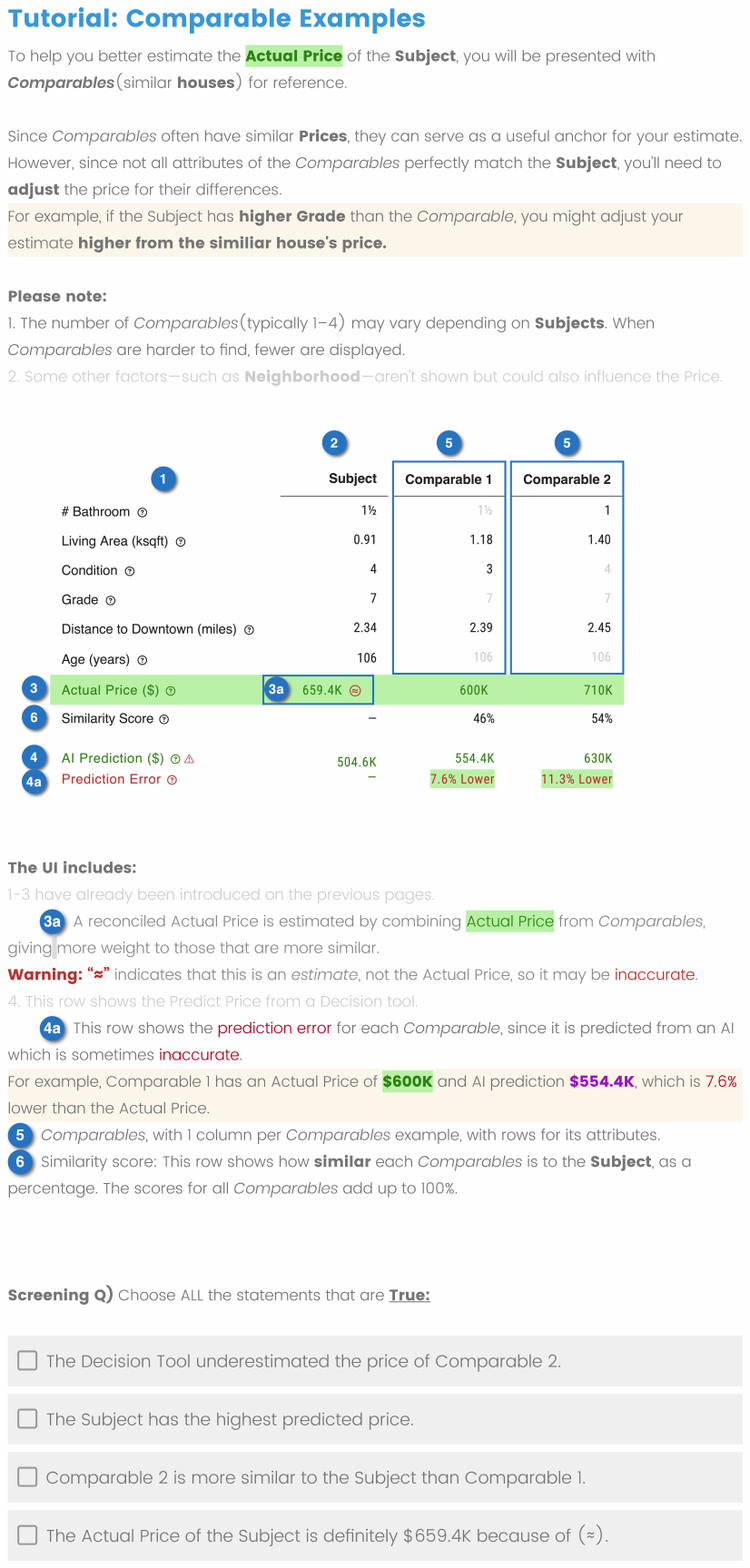}
    \caption{Tutorial on Comparables Only and Screening Question. }
    \Description{
    This figure shows the tutorial screen for Comparable Examples in the House Price task. The interface introduces how participants use comparable houses as reference points to estimate the price of a subject house. 
    The top section explains that comparables often share similar prices but differ in attributes, so adjustments may be needed. It also notes that the number of comparables varies from one to four depending on availability, and some factors like neighborhood are not shown. 
    The main table displays the subject house in one column and two comparables in adjacent columns. Attributes include number of bathrooms, living area, condition, grade, distance to downtown, and age. The subject has 1.5 bathrooms, 0.91 thousand square feet, condition 4, grade 7, distance 2.34 miles, and age 106 years. Comparable 1 is similar but has slightly larger living area and lower condition, while Comparable 2 differs more with larger living area and slightly greater distance. 
    Rows below show outcomes. The reconciled actual price for the subject is estimated as 659.4K dollars. Comparable 1 has an actual price of 600K and a predicted price of 554.4K, noted as 7.6\% lower than its actual price. Comparable 2 has an actual price of 710K and a predicted price of 630K, 11.3\% lower than actual. The subject’s AI-predicted price is 504.6K. Similarity scores indicate Comparable 1 is 46\% similar and Comparable 2 is 54\% similar to the subject. 
    At the bottom, a screening question asks participants to select all true statements, with answer options such as whether the decision tool underestimated Comparable 2, whether the subject has the highest predicted price, and whether Comparable 2 is more similar to the subject than Comparable 1. 
    }
    \label{fig:tut_comp_only}
\end{figure*}

\begin{figure*}[t]
    \centering
    \includegraphics[scale=0.66]{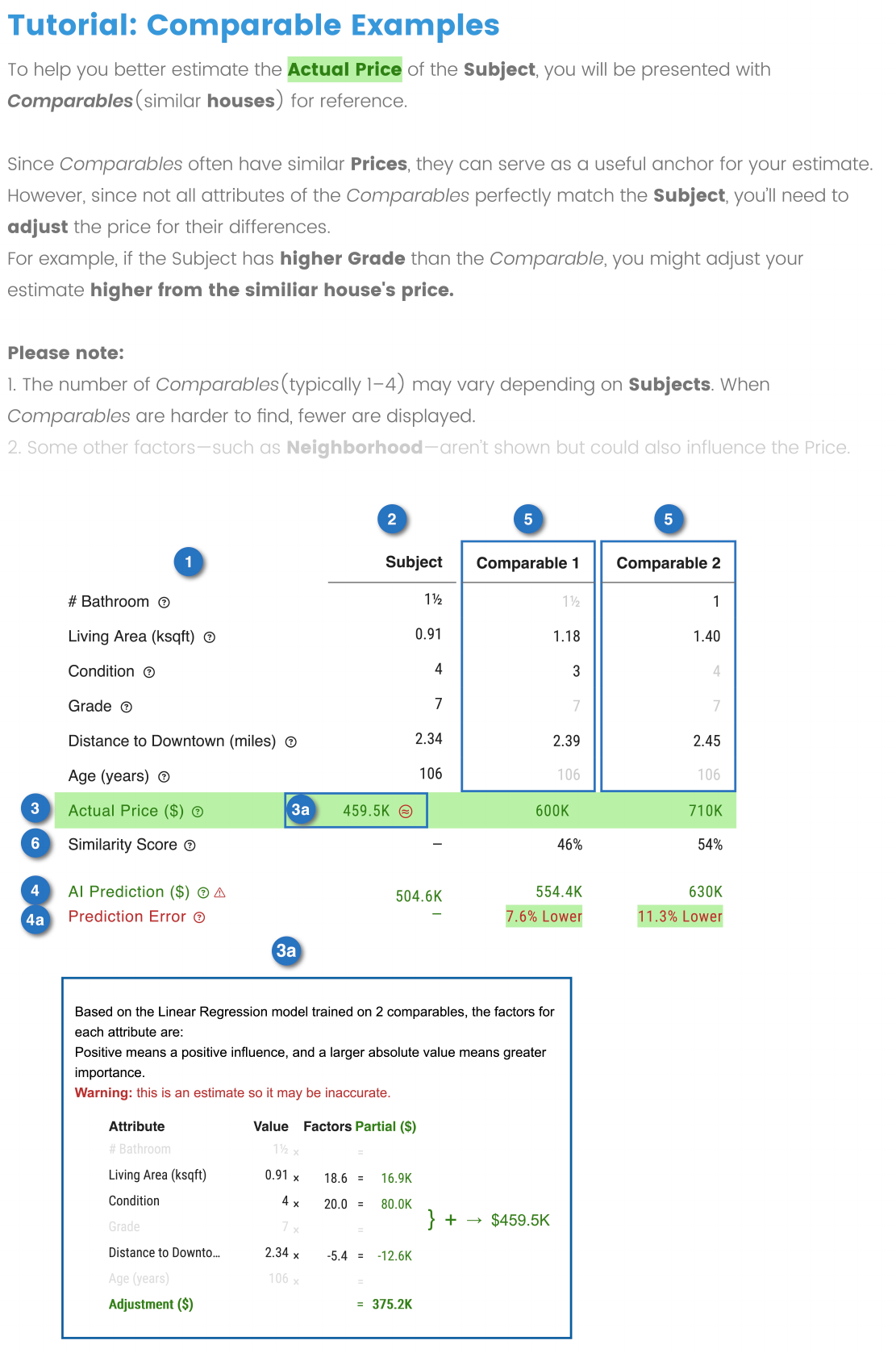}
    \caption{Tutorial on Comparables with Linear Regression and Screening Questions.}
    \Description{
    This figure shows part one of the tutorial for Linear Regression explanations in the House Price task. The interface explains that regression identifies how each attribute influences the price by assigning a factor with a positive or negative sign and a magnitude. The top section introduces the idea that regression summarizes attribute importance across comparables, rather than focusing on individual examples. 
    The table shows the subject house in one column and several comparables in adjacent columns. Attributes include bathrooms, living area, condition, grade, distance to downtown, and age. Below the table, a panel introduces regression factors: for example, living area has a positive weight, condition has a smaller positive weight, and distance to downtown has a negative weight. A note emphasizes that these weights represent general trends and not exact dollar values. 
    At the bottom, a comprehension question asks participants to interpret whether more bathrooms should increase or decrease the predicted price, with multiple-choice options.
}
    \label{fig:tut_linear_reg}
\end{figure*}

\begin{figure*}[t]
    \centering
    \includegraphics[scale=0.6]{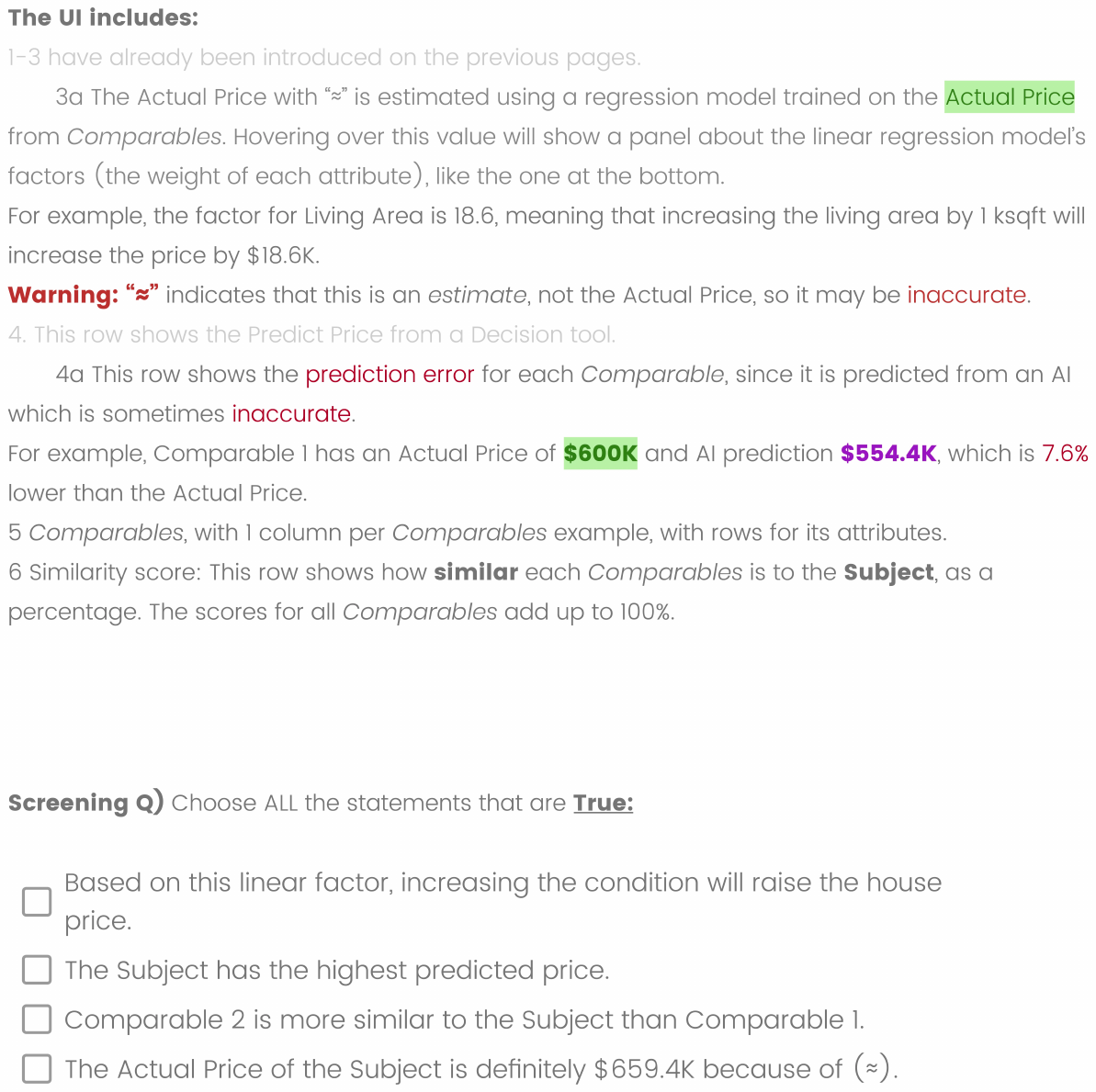}
    \caption{Tutorial on Comparables with Linear Regression and Screening Questions.}
    \Description{
    This figure shows part two of the tutorial for Linear Regression explanations in the House Price task, continuing from part one. The interface demonstrates how regression factors combine to produce an adjusted prediction for the subject house. 
    The table again shows the subject house and its comparables. Below, a panel illustrates how attribute differences between the subject and each comparable are multiplied by the regression factors to calculate adjustments. The resulting adjustments are summed to yield a reconciled regression-based estimate of price. 
    For example, a larger living area in the subject compared to a comparable results in a positive adjustment, while greater distance to downtown produces a negative adjustment. The adjusted comparable values are displayed alongside the subject, illustrating how the regression explains differences. 
    At the bottom, a comprehension question asks participants to choose which attribute contributed most to raising the subject’s estimated price, with multiple-choice options. 
    }
    \label{fig:tut_linear_reg_2}
\end{figure*}

\begin{figure*}[t]
    \centering
    \includegraphics[scale=0.46]{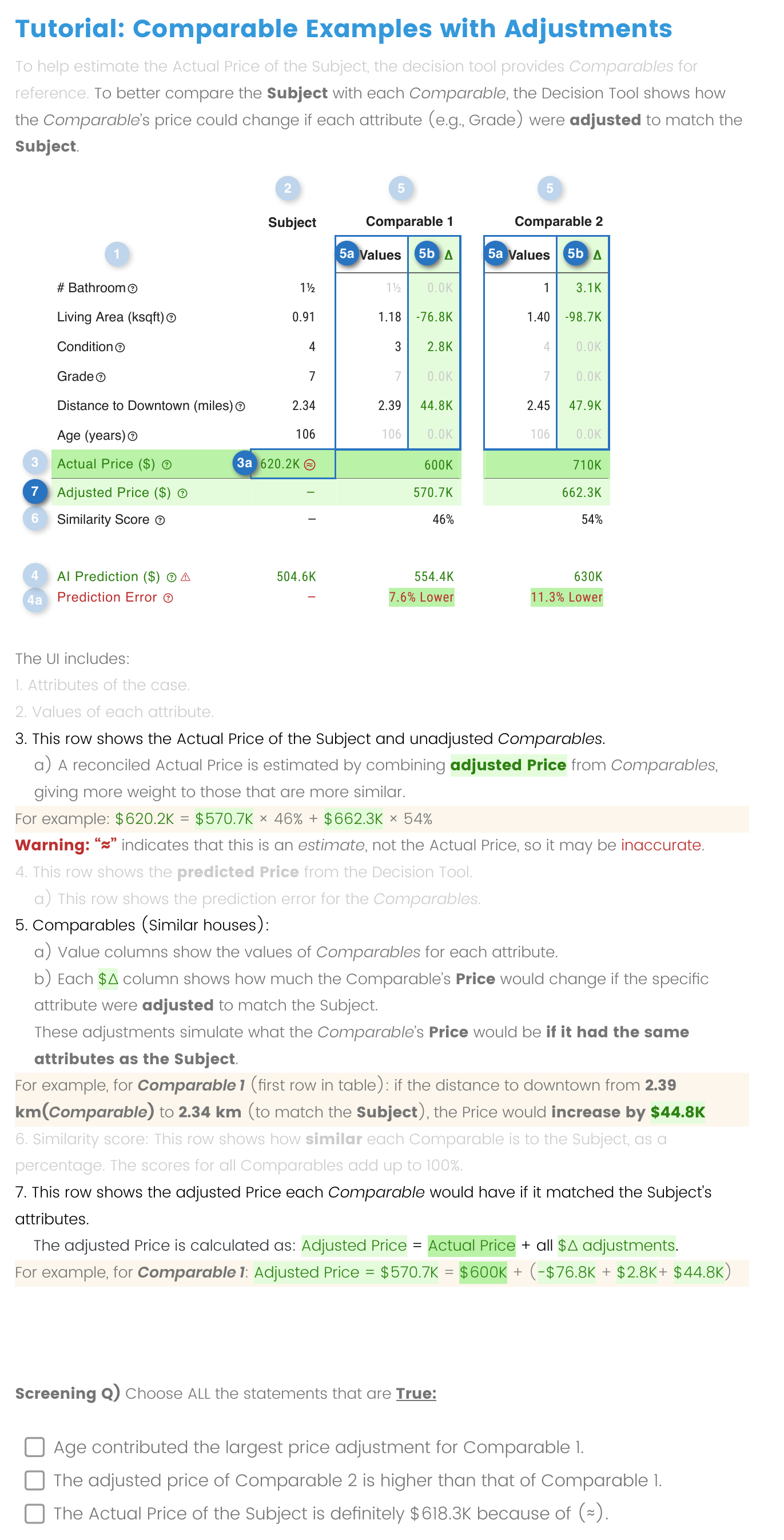}
    \caption{Tutorial on Comparables with Linear Adjustments and Screening Questions.}
    \Description{
    This figure shows the tutorial for Comparable Examples with Linear Adjustments in the House Price task. As in the earlier figure, the interface illustrates how the Decision Tool modifies comparable prices by adjusting attribute values to match the subject. 
    The table again lists the subject house in one column and two comparables in adjacent columns, with attributes for bathrooms, living area, condition, grade, distance to downtown, and age. Adjustment columns show how each comparable’s price would change if an attribute matched the subject. For example, Comparable 1 has a distance of 2.39 miles compared to the subject’s 2.34 miles, producing a +44.8K dollar adjustment. Age differences produce the largest negative adjustment, reducing Comparable 1’s price substantially. 
    The Actual Price row shows 600K dollars for Comparable 1 and 710K dollars for Comparable 2. After adjustments, Comparable 1 reduces to 570.7K and Comparable 2 reduces to 662.3K. A reconciled subject price of 620.2K is computed as a weighted average of these adjusted prices, with weights 46\% and 54\% according to similarity scores. A warning symbol (≈) indicates that this reconciled price is an estimate, not the actual market value. 
    The table also includes AI Predictions for each case and prediction errors, reinforcing that model estimates may differ from actual and adjusted values. At the bottom, a comprehension question asks participants to evaluate statements such as which attribute contributed the largest adjustment or which comparable had the higher adjusted price. 
    }
    \label{fig:tut_linear_adjustments}
\end{figure*}

\begin{figure*}[t]
    \centering
    \includegraphics[scale=0.5]{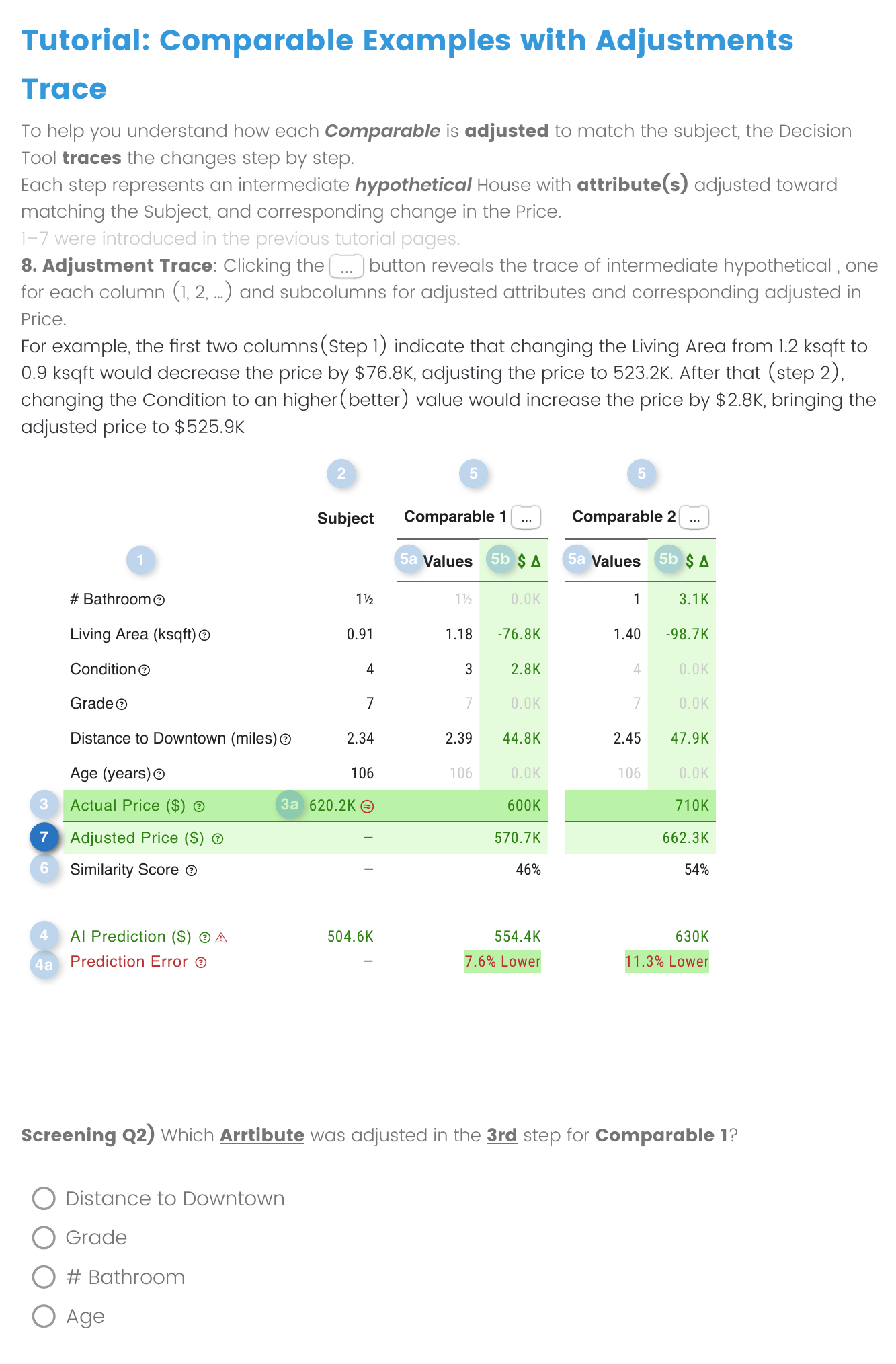}
    \caption{Tutorial on Comparables with Trace Adjustments and Screening Questions.}
    \Description{
    This figure shows the tutorial screen for Comparable Examples with Trace Adjustments in the House Price task. The interface explains how the Decision Tool traces adjustments step by step to illustrate how each Comparable’s price is transformed to match the subject. 
    The main table displays the subject house in one column and two comparables in adjacent columns, with attributes including bathrooms, living area, condition, grade, distance to downtown, and age. For each comparable, the table includes a trace panel where intermediate steps are shown. Each step corresponds to adjusting one attribute toward the subject’s value and recalculating the price. 
    For example, the first step for Comparable 1 changes the living area from 1.2 to 0.9 thousand square feet, decreasing the price by 76.8K to 523.2K. The second step changes the condition from 3 to 4, increasing the price by 2.8K to 525.9K. The trace continues in this fashion, with each adjustment incrementally altering the comparable’s price until it fully matches the subject’s attributes. 
    The rows below summarize the results. Comparable 1 adjusts from 600K to 570.7K, while Comparable 2 adjusts from 710K to 662.3K. A reconciled subject price is computed by combining these adjusted prices weighted by similarity scores of 46\% and 54\%, yielding an estimate of 618.3K. A warning symbol (≈) indicates that this reconciled price is an estimate rather than an actual market value. AI predictions and prediction errors are also shown. 
    At the bottom, a comprehension question asks participants which attribute was adjusted in the third step for Comparable 1, reinforcing understanding of the trace process. 
    }

    \label{fig:tut_trace}
\end{figure*}

\begin{figure*}[t]
    \centering
    \includegraphics[scale=0.46]{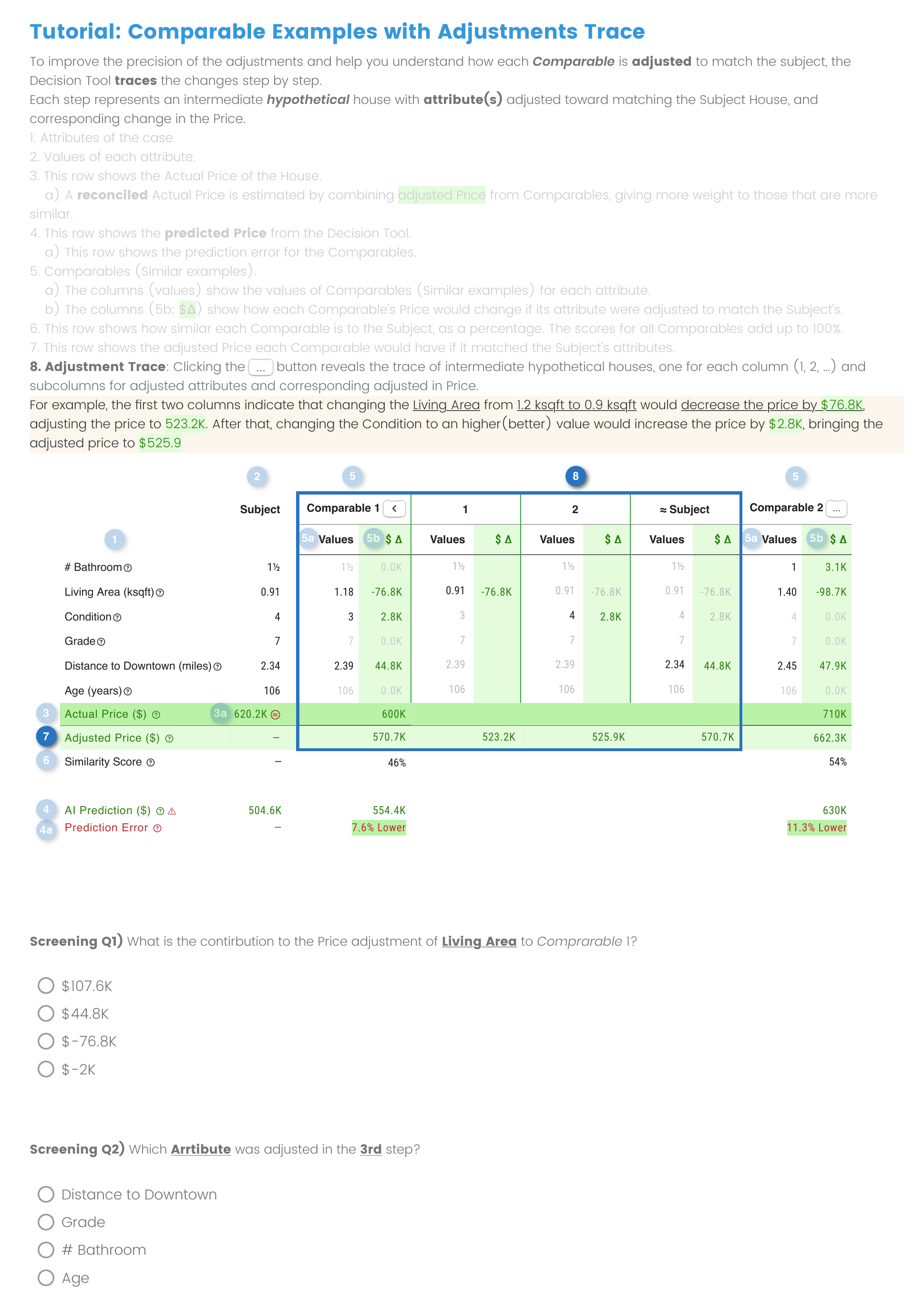}
    \caption{Tutorial on Trace Details and Screening Questions.}
    \Description{
    This figure shows a tutorial screen for Comparable Examples with Trace Adjustments in the House Price task. The interface demonstrates how the Decision Tool produces a step-by-step trace of adjustments, creating intermediate hypothetical houses until each comparable matches the subject. 
    The table lists the subject house in one column and two comparables in adjacent columns, with attributes including bathrooms, living area, condition, grade, distance to downtown, and age. For each comparable, trace subcolumns expand the adjustment process. Each column represents a step where one attribute is changed and the price is updated accordingly. 
    For example, the first step for Comparable 1 changes the living area from 1.2 to 0.9 thousand square feet, decreasing the price by 76.8K to 523.2K. In the next step, the condition is improved from 3 to 4, increasing the price by 2.8K to 525.9K. Subsequent steps adjust other attributes such as bathrooms and distance to downtown, gradually shifting the comparable’s attributes and price to align with the subject. 
    Rows below the attributes show the actual prices of the comparables, AI predictions, prediction errors, and adjusted prices after all steps are completed. Comparable 1 adjusts from 600K to 570.7K, and Comparable 2 adjusts from 710K to 662.3K. A reconciled subject price is then estimated by combining these adjusted values with similarity scores of 46\% and 54\%. A warning symbol (≈) indicates this reconciled price is only an estimate. 
    At the bottom, screening questions test comprehension, asking participants to identify the contribution of living area to Comparable 1’s adjustment and which attribute was changed in the third step. 
}

    \label{fig:tut_trace_details}
\end{figure*}

\begin{figure*}[t]
    \centering
    \includegraphics[scale=0.5]{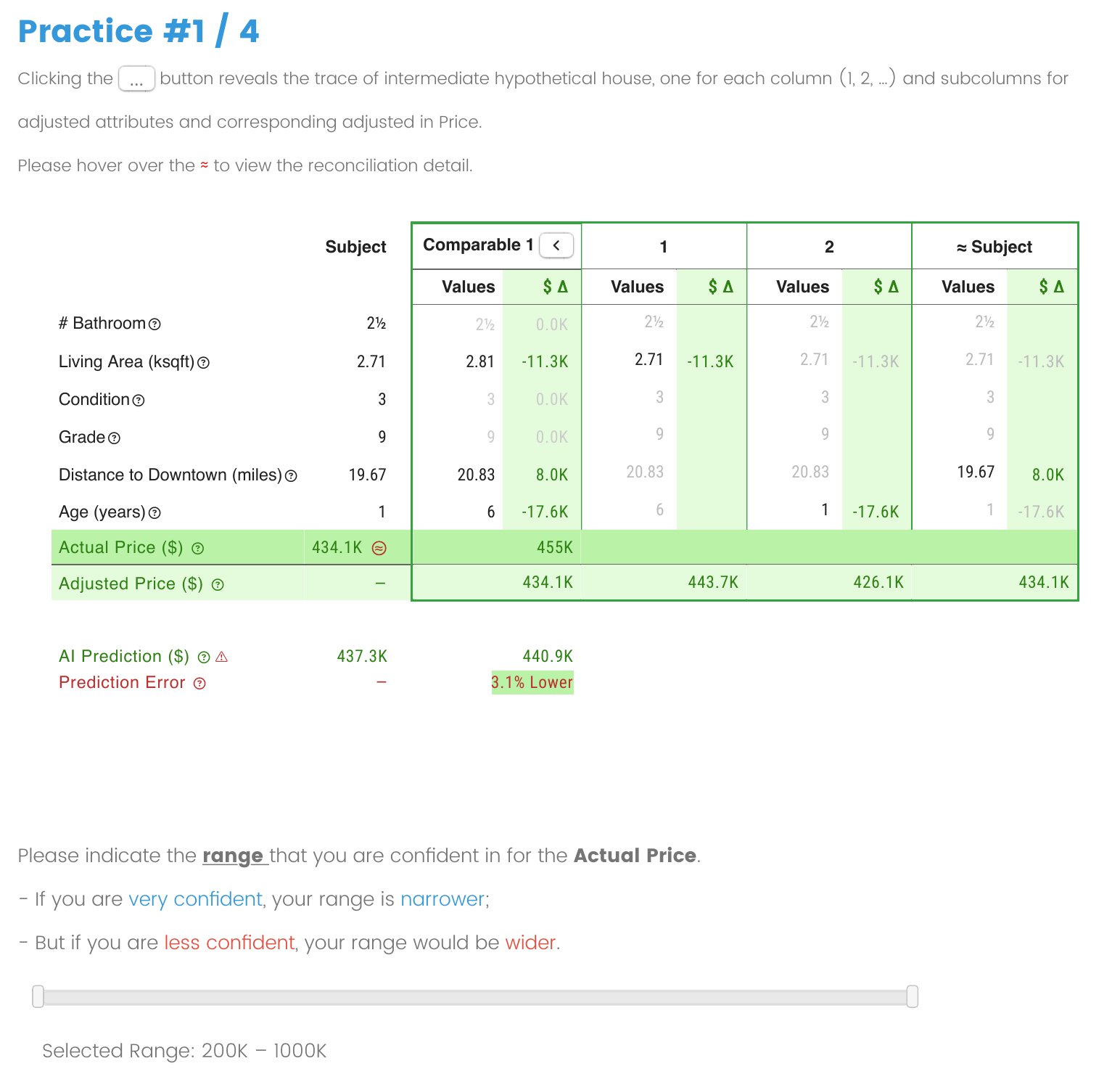}
    \caption{Practice session. The UI varies with the experimental condition. A slider bar below the UI allows users to indicate their response interval.}
    \Description{
    This figure shows a practice task screen for Trace Adjustments in the House Price domain. The subject house has six attributes: 2.5 bathrooms, 2.71 thousand square feet of living area, condition 3, grade 9, distance to downtown 19.67 miles, and age 1 year. 
    The table presents Comparable 1 alongside the subject. Adjustment subcolumns trace the step-by-step process of aligning Comparable 1’s attributes with those of the subject. For example, Comparable 1 has 2.81 thousand square feet of living area compared to the subject’s 2.71, producing a negative adjustment of –11.3K dollars. Its distance to downtown is 20.83 miles, farther than the subject’s 19.67 miles, yielding a positive adjustment of +8.0K. Age also differs, with Comparable 1 being 6 years old compared to 1 year for the subject, producing a negative adjustment of –17.6K. 
    The Actual Price row shows 434.1K dollars for the subject and 455K for Comparable 1. After adjustments, Comparable 1’s price moves stepwise across intermediate columns (434.1K, 443.7K, 426.1K) until it converges with the subject at 434.1K. 
    The AI Prediction row shows values of 437.3K for the subject and 440.9K for Comparable 1. The prediction error row highlights that Comparable 1’s AI prediction is 3.1\% lower than its actual price. 
    At the bottom, instructions ask participants to indicate a confidence range for the subject’s actual price using a slider. The text explains that a narrower range reflects higher confidence while a wider range reflects lower confidence. The slider is shown with the currently selected range spanning 200K to 1000K. 
}
    \label{fig:trace_practice}
\end{figure*}

\begin{figure*}[t]
    \centering
    \includegraphics[scale=0.5]{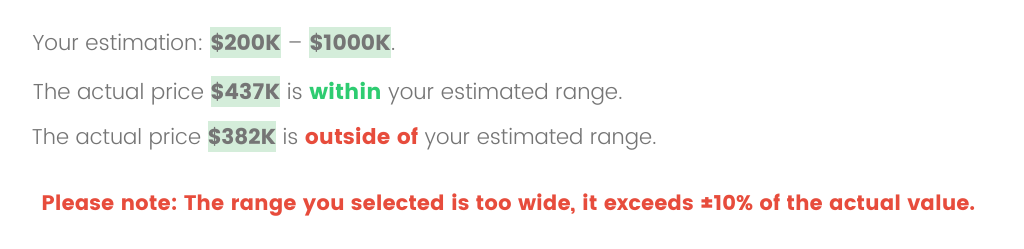}
    \caption{Practice session: feedback screen after each case. The participant’s estimated price range is compared with actual values. The system highlights when values are within or outside the range and warns if the selected range is too wide (exceeding $\pm$10\% of the actual price).}
    \Description{
    This figure shows the feedback screen following a practice task in the House Price domain. The participant’s estimated range for the subject’s actual price is displayed as 200K to 1000K. Below, the system provides evaluation feedback. 
    The actual price of 437K is reported to be within the participant’s estimated range, highlighted in green with the word “within.” A second reference value of 382K is shown to be outside the estimated range, highlighted in red with the word “outside.” 
    At the bottom, a message in bold red text warns that the selected range is too wide, exceeding plus or minus 10\% of the actual value. 
    }
    \label{fig:ans}
\end{figure*}

\begin{figure*}[t]
    \centering
    \includegraphics[scale=0.9]{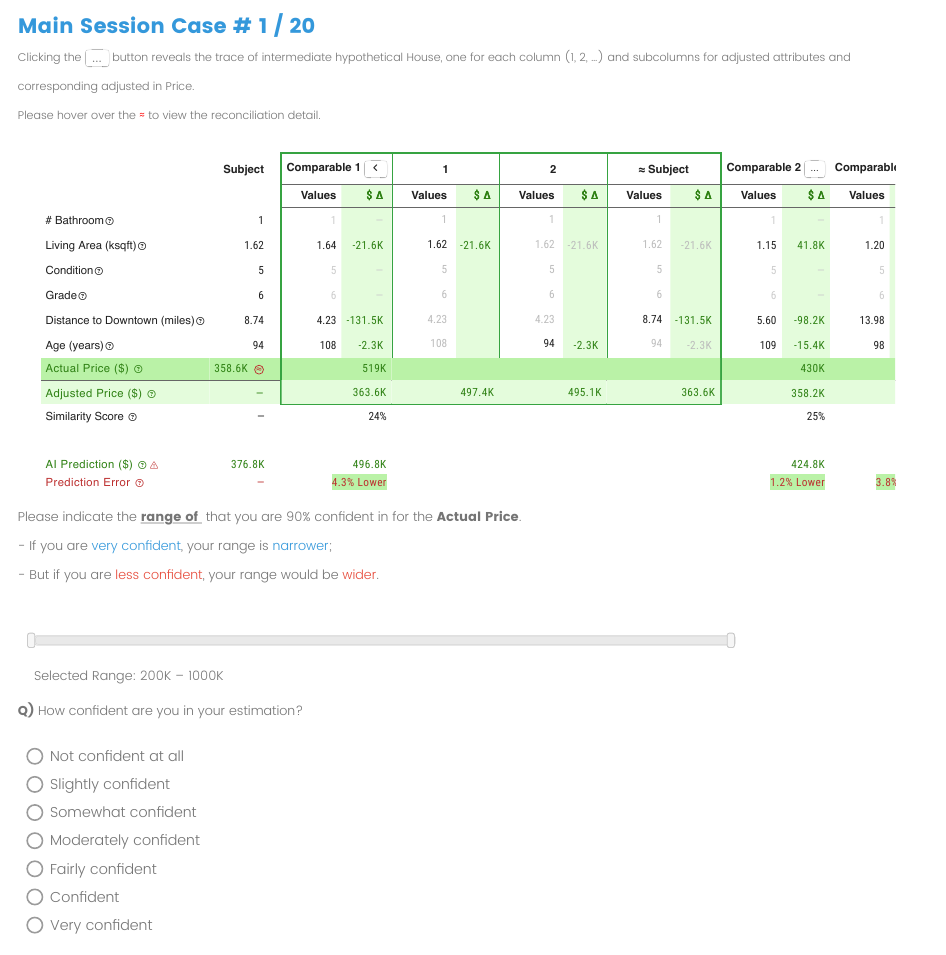}
    \caption{Main session. The UI varies with the experimental condition. The participant is asked to input their response interval and indicate their confidence level. }
    \Description{
    This figure shows a main session case screen for the House Price task using Trace Adjustments. The interface displays the subject house and three comparables, each shown with intermediate trace columns where attribute values are progressively adjusted to match the subject. Each step represents a hypothetical intermediate house and the corresponding price update. 
    The subject house has 1 bathroom, 1.62 thousand square feet of living area, condition 5, grade 6, distance to downtown 8.74 miles, and age 94 years. Comparables differ across these attributes, and the trace columns record incremental adjustments. For example, changing the distance to downtown from 4.23 miles to 8.74 miles produces a –131.5K adjustment, while reducing age from 109 to 94 years produces a –15.4K adjustment. Adjustments accumulate step by step, with adjusted prices displayed in the table. 
    Rows below the attributes list actual prices, adjusted prices, similarity scores, AI predictions, and prediction errors. AI predictions are shown for each comparable, with errors such as 4.3\% lower, 1.2\% lower, and 3\% higher than actual prices. The reconciled subject price is indicated with the symbol (\approx), marking it as an estimate. 
    At the bottom, participants are prompted to indicate a 90\% confidence range for the subject’s actual price. The instruction notes that narrower ranges reflect greater confidence and wider ranges reflect less confidence. A scale of response options from “Not confident at all” to “Very confident” is displayed, allowing participants to report their confidence level. 
    }
    \label{fig:trace_main}
\end{figure*}

\begin{figure*}[t]
    \centering
    \includegraphics[scale=0.3]{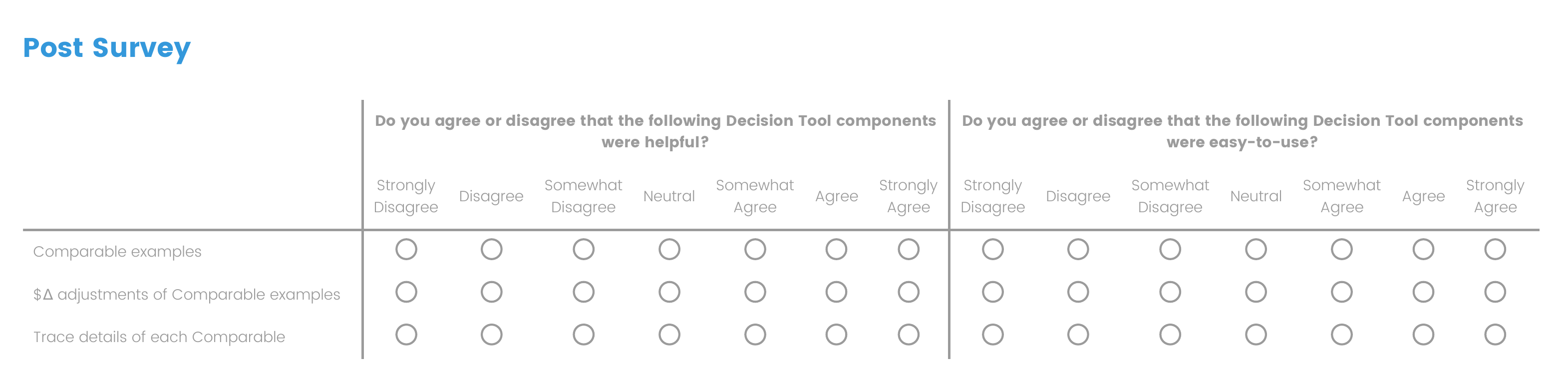}
    \caption{Post-survey subjective ratings of the helpfulness and ease of use of different components.}
    \Description{
    This figure shows the post-survey interface where participants evaluate components of the Decision Tool. The table lists three components in rows: Comparable examples, dollar adjustments of comparable examples, and trace details of each comparable. 
    Two sets of questions form the columns. The first set asks whether participants agree or disagree that the components were helpful. The second set asks whether participants agree or disagree that the components were easy to use. 
    For each question, a seven-point Likert scale is provided with options: Strongly Disagree, Disagree, Somewhat Disagree, Neutral, Somewhat Agree, Agree, and Strongly Agree. Each cell contains a circular radio button to record the response. }
    \label{fig:post_survey}
\end{figure*}

Figures~\ref{fig:survey_application}--\ref{fig:post_survey} depict the Qualtrics survey with each of the four XAI types for the summative user study.
\label{sec:survey}
\label{sec:application_scenario}

\end{document}